\documentclass[12pt]{article}
\usepackage{setspace}
\usepackage{amsmath}
\usepackage[top=1in,bottom=1in,right=1in,left=0.5in]{geometry}
\usepackage{caption}
\usepackage{psfrag}
\usepackage{graphicx}
\usepackage{hyperref}
\usepackage{subfigure}
\usepackage{authblk}
\usepackage{verbatim}
\usepackage{float}
\usepackage{bibcheck}
\usepackage[natbib=true,sorting=none,style=nature]{biblatex}
\title{Hydrodynamical properties of back reacted thermal plasma with finite 't Hooft coupling correction}
\author{Rishi Pokhrel\thanks{E-Mail: rishipokhrel.smit@gmail.com and rishi\_20211037@smit.smu.edu.in} }
\author{Karma P. Sherpa\thanks{E-Mail:sherpa.karma.pincho@gmail.com and karma\_202310028@smit.smu.edu.in} }
\author{Indra K. P. Chettri\thanks{E-Mail:indrapandey.smit@gmail.com and indra\_202410054@smit.smu.edu.in} }
\author{Tanay K. Dey\thanks{E-mail: tanay.dey@gmail.com and tanay.d@smit.smu.edu.in}}
\affil{Department of Physics, Sikkim Manipal Institute of Technology, Sikkim Manipal University, Majitar, Sikkim-737136, India.}

\date{}
\begin{document}
	\maketitle

	%%------------------ABSTRACT------------%%
\begin{abstract}
In this work, holographic approach has been used to analyse the hydrodynamical properties of $\mathcal{N}=4$ Super Yang-Mills thermal plasma with finite 't Hooft coupling correction and flavour quarks dual to the AdS Gauss-Bonnet gravity in the presence of string cloud. The drag exerted on an external probe quark while translating through the thermal plasma enhances with probe velocity, flavour quark density, temperature and finite coupling correction. The jet quenching parameter gets enhanced on increase of flavour quark density, temperature and finite coupling correction.  Quark-antiquark screening length is observed to reduce with rapidity parameter, flavour quark density, finite coupling correction and temperature which suggests an early transition to the thermal plasma phase of QGP. The screening length is found to be larger for the parallel orientation compared to the perpendicular configuration. Finally, the rotational dynamics of the heavy probe quark is studied. The rotational and drag energy loss increases with angular frequency and quark density, but is almost independent of the 't Hooft coupling.  
 
\end{abstract}

\clearpage

%%%%%%%%%%%%%%%%%%%%%%%%%%%%%%%%%%%%%%%%%%%%%%%%%%%%%%%%%%%%%%%%%%%%%%%%%%%%%%%%%%%%%%%%%%%%%%%%%%%%%%%%%%%%%%%%%%%%%%%%%

\section{Introduction}\label{sec_intro}
The AdS/CFT duality, describes a correspondence between a weakly coupled gravity theory in $(n+1) $ dimensional AdS spacetime and a strongly coupled Conformal Field Theory (CFT) on the $n$ dimensional boundary of the AdS spacetime \cite{Maldacena1998,Maldacena1999,Witten1998}. The AdS/CFT correspondence has been used to study various properties of strongly coupled gauge theory dual to the weakly coupled gravity theory. The creation of the deconfined state of quarks and gluons known as the Quark-Gluon Plasma (QGP) is suggested by the Large Hadron Collider (LHC) and Relativistic Heavy Ion Collider (RHIC) \cite{Zajc_2008,muller2007,Shuryak_2007,d_Enterria_2007, salgado2006, Shuryak_2007_report,Tannenbaum_2006,Muller_2006,Gyulassy_2005,Adcox_2005,Arsene_2005,Back_2005,Adams_2005}. The interaction between QGP and an external probe quark is considered to be a strongly coupled system \cite{Baier_1997,Eskola_2005}. Through the conventional field theoretic approach, the analysis of different properties such as the drag force, jet quenching, screening length, radial profile and energy loss etc. associated with a probe quark is difficult in the strongly coupled regime. Whereas the AdS/CFT correspondence provides a holographic approach to study the $n$ dimensional strongly coupled systems if the weakly coupled dual $(n+1)$ dimensional gravity description is known. Using the AdS/CFT duality, an external probe quark in the boundary gauge theory is introduced by considering a string in the dual bulk AdS spacetime with its end point attached to the boundary. The string's body is suspended in the AdS bulk which corresponds to the gluonic field between the quark-antiquark pair($q\bar{q}$). The motion of the endpoint restricts the body of the probe string to lag behind, generating a drag force. The drag exerted on an external probe quark has been discussed in \cite{Herzog_2006,Gubser_2006,Caceres2006} and further generalisation is presented in \cite{Casalderrey_Solana_2006,Matsuo_2006,Talavera_2007,Nakano_2007,Chen_2024}. The jet quenching parameter($\hat{q}$) is calculated using the approach of \cite{Liu:2006ug} for various systems \cite{Caceres2006a,Nakano_2007,Bertoldi_2007,Cotrone_2007,Bigazzi_2009,Bigazzi_2011,Chen_2024}. The screening length of moving heavy probe meson has been studied in \cite{Chernicoff_2006,Chernicoff2008,Liu_2007a,Zhu_2024}. Considering rotating probe quark, the radial profile and energy loss of the corresponding probe string has been studied in \cite{Fadafan_2009,Athanasiou_2010,Herzog_2006,Herzog_2007,Mikhailov2003,Hou_2021}. In \cite{Fadafan_2008,Atashi_2020}, hot plasma at finite coupling is considered to study the drag force and energy loss. 

In the above works, back reaction effect due to the presence of large number of flavour quarks in the thermal plasma is not considered. Following bottom up approach of \cite{DeWolfe:2010he}, a holographic gravity model dual to the thermal plasma with large number of external heavy quarks (flavour quarks say charm) has been considered in \cite{Chakrabortty2011a}. The flavour quarks are introduced in the thermal plasma by considering a cloud of infinitely long strings elongated from boundary to center/horizon of the spacetime. The flavour quarks in the gauge theory is represented by the attached end points of the strings hanging from the boundary. Due to the presence of the large number of infinitely long strings in the bulk, the spacetime is back reacted and deformed from AdS configuration. But asymptotically it still provides AdS configuration. The dual gauge theory is named as back reacted $\mathcal{N}=4$ SYM thermal plasma. Using holographic technique, the drag force exerted on the probe quark in the dual gauge theory is also studied by introducing a probe string in the background of AdS spacetime with string cloud. Further in \cite{Chakrabortty2016a},  for the back reacted $\mathcal{N}=4$ SYM thermal plasma theory, the  energy loss, the jet quenching parameter and the screening length has been studied. In \cite{Pokhrel_2025}, we considered thermal plasma with baryon and flavour quark density dual to charged AdS with cloud of string background \cite{Pokhrel2023} and studied the hydrodynamical properties in terms of flavour quark density, potential and temperature.
  
In this work, finite correction in the 't Hooft  coupling  for back reacted $\mathcal{N}=4$ SYM thermal plasma theory is considered. The higher derivative Gauss-Bonnet(GB) term in the bulk theory is the gravity dual of the finite correction in the 't Hooft  coupling. In \cite{Dey:2020yzl}, the AdS background with higher derivative correction in presence of string cloud is studied. For this background, the famous $\frac{\eta}{s}$ ratio concerning shear viscosity and entropy density is calculated in \cite{Sadeghi:2022kgi}. The ratio becomes less than the Kovtun-Son-Starinets(KSS) bound $\frac{1}{4\pi}$ \cite{Kovtun:2004de,Son:2007vk} which is supported by the recent RHIC  data \cite{Romatschke:2007mq,Song:2007fn,PHENIX:2006iih,Romatschke:2007eb,Dusling:2007gi}. This indicates the strongly coupled QGP can be treated as perfect fluid with very small viscosity relative to the entropy density and getting closer to the heavy ion collisions. In this work, we discuss the range of the GB coefficient and string density for which the ratio to be greater than zero. Then we study the hydrodynamical properties of the strongly coupled back reacted finite temperature QGP with finite 't Hooft coupling correction dual to the AdS gravity with higher derivative correction in presence of string cloud. The drag force has been studied and it has been observed that the drag force enhances with the increasing values of temperature, flavour quark density,  GB coefficient and probe quark velocity.

The $q\bar{q}$ pair separation distance has been studied with the consideration of the probe $q\bar{q}$ pair moving in the back reacted thermal plasma with finite correction in the 't Hooft  coupling. The separation distance has been analysed for both perpendicular and parallel configurations of the $q\bar{q}$ pair axis. The separation distance between the probe $q\bar{q}$ pair initially increases, reaches a maximum value (which is termed as the screening length for which the $q\bar{q}$ pair are in stable bound state configuration) and then falls off with respect to the constant of the equation of motion. Beyond the screening length, the $q\bar{q}$ pair gets separated or are screened in the plasma medium with no binding energy. The screening length reduces with the enhancement of rapidity parameter, flavour quark density, GB coefficient and temperature for both perpendicular and parallel orientations. Comparing the $q\bar{q}$ pair separation distance for both the configurations, the screening length is observed to be larger for parallel orientation than perpendicular orientation of the $q\bar{q}$ pair, signifying stable bound state configuration of the $q\bar{q}$ pair in the parallel orientation than the perpendicular configuration.

The jet quenching parameter which is related to the suppression of high transverse momentum of probe quark in the thermal medium and its energy loss has been studied as a function of GB coefficient, temperature and flavour quark density. The quenching parameter enhances with the increase of any mentioned parameter, suggesting increment in energy loss as a result of high transverse momentum suppression.

The radial profile and energy loss of a rotating heavy probe quark against its velocity for different set of parameters has been studied. The radial profile remains constant from the boundary to the horizon if the angular frequency is less than unity. Otherwise, it goes up which indicates larger radius towards the horizon than at the boundary. Furthermore, the radial profile reduces with the increase in string density. But it is almost independent on the GB coefficient. The rotational energy loss increases with increase in string density and angular frequency. Whereas, it slightly decreases with the increase of GB coefficient. The qualitative nature of drag energy loss is similar to that of rotational loss. The vacuum energy loss depends only on the probe quark's velocity and angular frequency. The rotational to vacuum energy loss ratio indicates that for low probe velocity rotational loss dominates and for high probe velocity the rotational and vacuum loss tends to be equal and the ratio approaches towards unity suggesting a destructive interference between the two radiations. The rotational dominance is high for higher value of GB coefficient, string density and low for higher value of angular frequency. Similar conclusion is observed for the ratio of the drag to the vacuum energy loss. Whereas, the rotational to the drag energy loss ratio tends toward unity for smaller value of probe velocity. It increases and becomes more than unity when the probe velocity reaches toward the speed of light, indicating a constructive interference between rotational and drag energy loss.

This work is organised as follows: in section \ref{sec_setup}, we summarise gravity dual to the back reacted thermal plasma with finite correction in 't Hooft  coupling. Then the drag force is analysed in section \ref{sec_drag_force}.  In section \ref{sec_screening_length}, we discussed the screening length of the $q\bar{q}$ pair. In section \ref{sec_jet_quenching}, we study the jet quenching parameter. In section \ref{sec_radial_profile}, the rotating probe string's radial profile has been analysed while the corresponding energy loss has been discussed in section \ref{sec_energy_loss}. Finally, we summarise and discuss our work in section \ref{sec_conclusion}.
%%%%%%%%%%%%%%%%%%%%%%%%%%%%%%%%%%%%%%%%%%%%%%%%%%%%%%%%%%%%%%%%%%%%%%%%%%%%%%%%%%%%%%%%%%%%%%%%%%%%%%%%%%%%%%%%%%%%%%%%%%%%%%%%%%%%%%%%%%%%%%%%%%%%%%%%%%%%%%%%%%%%%%%%%%%%%%%%%%%%%%%%%%%%%

%%%%%%%%%%%%%%%%%%%%%%%%%%%%%%%%%%%%%%%%%%%%%%%%%%%%%%%%%%%%%%%%%%%%%%%%%%%%%%%%%%%%%%%%%%%%%%%%%%%%%%%%%%%%%%%%%%%%%%%%%%%%%%%%%%%%%%%%%%%%%%%%%%%%%%%%%%%%%%%%%%%%%%%%%%%%%%%%%%%%%%%%%%%%%
\section{Dual Gravity}
\label{sec_setup}
\begin{figure}[!h]
\centering
\includegraphics[width=0.4\linewidth]{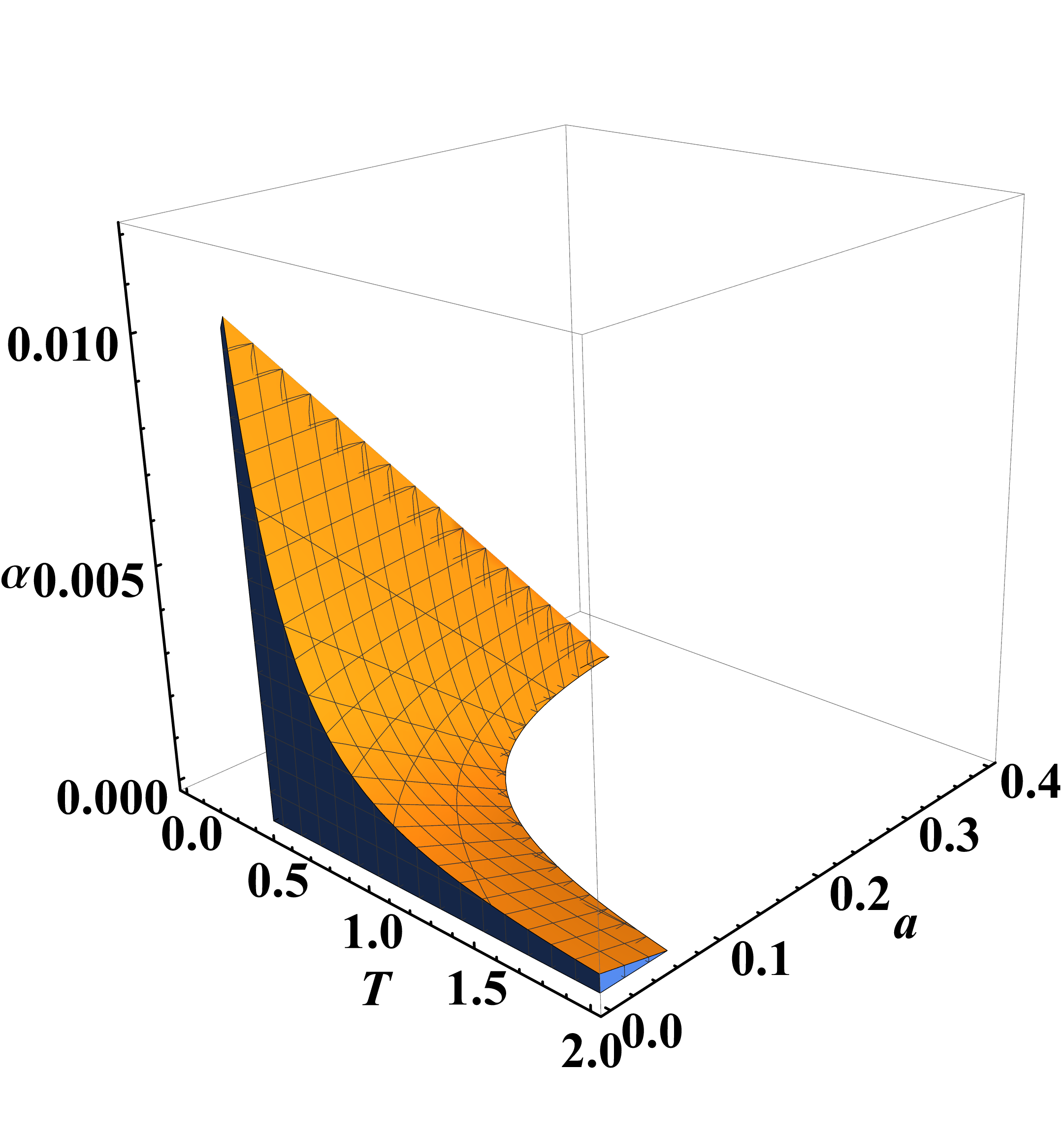}
\caption{Plot of forbidden region where $\frac{\eta}{s}$ is not positive in $T,\,a,\,\alpha $ space.}
\label{region3D}
\end{figure}

We introduce the $5$ dimensional bulk gravity dual to the back reacted thermal plasma with finite correction in 't Hooft  coupling by the following action  \cite{Dey:2020yzl},
\begin{equation}
	S = -\frac{1}{16 \pi G_{5}} \int_M d x^{5} \sqrt{-g} \left(R - 2 \Lambda + \alpha L_{GB}\right) + S_{SC},
	\label{eq_total_bulk_action}
\end{equation} 
where $g_{\mu\nu}$, $G_5$, $R$, $\Lambda$ and $L_{GB}$ are spacetime metric tensor, gravitational constant, Ricci scalar, cosmological constant and Gauss-Bonnet term respectively.  The GB term is given as, 
\begin{equation}
	L_{GB} = R^2 - 4R_{\mu\nu}R^{\mu\nu} + R_{\mu\nu\rho\sigma}R^{\mu\nu\rho\sigma}.
	\label{eq_L_GB}
\end{equation}
The finite correction in the boundary gauge theory is incorporated in gravity by the GB coupling $\alpha$. In this work, $\alpha$ is considered to be very small so that the $R^4$ term in the action can be neglected. $S_{SC}$ is the action due to the presence of cloud of strings which are elongated from the boundary to the centre/horizon of the bulk. The end points of these strings represent the quarks/antiquarks in the boundary gauge theory. The action is defined in the following way, 
\begin{equation}
	S_{SC} = -\frac{1}{2}\sum_i T_i\int d^2\xi \sqrt{-\gamma} \gamma^{pq} \partial_p X^\mu \partial_q X^\nu g_{\mu\nu}.
	\label{eq_action_SC_term}
\end{equation}
Here, $\gamma^{pq}$ is the induced metric on the string world sheet and $p, q$ represent the world sheet coordinates. $T_i$ is the tension of the $i'$th string hanging from the boundary to the center or horizon of the spacetime. 
The spacetime metric solution of the action (\ref{eq_total_bulk_action}) can be represented as,
\begin{equation}
	ds^2 = -V(r)dt^2 + \frac{dr^2}{V(r)} + r^2 d\Omega_{3}^2,
	\label{eq_metric_ansatz}
\end{equation}
where $d\Omega_{3}^2$ is the line element of the unit $3-$sphere and $V(r)$ is,
\begin{equation}\label{eq_vr}
	V(r) = 	1 + \frac{r^2}{4 \alpha } \left(1-\sqrt{1+\frac{32 \alpha  m}{r^4}-\frac{8 \alpha }{l^2}+\frac{16 a \alpha }{3 r^3}}\right).
\end{equation}
Here, $l$ is the AdS radius connected to cosmological constant as $\Lambda = -\frac{6}{l^2}$, $m$ is the constant of integration related to the mass of the black hole and $a$ is the string density defined as,
\begin{equation}
	a(x) = T\sum_i\delta_i^3(x-X)  \quad \quad  {\rm with} \quad \quad a\ge 0.
\end{equation}
Notice that the metric $V(r)$ becomes real if the GB coupling belongs to the range $ 0\le\alpha \le \frac{l^2}{8}$.
Using the parametrisation, $r=\frac{l^2}{u}$, the metric solution in terms of $u$ coordinate can be written as,
\begin{equation}
	V(u) = 1+\frac{l^4}{4 \alpha  u^2}	-\frac{l^4}{4 \alpha  u^2} \sqrt{1+\frac{32 \alpha  m u^4}{l^8}-\frac{8 \alpha }{l^2}+\frac{16 a \alpha  u^3}{3 l^6}} = f(u) h(u),
	\label{eq_metric_solution_V1u}
\end{equation}
where $f(u)$ and $h(u)$ are given by,
\begin{equation}
	f(u) = \frac{l^2}{u^2},
	\label{eq_f1u}
\end{equation}
and
\begin{equation}\label{hu}
	h(u) = \frac{u^2}{l^2}+\frac{l^2}{4 \alpha }-\frac{l^2}{4 \alpha } \sqrt{1+\frac{32 \alpha  m u^4}{l^8}-\frac{8 \alpha }{l^2}+\frac{16 a \alpha  u^3}{3 l^6}}.
\end{equation}
Further, the mass parameter $m$ can be calculated using the fact that $V(u=u_h) = 0$, where $u_h$ is the horizon radius and the mass parameter $m$ is obtained as,
\begin{equation}
	m = \frac{3 l^6 + 3 l^4 {u_h}^2 +6 \alpha  {u_h}^4 - 2 a l^2 {u_h}^3}{12 {u_h}^4}.
\end{equation}
The black hole temperature is calculated and obtained as,
\begin{equation}
	T = \frac{ 6 l^4 + 3 l^2 {u_h}^2- a {u_h}^3 }{6 \pi  l^4 {u_h}+24 \pi  \alpha  {u_h}^3}.
	\label{eq_T1_temp}
\end{equation}
The viscosity to entropy density ratio has been calculated for this background in \cite{Sadeghi:2022kgi}, which can be written as,
\begin{equation}
	\frac{\eta}{s} = \frac{1}{4\pi}(1-\frac{8\alpha}{l^2}-\frac{4 \alpha a u_h^3}{3 l^6}).
\end{equation}
The ratio is less than the KSS bound $\frac{1}{4\pi}$ which indicates QGP approaching towards perfect fluid with very small viscosity relative to the entropy density. The ratio will be greater than or equal to zero $(\frac{\eta}{s}\ge 0)$, if the following constraint on the GB coefficient, string density and horizon radius are satisfied,
\begin{gather}
4\alpha a  u_h^3 + 24 \alpha l^4 -3l^6 \le 0.
\end{gather}
The discriminant of this cubic equation $ \Delta = -3888\alpha^2 a^2 l^8(8\alpha -l^2)^2$ is negative which leads to have only one black hole solution in the bulk theory. In \cite{Dey:2020yzl}, one or three black hole regions are discussed in details in $T,\,a,\,\alpha$ space. Here, in figure \ref{region3D}, we have shown the forbidden region for one black hole solution. There is a lower cut off on the GB coupling as well as string density for a fixed temperature and it reduces once the temperature increases.
\\

Now, for this set up, we study the hydrodynamical properties of the back reacted thermal plasma with finite 't Hooft coupling correction in the permitted region of $T,\,a,\,\alpha$ space. 
%%%%%%%%%%%%%%%%%%%%%%%%%%%%%%%%%%%%%%%%%%%%%%%%%%%%%%%%%%%%%%%%%%%%%%%%%%%%%%%%%%%%%%%%%%%%%%%%%%%%%%%%%%%%%%%%%%%%%%%%%%%%%%%%%%

%%%%%%%%%%%%%%%%%%%%%%%%%%%%%%%%%%%%%%%%%%%%%%%%%%%%%%%%%%%%%%%%%%%%%%%%%%%%%%%%%%%%%%%%%%%%%%%%%%%%%%%%%%%%%%%%%%%%%%%%%%%%%%%%%%%
\section{Drag Force}\label{sec_drag_force}
\begin{figure}[!h]
	\centering
	\subfigure[]{\includegraphics[width=0.32\linewidth]{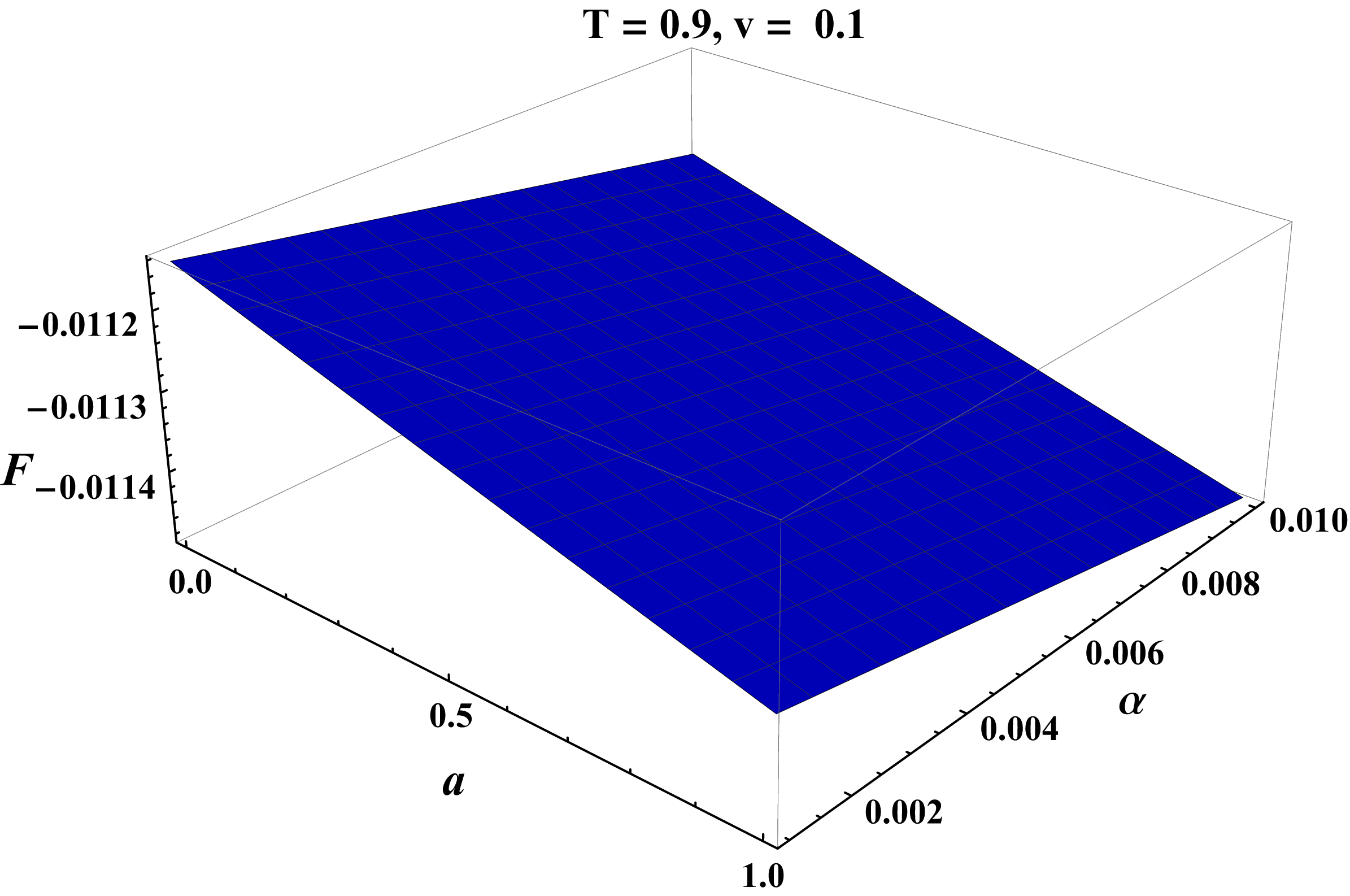}}
	\subfigure[]{\includegraphics[width=0.32\linewidth]{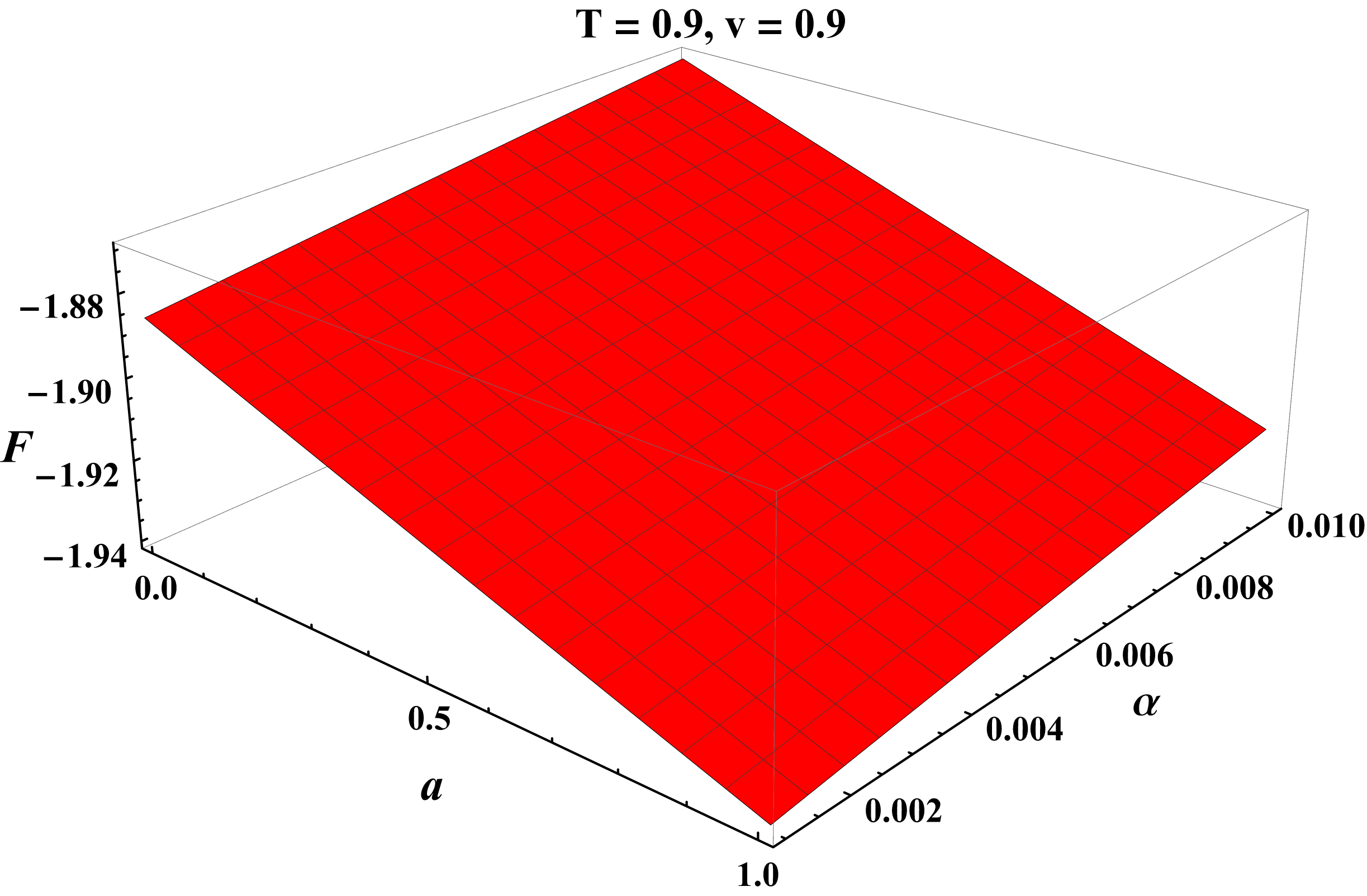}}
	\subfigure[]{\includegraphics[width=0.32\linewidth]{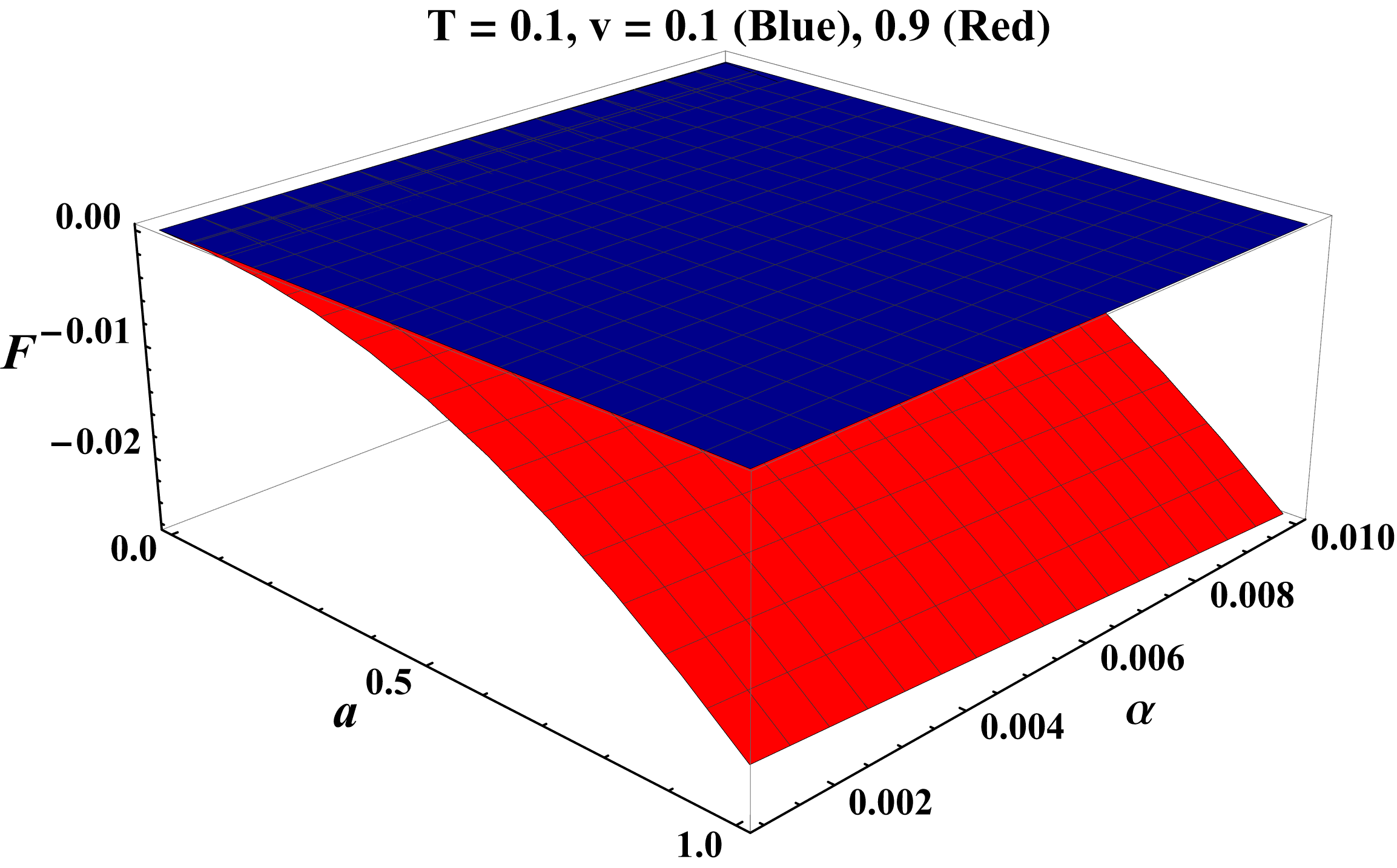}}
	\caption{Plot of drag force $F$ with respect to string density $a$ and GB coefficient $\alpha$ for different set of $T$ and velocity $v$.}
	\label{fig_drag3D}
\end{figure}

The drag force exerted on an external probe quark translating through the thermal plasma with velocity $v$ dual to AdS GB black hole with string cloud background is computed in this section. The external quark is realised by an external probe string in the AdS bulk, having its one end point attached to the boundary and the other one hanging up to the horizon of the background spacetime. The probe string's motion is represented by the standard Nambu-Goto action. The string is constrained to move with the gauge,
\begin{equation}
	X^\mu(\tau, \sigma) = (t = \tau, u = \sigma, x = vt + \xi(\sigma), y = 0, z = 0).
	\label{gaugefixing} 
	\end{equation}
Here, $\xi(\sigma)$ is related to the drag effect on the probe string along the radial direction. According to the AdS/CFT correspondence, the conjugate momentum corresponding to $\xi(u)$, flowing through the string from the boundary to the horizon amounts to the loss of momentum of the boundary probe quark. Therefore, the drag force exerted on the probe quark can be obtained from the calculation of conjugate momentum for $\xi(u)$ and it takes the following form \cite{Pokhrel_2025}\footnote{The detailed calculation of the hydrodynamical properties are given in \cite{Pokhrel_2025,Chakrabortty2016a} for the generic background metric. Therefore, instead of repeating the calculation once again here, we have directly mentioned the formulas. Interested reader may go through the calculation done there.}
\begin{equation}
	F =-\frac{\sqrt{\lambda}}{2\pi}  \left(\frac{v}{u_v^2(T,a,\alpha)}\right),
	\label{eq_drag_force}
\end{equation}
where, $\lambda$ is the 't Hooft coupling of the boundary gauge theory and $u_v(T,a,\alpha)$ is the solution of the equation,
\begin{equation}
	h(u_v) - v^2=0.
	\label{eq_huv_solution}
\end{equation}
 The drag force obtained using equation (\ref{eq_drag_force}), is plotted as a function of string density and GB coupling for different set of parameters in the figure (\ref{fig_drag3D})
\footnote{In this work, we have used the following dimensionless parameters for plotting. $$a\rightarrow \frac{a}{l},\,  u_h\rightarrow \frac{u_h}{l},\, \alpha\rightarrow \frac{\alpha}{l^2},$$ such that $$T(u_h,a,\alpha)\rightarrow T\left(\frac{u_h}{l},\frac{a}{l},\frac{\alpha}{l^2}\right).$$ The AdS radius has been set to unity.}.
 It is observed that at high temperature and velocity drag force decreases slightly with GB coupling but increases with string density. For other combination of the temperature and velocity, the drag force enhances for both the string density and GB coupling.
%%%%%%%%%%%%%%%%%%%%%%%%%%%%%%%%%%%%%%%%%%%%%%%%%%%%%%%%%%%%%%%%%%%%%%%%%%%%%%%%%%%%%%%%%%%%%%%%%%%%%%%%%%%%%%%%%%%%%%%%%%%%%%%%%%

%%%%%%%%%%%%%%%%%%%%%%%%%%%%%%%%%%%%%%%%%%%%%%%%%%%%%%%%%%%%%%%%%%%%%%%%%%%%%%%%%%%%%%%%%%%%%%%%%%%%%%%%%%%%%%%%%%%%%%%%%%%%%%%%%%
\section{Screening Length}\label{sec_screening_length}
This section deals with the study of screening length of quark-antiquark pair which is defined as the maximum separation between the pair for which they are in stable bound state configuration. The meson is realised as an external string with their endpoints attached to the boundary and the string's body hanging towards the horizon with a turning point in between. Considering the gravity dual as the AdS GB black hole spacetime with cloud of strings, we study the screening length of the $q\bar{q}$ pair in both perpendicular and parallel orientations \cite{Pokhrel_2025}. The orientations are defined based on the alignment of the axis of $q\bar{q}$ pair with its direction of motion. In order to study the screening length we consider the boost in the dual gravity as below,
\begin{equation}
	\begin{split}
	dt= cosh(\beta) dt^* - sinh(\beta) dz^*\\
	dz = -sinh(\beta) dt^* + cosh (\beta) dz^*,
	\end{split}
\end{equation}
with $\beta = tanh^{-1} v$ as the rapidity parameter which measures the relative velocity of the $q\bar{q}$ pair with respect to the surrounding thermal plasma. 
\subsection{Perpendicular:}
\begin{figure}[h]
	\centering
	\subfigure[]{\includegraphics[width=0.315\linewidth]{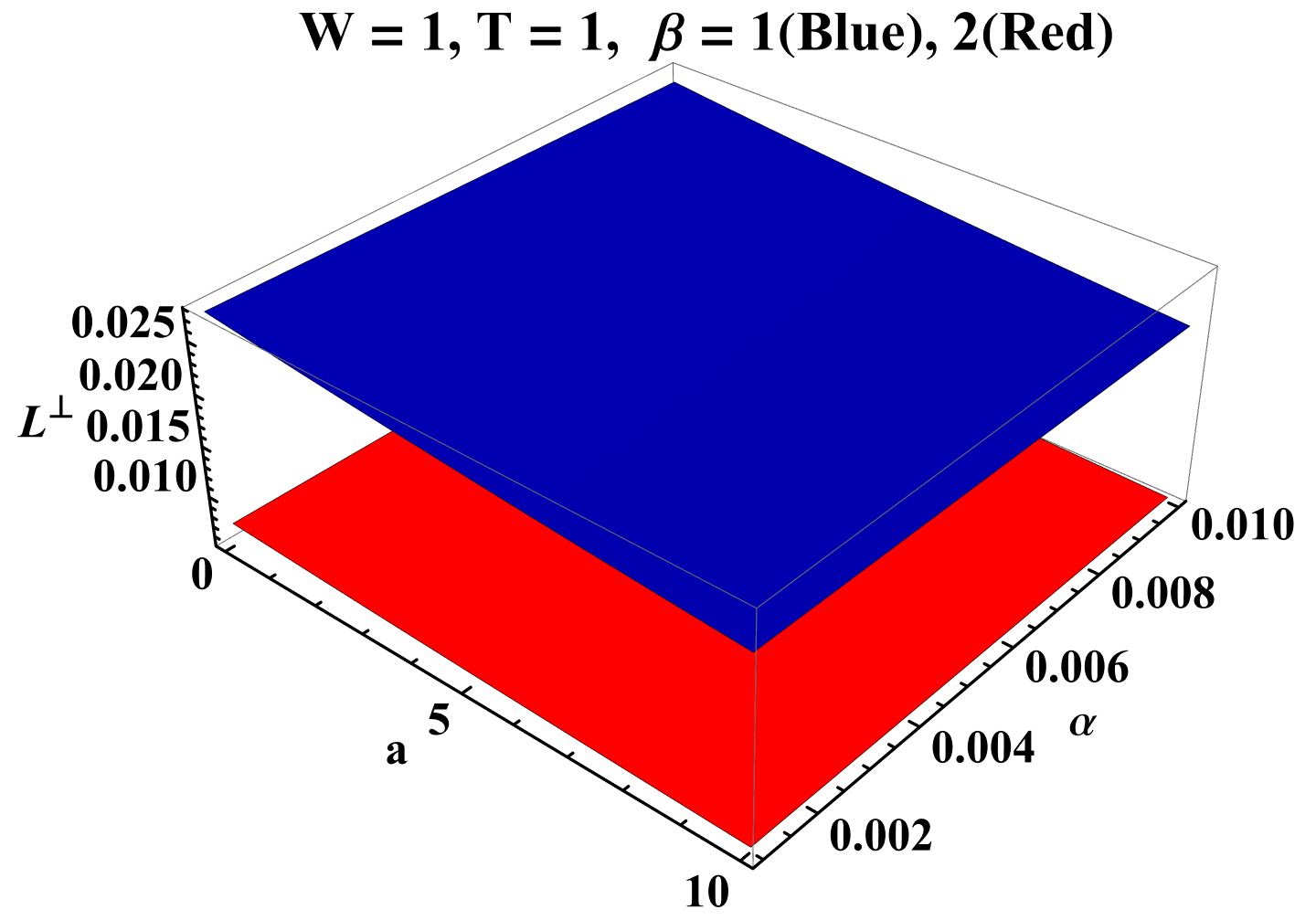}}
	\subfigure[]{\includegraphics[width=0.315\linewidth]{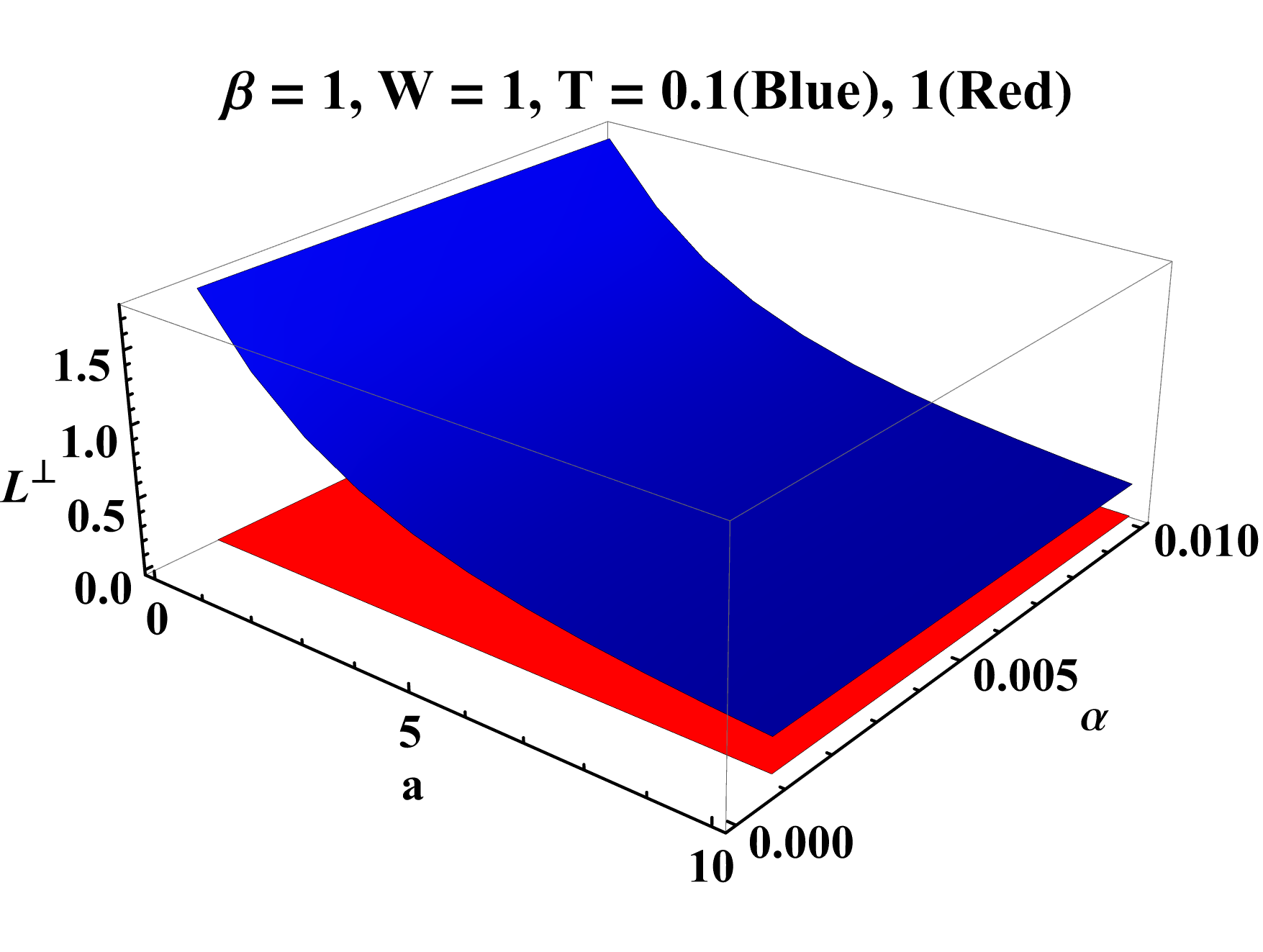}}
	\subfigure[]{\includegraphics[width=0.33\linewidth]{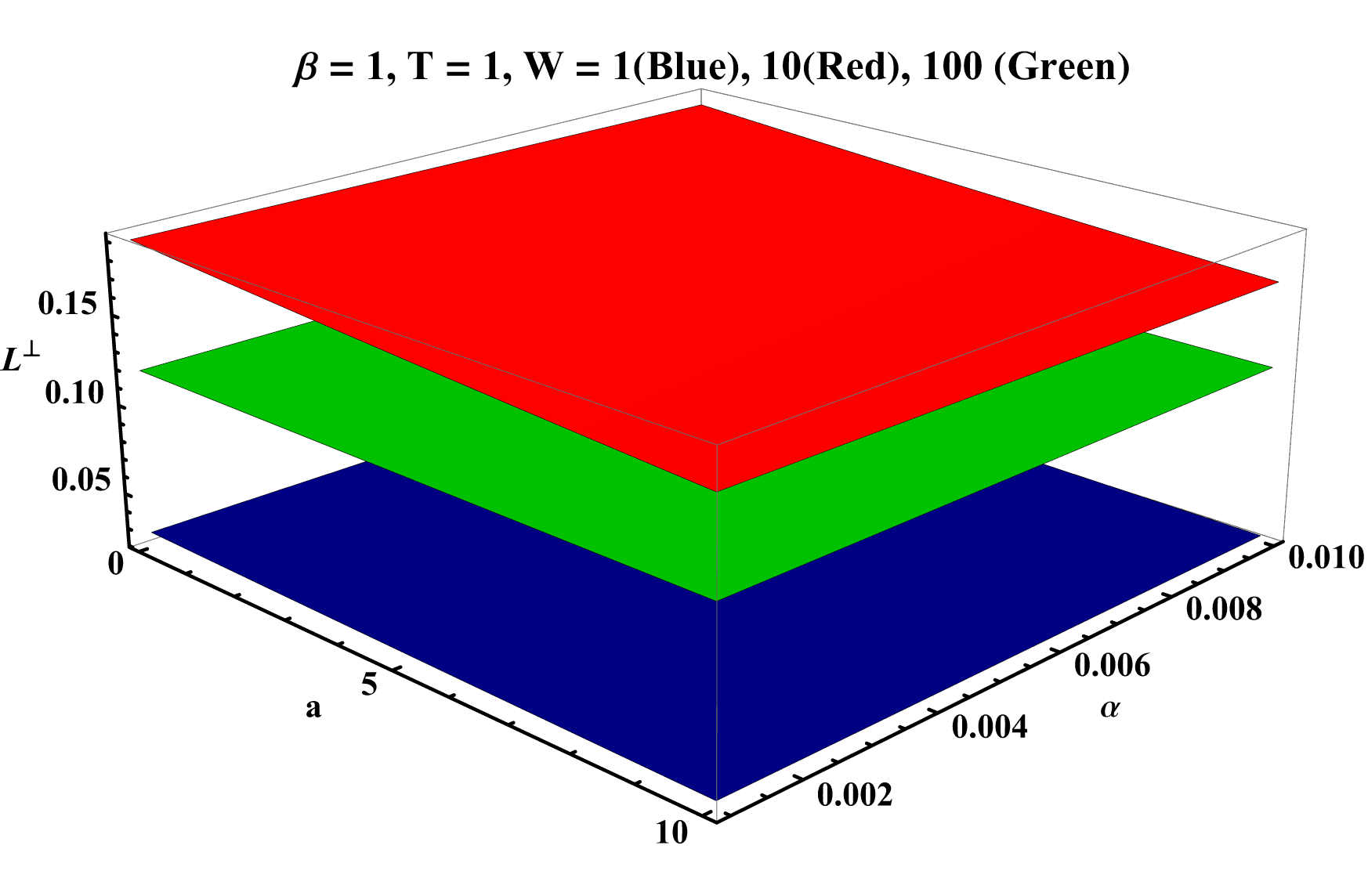}}
	\caption{$L^{\perp}$ vs $a$ and $\alpha$ for different set of values of $\beta$, $T$ and  $W$.}
	\label{fig_screening3D_perp}
\end{figure}

\begin{figure}[h]
	\centering
	\subfigure[]{\includegraphics[width=0.45\linewidth]{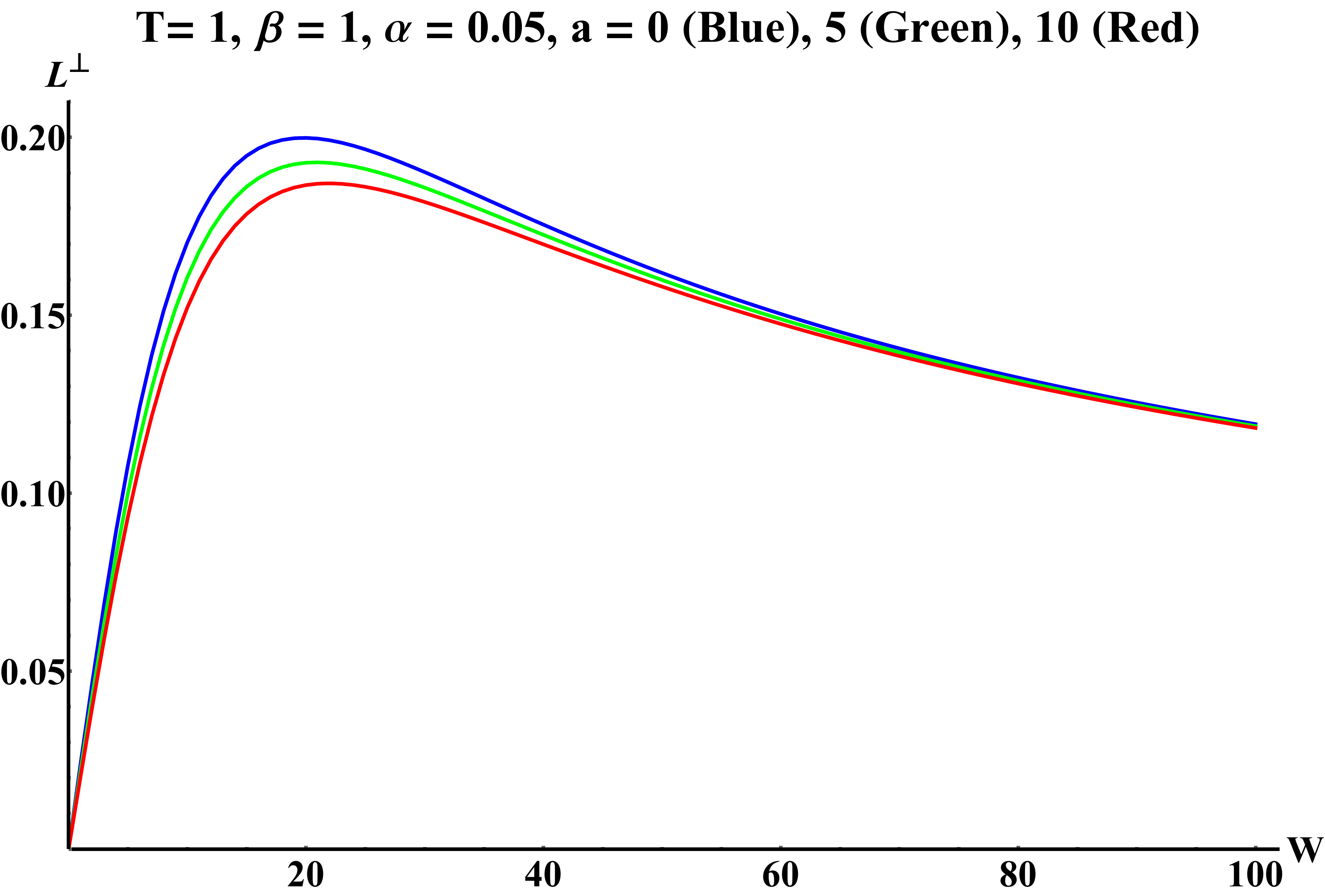}}
	\noindent\subfigure[]{\includegraphics[width=0.45\linewidth]{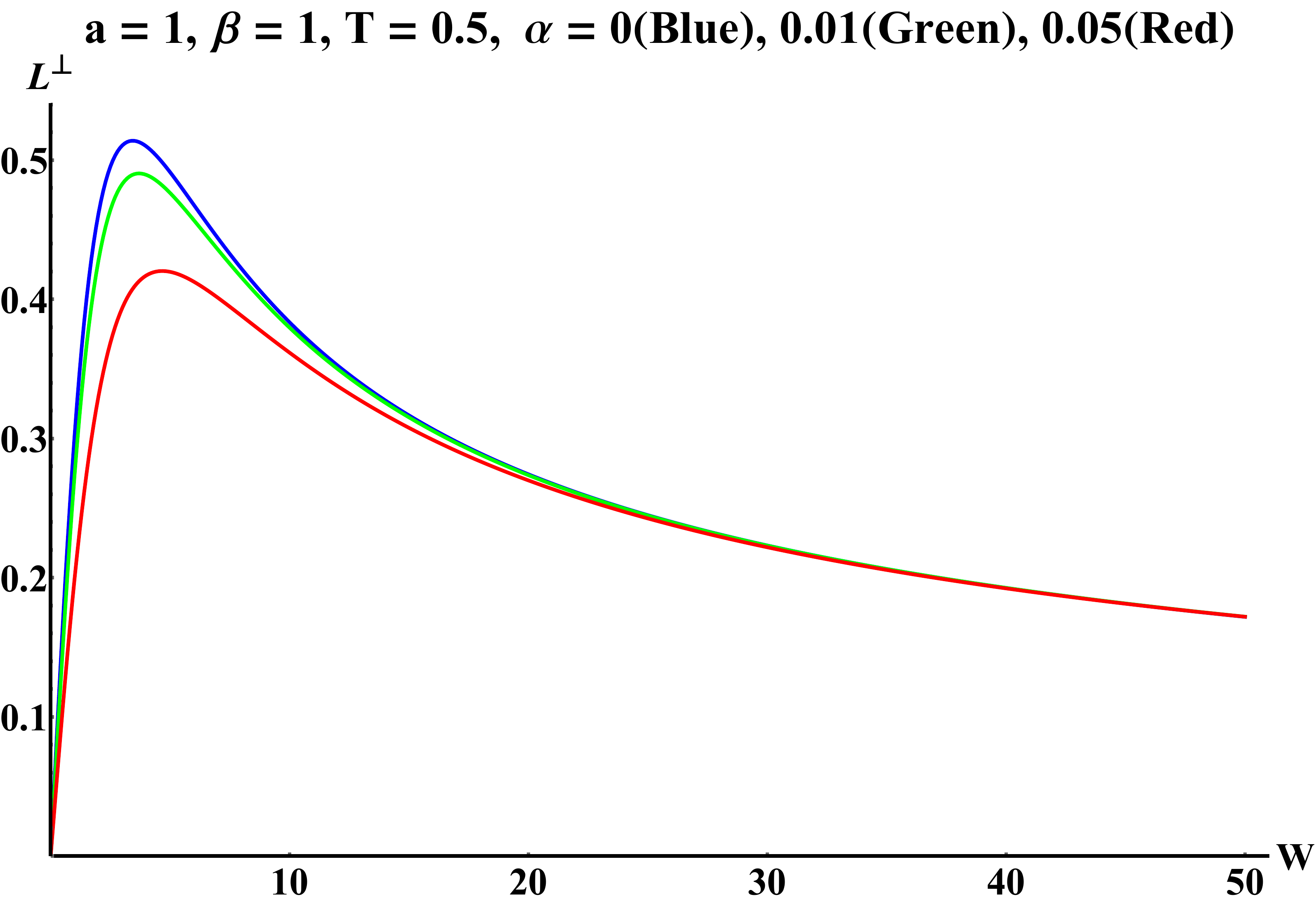}}
	\caption{$L^{\perp}$ vs $W$ for: (a) fixed parameters $T,\,\beta,\,\alpha$ and different $a$ and (b) fixed parameters $a,\,\beta,\,T$ and different $\alpha$.}
	\label{fig_L_vs_W_perp_case1}
\end{figure}

\begin{figure}[h]
	\centering
	\subfigure[]{\includegraphics[width=0.45\linewidth]{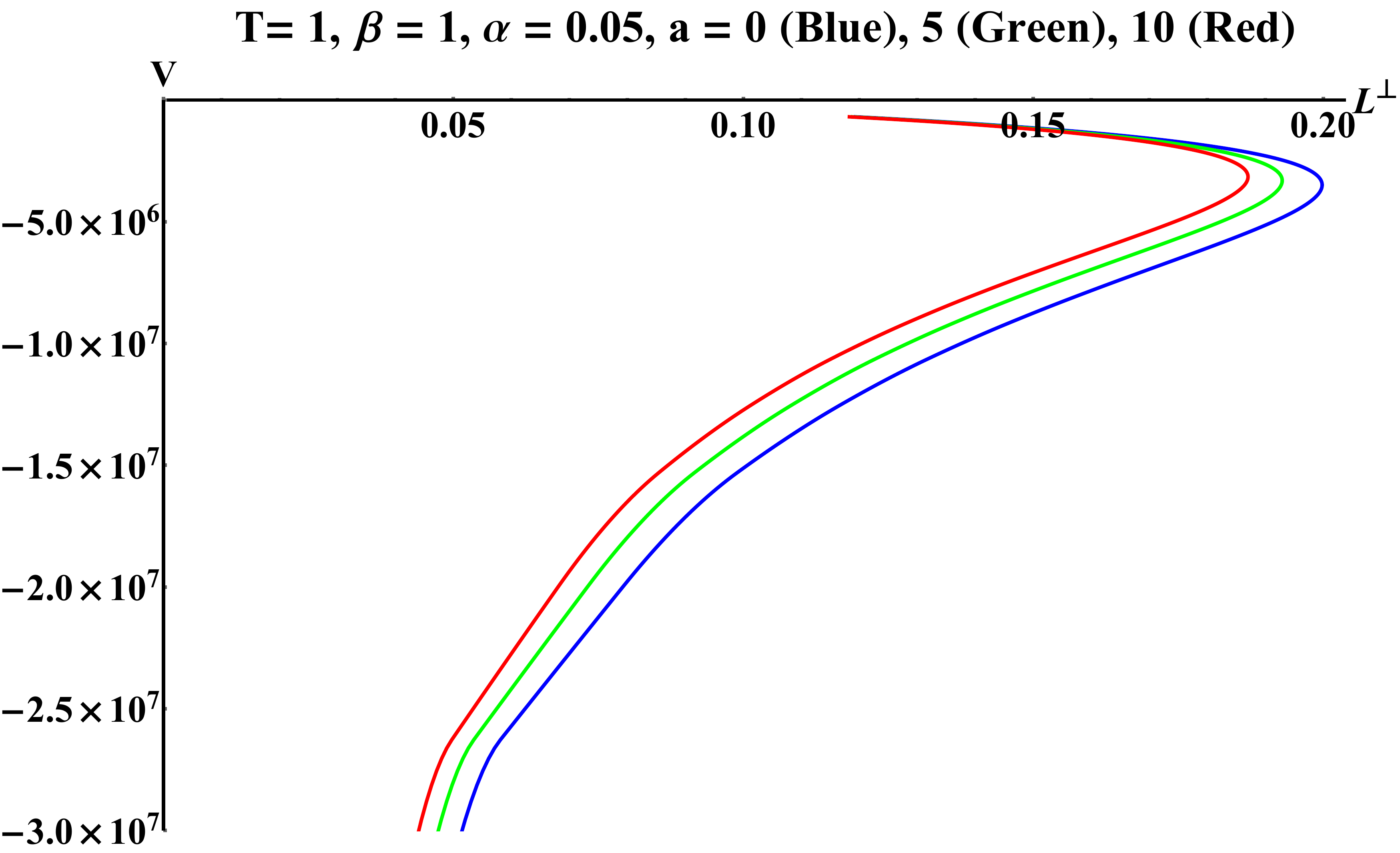}}
	\subfigure[]{\includegraphics[width=0.45\linewidth]{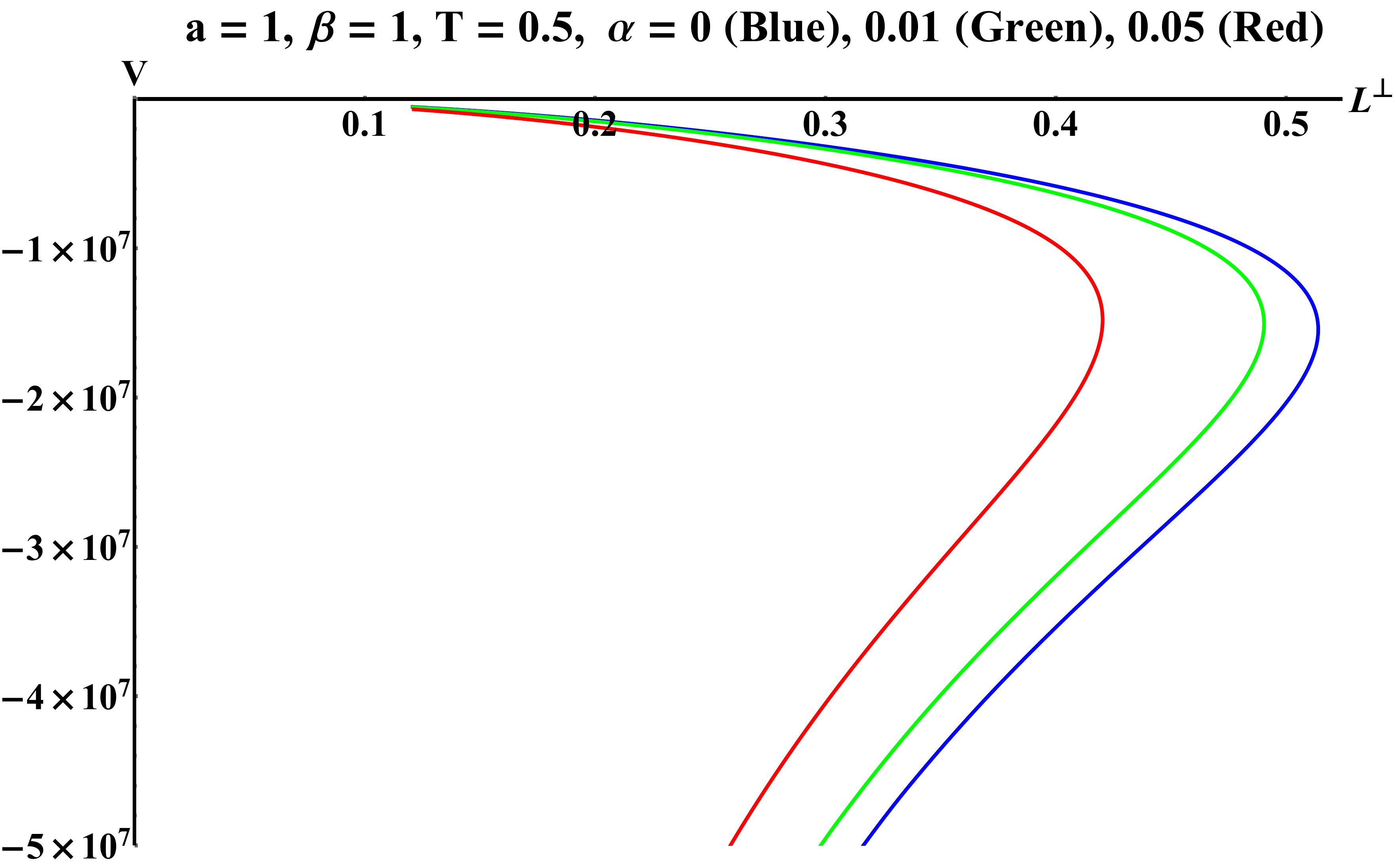}}
	\caption{$V$ vs $L^\perp$ for: (a) fixed parameters $T,\,\beta,\,\alpha$ and different $a$ and (b) fixed parameters $a,\,\beta,\,T$ and different $\alpha$.}
	\label{fig_V_vs_L_perp_case1}
\end{figure}

In this subsection, the separation between the $q\bar{q}$ pair in perpendicular orientation has been studied. The axis of the pair is perpendicular to the direction of motion of the pair. For the simplicity of the computation, we work with the static gauge $\tau = t^*,\, \sigma = x,\, y=z^*=0$ and boundary condition as, 
\begin{equation}
	u(\sigma =\pm \frac{L}{2}) = 0,\, u(\sigma =0) = u_{ext},\, u'(\sigma=0) =0.
\end{equation}
For this gauge, the world-sheet action can be written in the following form,
\begin{equation}\label{perpenaction}
	S=-\frac{\mathcal{T}}{2\pi\alpha'}\int d\sigma\sqrt{f^2[1+(h-1)cosh^2(\beta)] + \frac{f^2}{h}[1+(h-1)cosh^2(\beta)]u'^2}.
\end{equation}
Here $\mathcal{T}$ is the temporal length of the Wilson loop traced out by the $q\bar q$ pair's motion in the boundary and $\alpha'$ is related to the string tension. The Lagrangian does not have an explicit dependence on $\sigma$. Therefore the Hamiltonian like equation can be written as,
\begin{equation}\label{perpenhamilton}
\frac{\partial \mathcal{L}}{\partial u'} u' - \mathcal{L} = W.
\end{equation}
Here $W$ is the constant of motion and $u'$ is the derivative of radial coordinate with respect to $\sigma$. The seperation length is obtained as in \cite{Pokhrel_2025},
\begin{equation}
	L^{\perp} =\int_{-\frac{L}{2}}^{\frac{L}{2}}dx=2\int_{0}^{u_t}\frac{du}{u'} =2\int_{0}^{u_c} \frac{ W du}{\sqrt{h}\sqrt{f^2(1+(h-1)cosh^2(\beta))-{W}^2}}.
	\label{screening_length_final_equation}
\end{equation}
The $u_t$ is the solution of equation $f^2[1+(h-1)cosh^2(\beta)]-{W}^2 = 0$.  From equation (\ref{screening_length_final_equation}), the screening length can be studied for different set of parameters as provided in the following plots (\ref{fig_screening3D_perp},\ref{fig_L_vs_W_perp_case1}). The separation length $L^{\perp}$ decreases with the increase of GB coupling, string density, temperature and rapidity parameter. The changes in $L^{\perp}$ is prominent for the change of rapidity parameter. For small string density there is a sharp change for temperature but for large string density the change becomes flatter. For the change of $W$ the separation length increase initially then decreases.
Further from figure (\ref{fig_L_vs_W_perp_case1}), it is observed that the separation length $L^{\perp}$ initiate with zero value and then reaches a maximum value known as the screening length ($L_S$), finally decreases and on further increase of $W$ goes towards zero. $L_S$ is the separation length between the $q\bar{q}$ pair beyond which they would separate into individual quark and antiquark.\\
Also noticed that, $W$ has two possible values for every value of $L^{\perp}<L_S$. Now, to have insight on the favoured value of $W$, we study the potential energy $V$ of $q\bar{q}$ pair with respect to the separation distance. Holographically, the potential has been calculated as \cite{Fadafan_2009},
\begin{equation}
	V = -\frac{S - S_0}{\mathcal{T}},
\end{equation}
where $S_0$ is the action for the straight string stretched from the boundary to the horizon. The straight string is produced when the tip of the string representing $q\bar q $  pair reaches the horizon of the black hole and breaks down. Using equation (\ref{perpenaction}, \ref{perpenhamilton}) and (\ref{screening_length_final_equation}), the potential energy $V$ can be expressed as function of separation length $L^{\perp}$. Finally, we have plotted $V$ against $L^{\perp}$ for different set of parameters as provided in the following plots (\ref{fig_V_vs_L_perp_case1}).

It is observed from figure(\ref{fig_V_vs_L_perp_case1}), that $L^{\perp}$ has a maximum value ($L_S$) corresponding to a $q\bar{q}$ potential $V$. For any value of $L^\perp < L_S$, $V$ has two possible values, larger value of $W$ corresponds to the larger value of potential and smaller value of $W$ corresponds to the smaller value of potential which is energetically favoured. Further, with increase in any parameter value, the potential increases, making the $q\bar{q}$ pair less stable in the confined state and more probable to transit to an unbound state.

\subsection{Parallel:}
In this subsection, the separation between the $q\bar{q}$ pair in parallel orientation has been studied. Therefore, we fix the static gauge in the following way, $\tau = t^*,\, \sigma = z^*,\, x=y=0$ and boundary condition on the probe string as,
\begin{equation}
	u(\sigma =\pm \frac{L}{2}) = 0,\, u(\sigma =0) = u_{ext},\, u'(\sigma=0) =0.
    \label{equation_screening_length_parallel_boundary_conditions}
\end{equation}
Following the computation of previous subsection, the screening length can be calculated as,
\begin{equation}
    L^\parallel = \int_{0}^{u_t} \frac{2 W  \sqrt{1 + (h-1) cosh^2(\beta)}}{h\sqrt{f^2 h - {W}^2}}.
\end{equation}
Here, $u_t$ is the solution of $ f^2 h - {W}^2 = 0$.
\begin{figure}[h]
	\centering
	\subfigure[]{\includegraphics[width=0.32\linewidth]{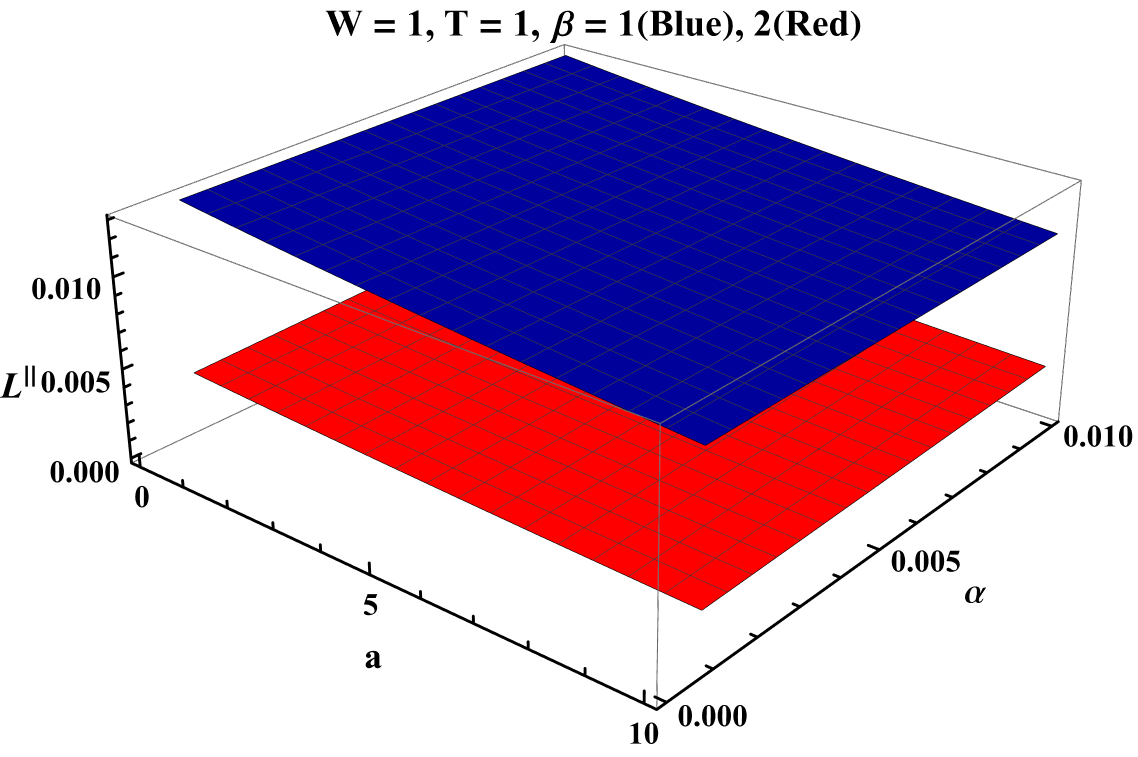}}
	\subfigure[]{\includegraphics[width=0.32\linewidth]{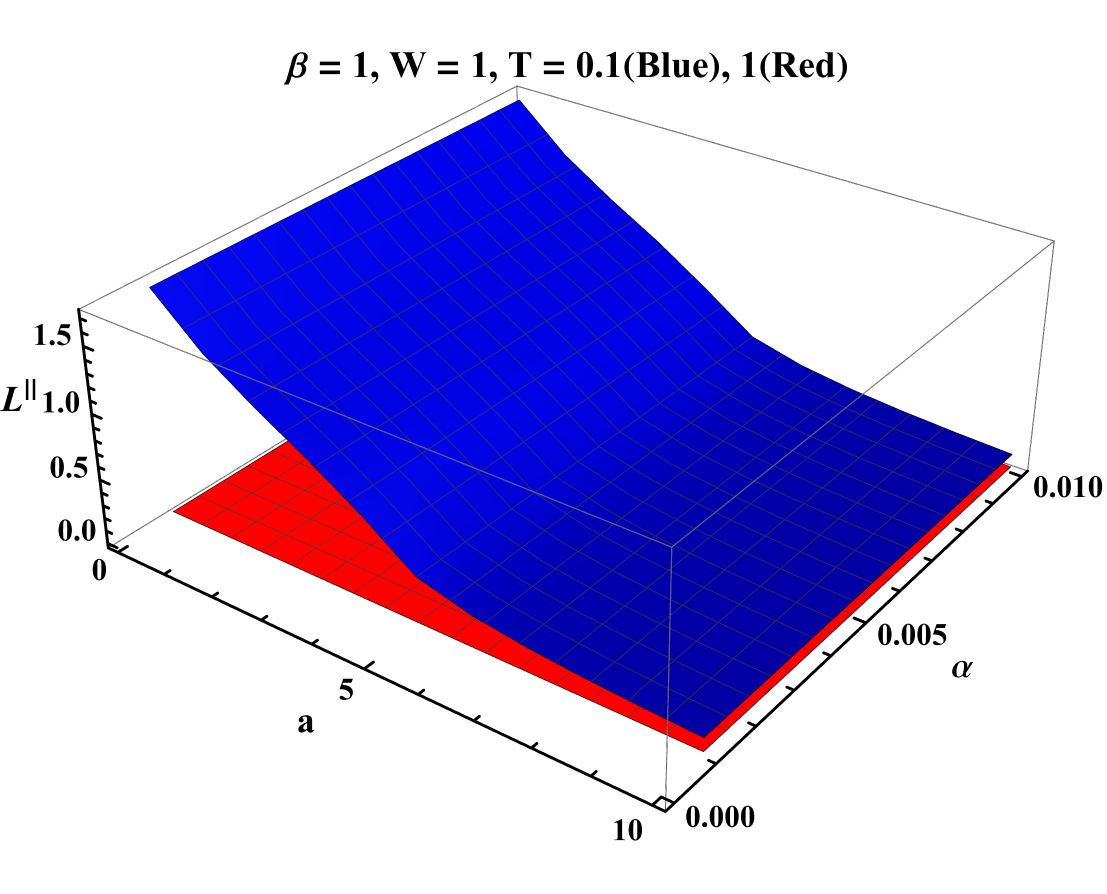}}
	\subfigure[]{\includegraphics[width=0.32\linewidth]{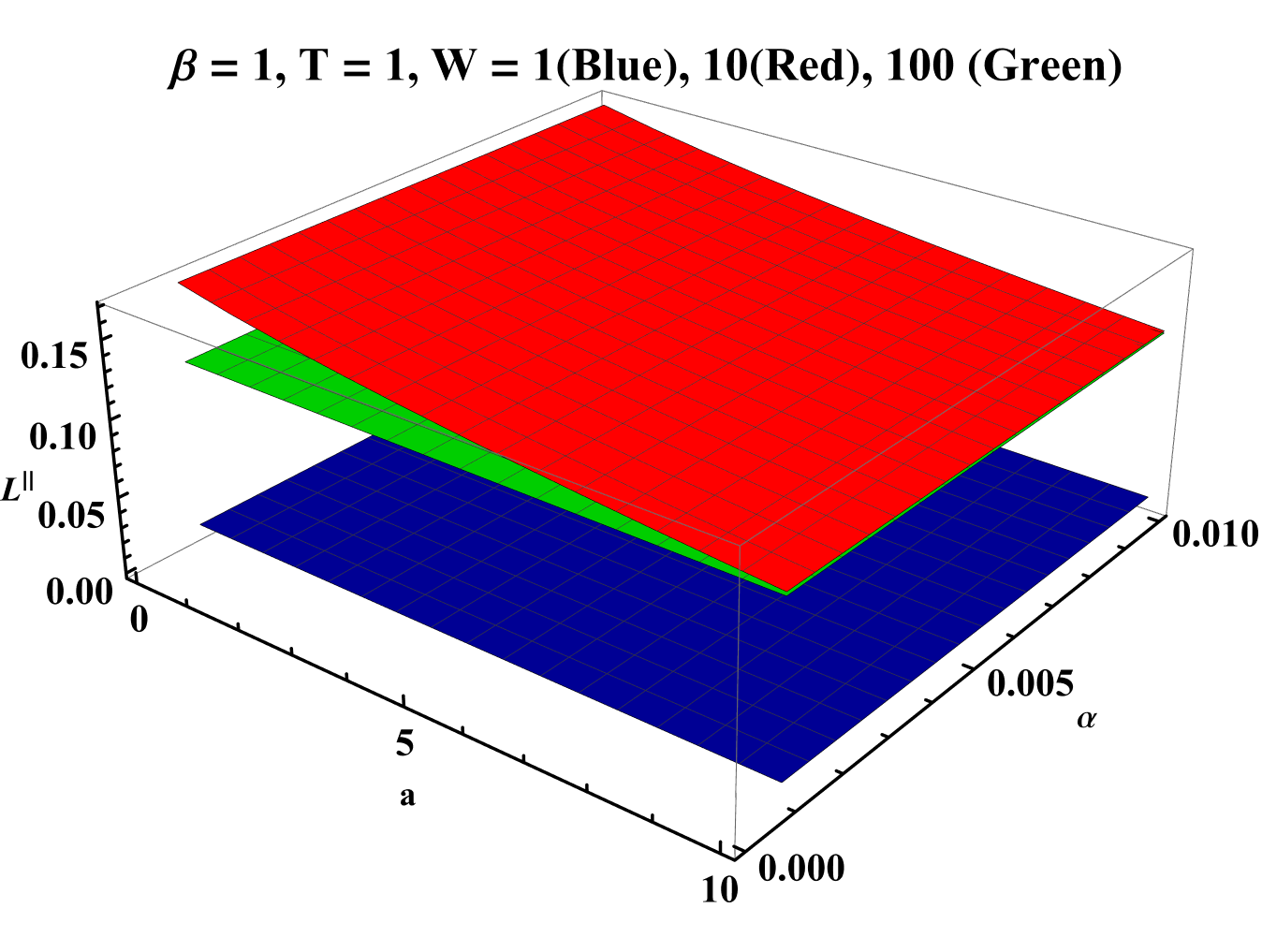}}
	\caption{$L^{||}$ vs $a$ and $\alpha$ for different values of  $\beta$,  $T$ and $W$.}
	\label{fig_screening3D_para}
\end{figure}
\begin{figure}[h]
	\centering
	\subfigure[]{\includegraphics[width=0.45\linewidth]{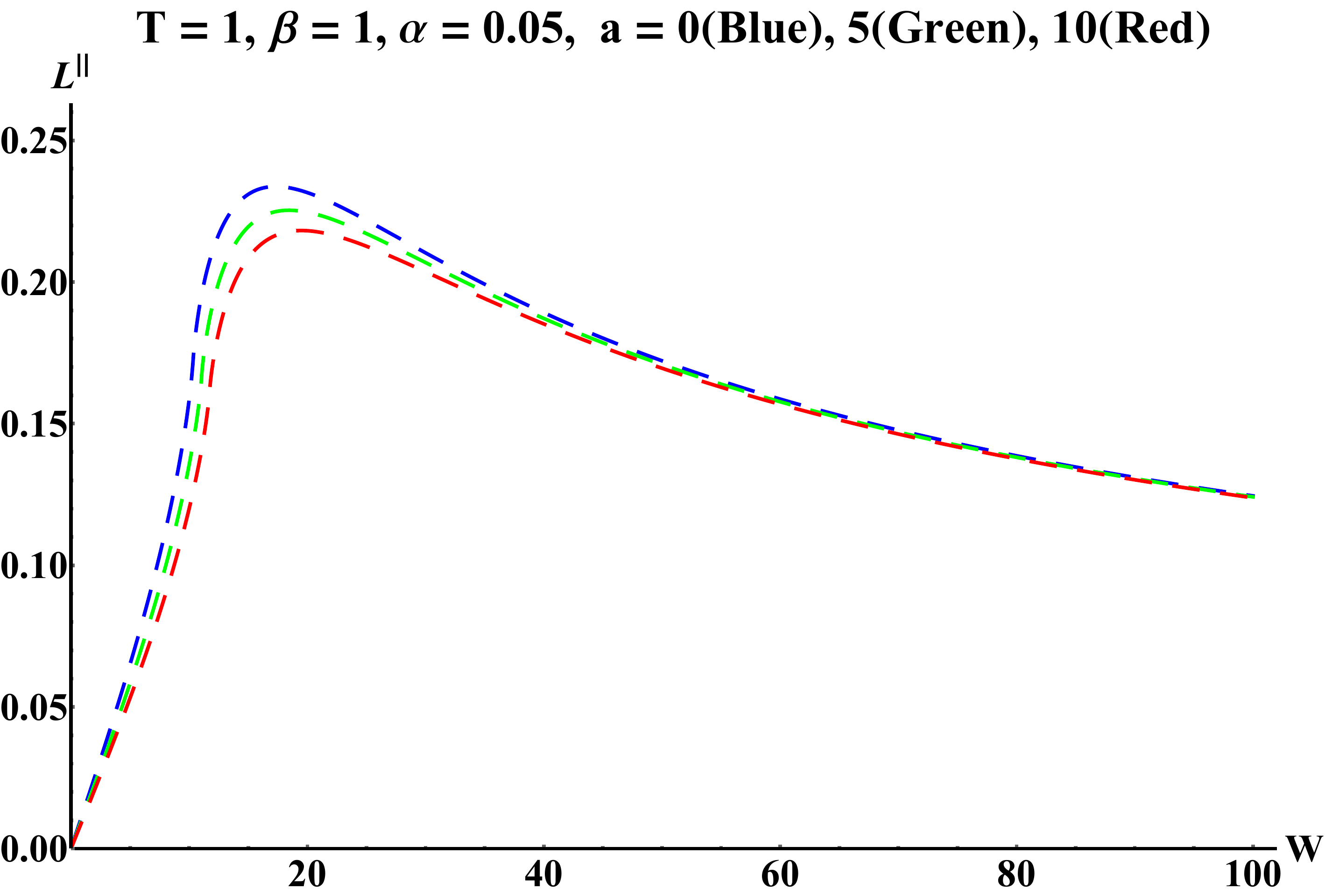}}
	\subfigure[]{\includegraphics[width=0.45\linewidth]{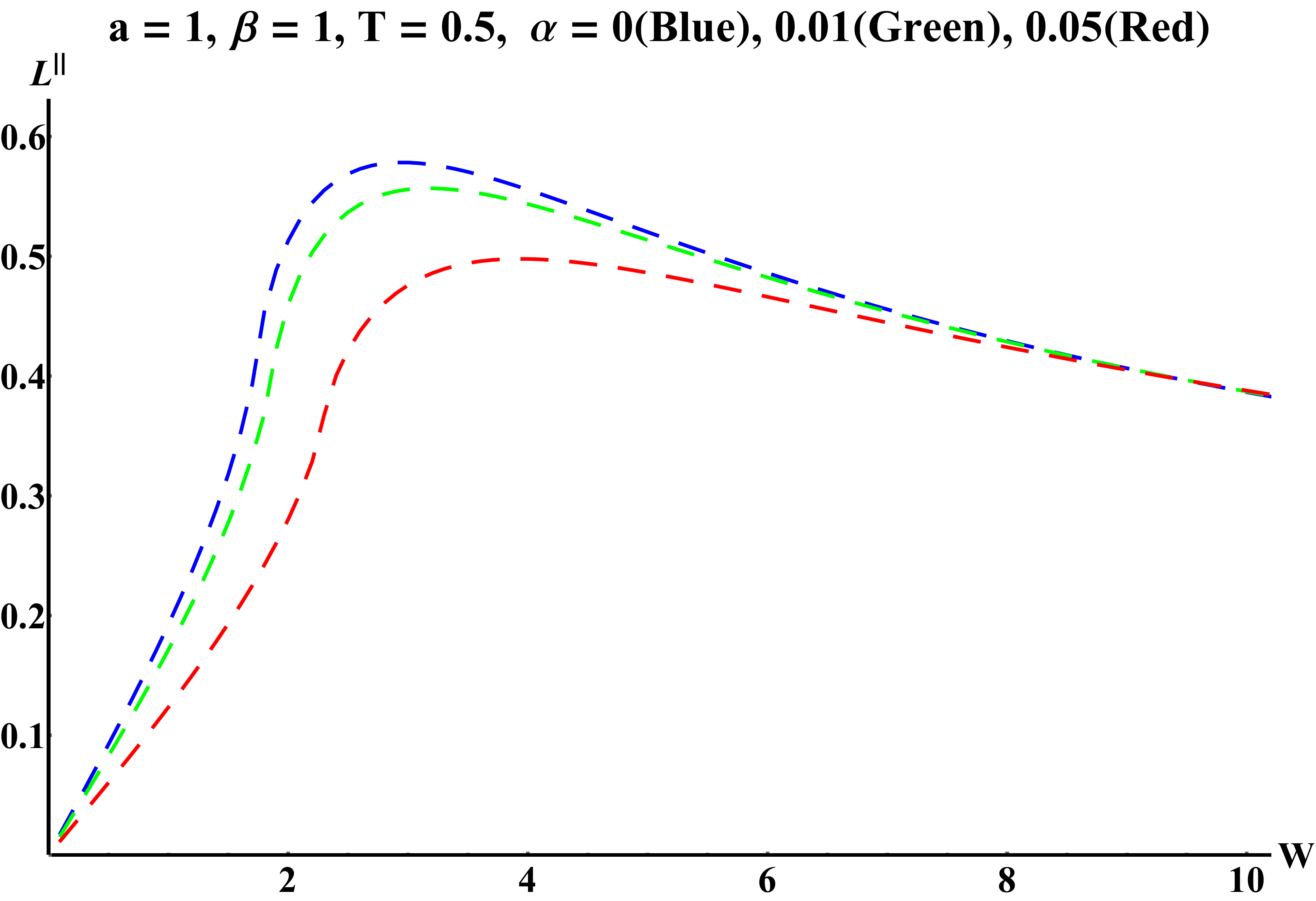}}
	\caption{$L^{||}$ vs $W$ for: (a) fixed parameters $T,\,\beta,\,\alpha$ and different $a$ and (b) fixed parameters $a,\,\beta,\,T$ and different $\alpha$.}
	\label{fig_L_vs_W_para_case1}
\end{figure}
From figure (\ref{fig_screening3D_para},\ref{fig_L_vs_W_para_case1}), it is observed that $L^{||}$ exhibits a similar nature to the perpendicular configuration, where $L_S$ decreases with increasing value of any parameters and the overall nature of the separation distance is also similar, which starts from zero then increases with $W$, reaches a maximum and then falls off with further increase in $W$.\\
Further, similar to the perpendicular case, the binding energy  for the parallel case has been plotted in the following figure (\ref{fig_V_vs_L_para_case1}).
\begin{figure}[h]
	\centering
	\subfigure[]{\includegraphics[width=0.45\linewidth]{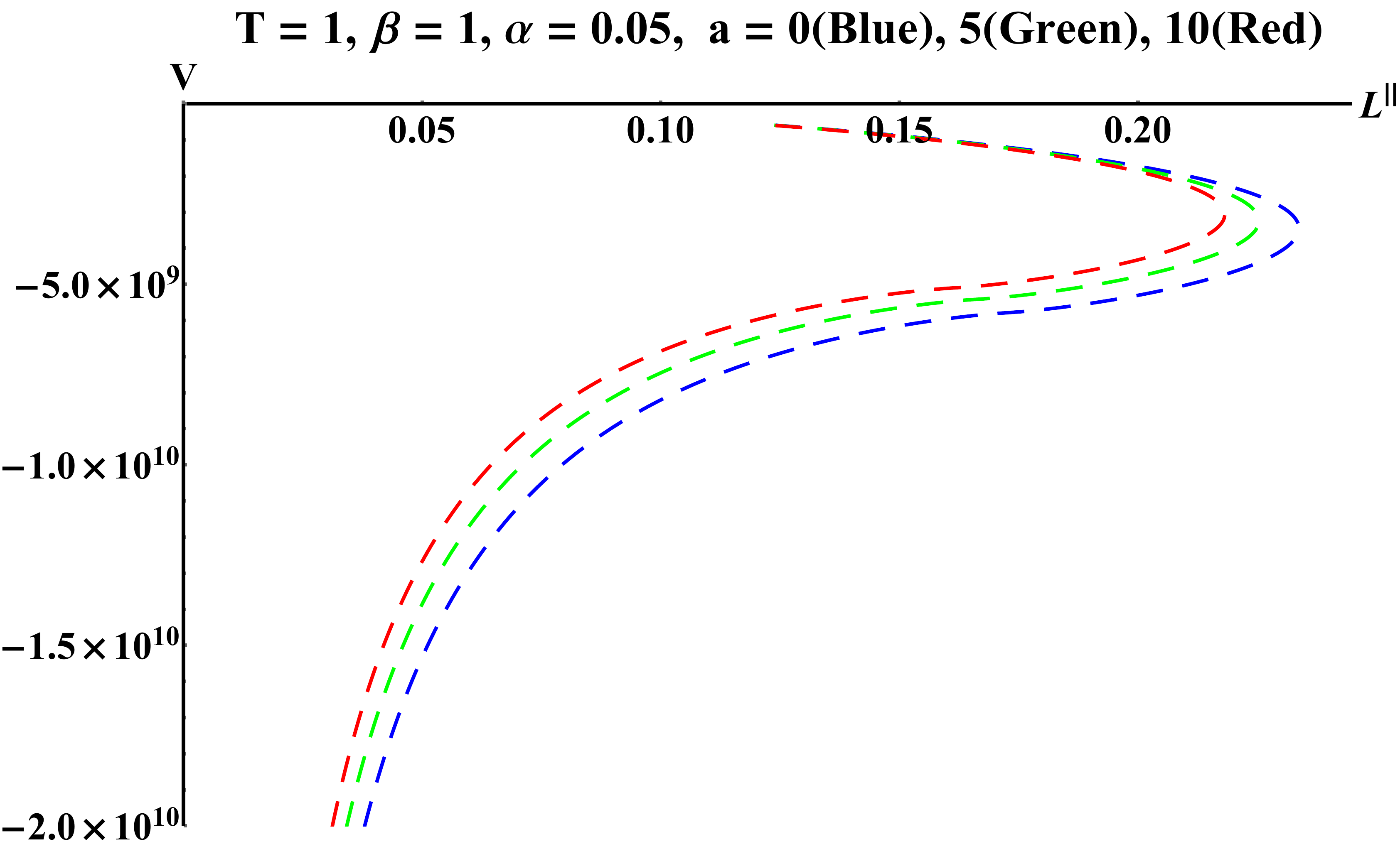}}
	\subfigure[]{\includegraphics[width=0.45\linewidth]{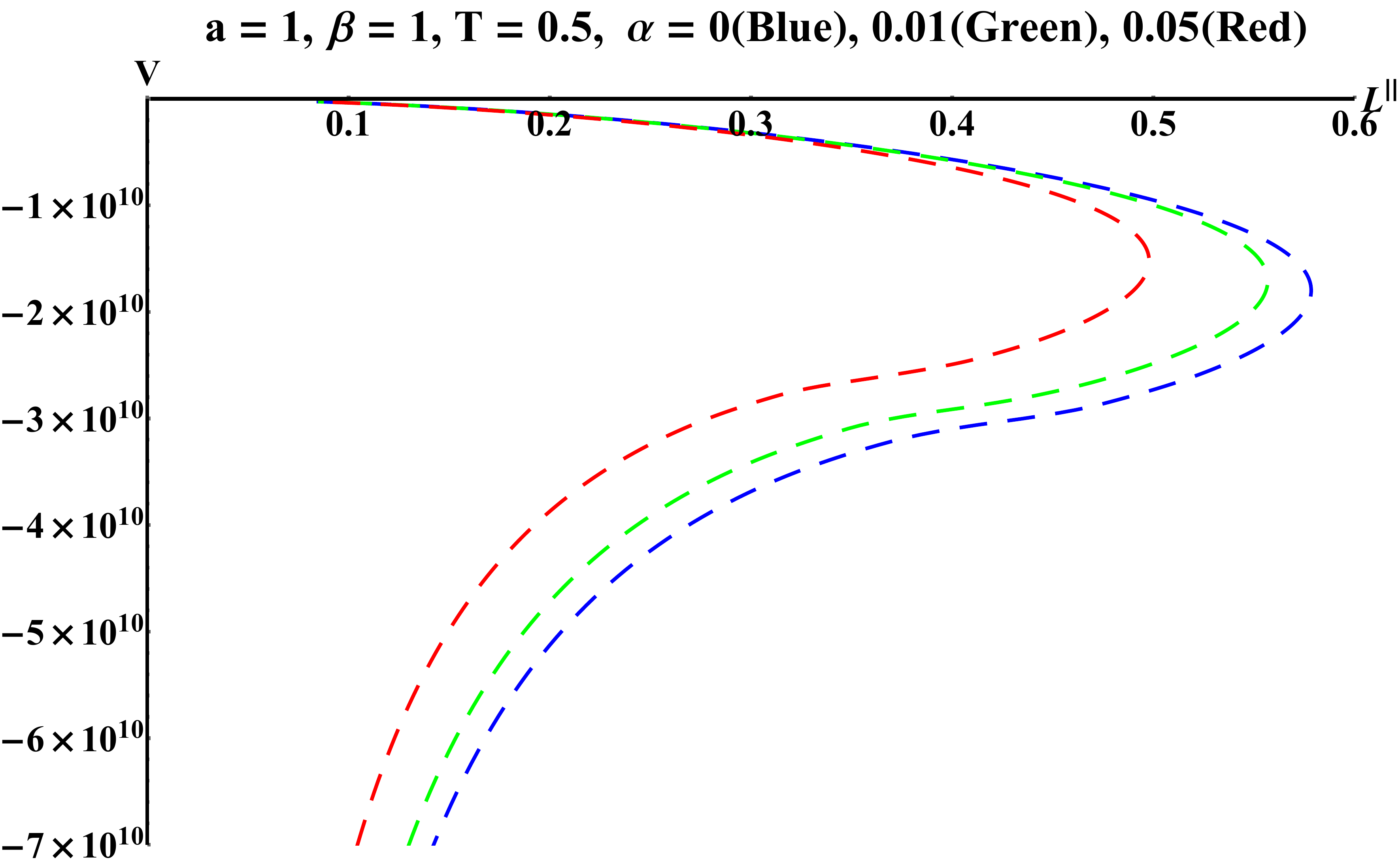}}
	\caption{$V$ vs $L^{||}$ for: (a) fixed parameters $T,\,\beta,\,\alpha$ and different $a$ and (b) fixed parameters $a,\,\beta,\,T$ and different $\alpha$.}
	\label{fig_V_vs_L_para_case1}
\end{figure}
The overall nature of the potential is similar to that of the perpendicular case, where $L^{||}$ has a maximum value ($L_S$) corresponding to some value of the potential and for any $L^{||}<L_S$ there are two probable values of the potential. Larger value of $W$, corresponds to larger potential and smaller $W$ corresponds to smaller potential, which is more stable and favoured.
\subsection{Comparison}
\begin{figure}[!h]
	\centering
	\subfigure[]{\includegraphics[width=0.45\linewidth]{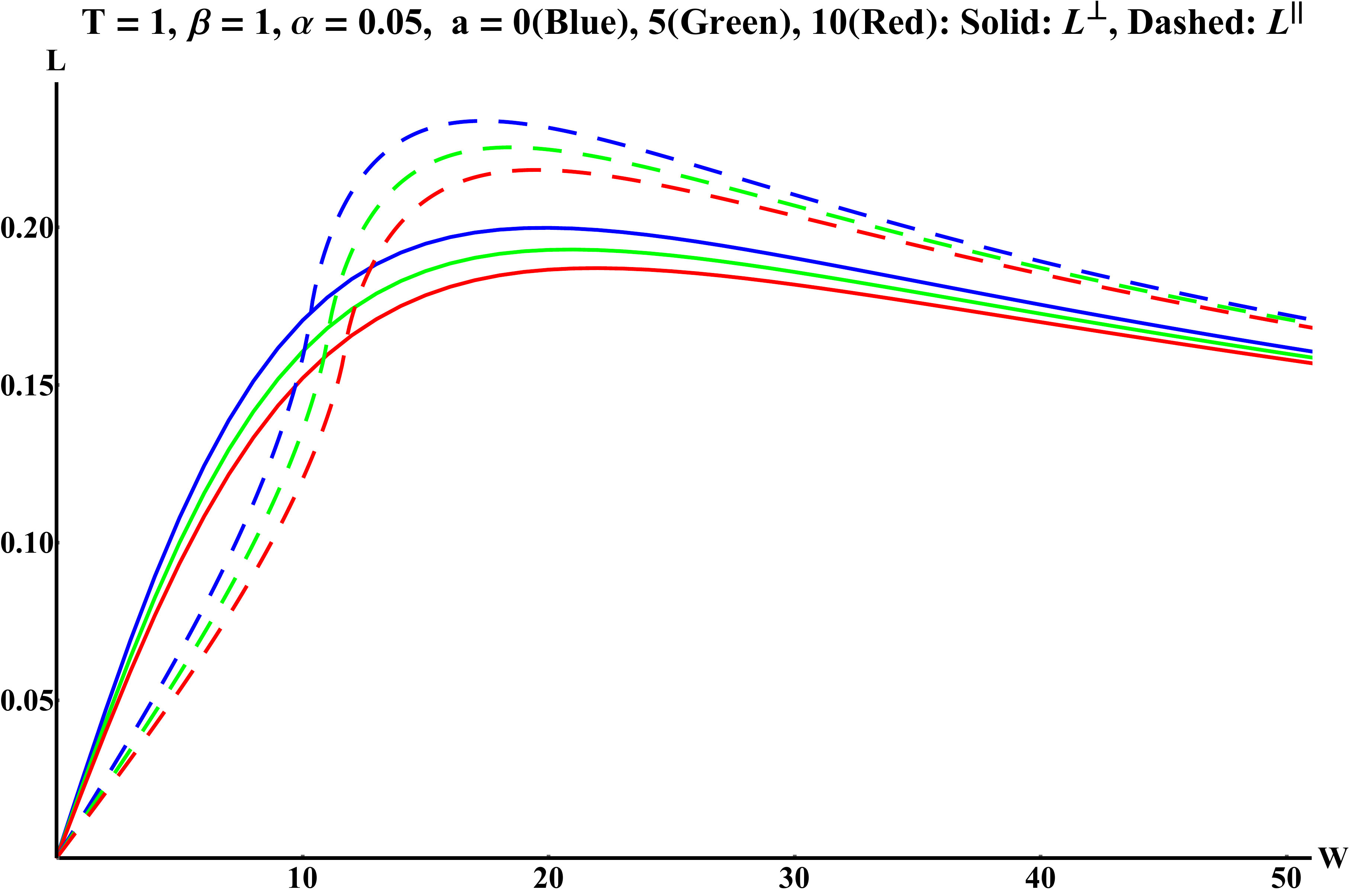}}
	\subfigure[]{\includegraphics[width=0.45\linewidth]{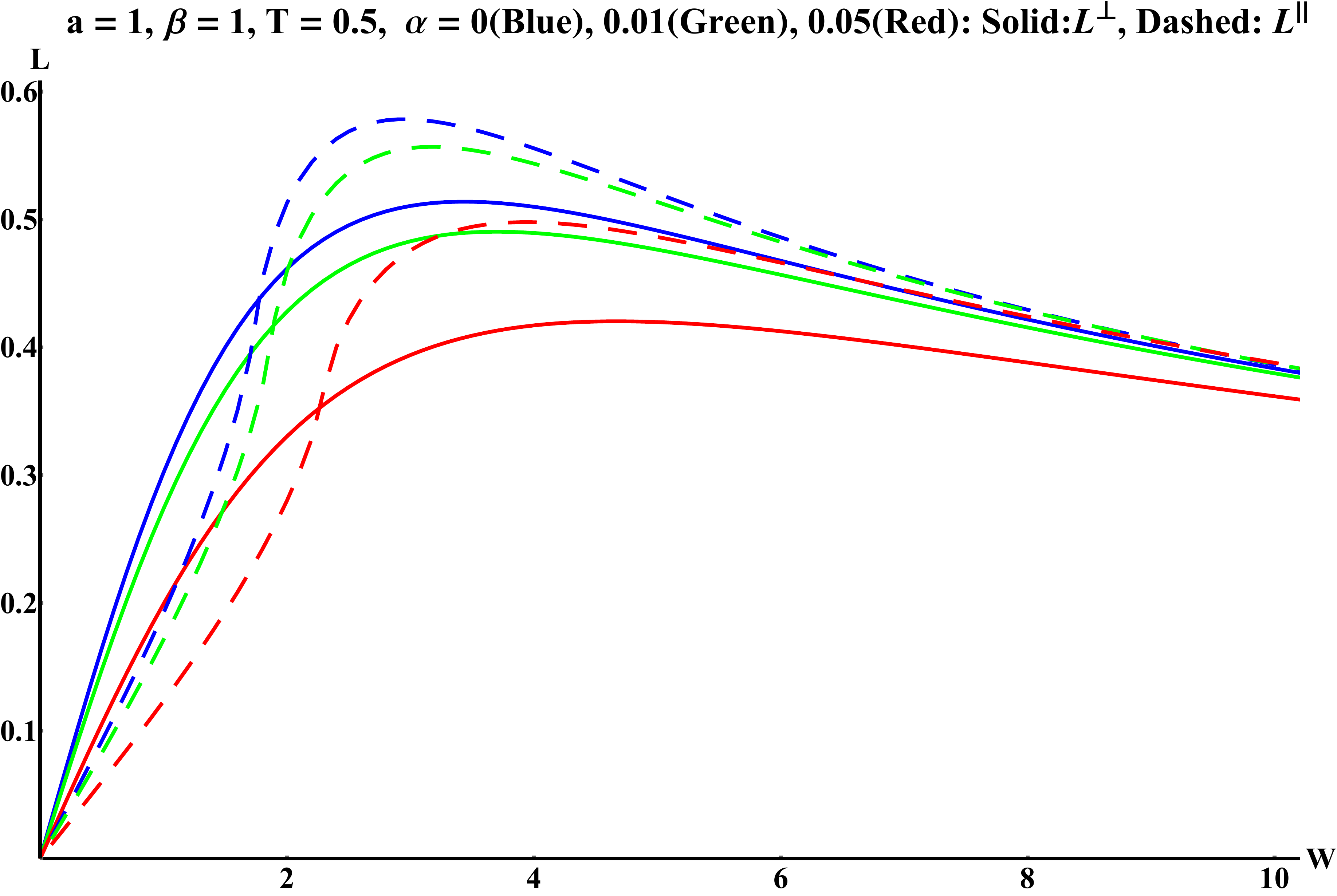}}
	\caption{Comparison of $L$ vs $W$ for: (a) fixed parameters $T,\,\beta,\,\alpha$ and different $a$ and (b) fixed parameters $a,\,\beta,\,T$ and different $\alpha$. Perpendicular case is represented by solid line and parallel case is represented by dashed line.}
	\label{fig_L_vs_W_Comp}
\end{figure}

Here, we compare the separation distance in both perpendicular and parallel orientations. The comparison is done for different set of parameters as provided in the following plots (\ref{fig_L_vs_W_Comp}).
The parallel orientation has larger value of the screening length as compared to the perpendicular configuration which suggests that in parallel orientation, the bound state will be more stable compared to the perpendicular orientation.
%%%%%%%%%%%%%%%%%%%%%%%%%%%%%%%%%%%%%%%%%%%%%%%%%%%%%%%%%%%%%%%%%%%%%%%%%%%%%%%%%%%%%%%%%%%%%%%%%%%%%%%%%%%%%%%%%%%%%%%%%%%%%%%%%%%%%%%%%%%%%%%%%%%%%%%%%%%%%%%%%%%%%%%%%%%%%%%%%%%%%%%%%%%%%%

%%%%%%%%%%%%%%%%%%%%%%%%%%%%%%%%%%%%%%%%%%%%%%%%%%%%%%%%%%%%%%%%%%%%%%%%%%%%%%%%%%%%%%%%%%%%%%%%%%%%%%%%%%%%%%%%%%%%%%%%%%%%%%%%%%%%%%%%%%%%%%%%%%%%%%%%%%%%%%%%%%%%%%%%%%%%%%%%%%%%%%%%%%%%%%%
\section{Jet Quenching Parameter}
\label{sec_jet_quenching}
In this section, holographically, we will analyse the jet quenching parameter ($\hat{q}$) which is related to the energy loss of the probe quark because of the suppression of the high transverse momentum in the thermal medium \cite{Liu:2006ug}. In order to compute $\hat{q}$, the field theoretic relation between $\hat{q}$ and expectation value of Wilson loop traced out by the $q\bar{q}$ pair with separation length $L$ and temporal time $\mathcal T$ along the light cone coordinate can be used. The jet quenching parameter can be written as a function of the regularised Nambu-Goto action $S-S_0$ of the fundamental probe string with two ends attached to the boundary representing the $q\bar q$ pair. Finally, by replacing the regularised action in terms of separation length of $q\bar q$ pair, the quenching parameter can be expressed  as,
\begin{equation}
	\hat{q} = \frac{\sqrt{\lambda}}{\pi I_1(T,a,\alpha)},
	\label{eq_jetq_formula}
\end{equation}
where,
\begin{equation}
	I_1 = \int_\delta^{u_h(T,a,\alpha)} \frac{u^2 du}{\sqrt{h(h-1)}}.
	\label{eq_jetq_I1_formula}
\end{equation}
Here, $u_h$ is the black hole horizon and $\delta$ represents the cut-off imposed close to the boundary to avoid the imaginary value of $u'$ in the radial direction. To obtain cut-off independent result, $\delta $ is set towards zero upon completion of the calculation.

\begin{figure}[!h]
	\centering
	\includegraphics[width=0.6\linewidth]{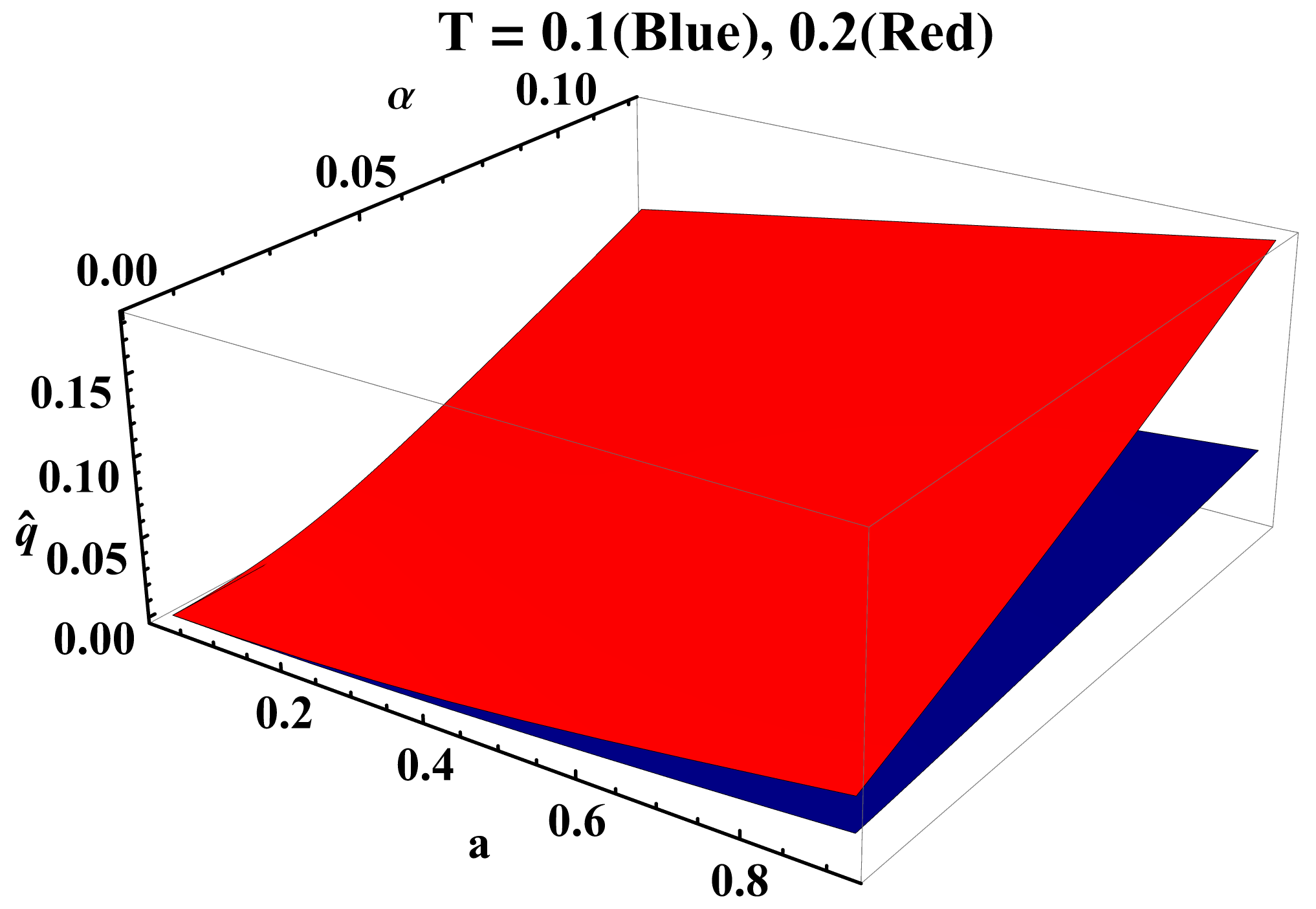}
	\caption{$\hat{q}$ vs $a$ and $\alpha$ for different $T$.}
	\label{fig_jetQ_3D}
\end{figure}

Considering the metric solution (\ref{eq_vr}), the nature of $\hat{q}$ with respect to different set of parameters is given in the following plot (\ref{fig_jetQ_3D}).
From the plot, it is concluded that $\hat{q}$ enhances with increase of temperature, string density and Gauss-Bonnet coefficient which suggests that the energy loss is increased in a medium with higher temperature and density. The nature is consistent with the expected behaviour of $\hat{q}$ in a strongly coupled plasma. It also suggests that the GB coefficient is also involved in modifying the probe quark's energy loss characteristics.
%%%%%%%%%%%%%%%%%%%%%%%%%%%%%%%%%%%%%%%%%%%%%%%%%%%%%%%%%%%%%%%%%%%%%%%%%%%%%%%%%%%%%%%%%%%%%%%%%%%%%%%%%%%%%%%%%%%%%%

%%%%%%%%%%%%%%%%%%%%%%%%%%%%%%%%%%%%%%%%%%%%%%%%%%%%%%%%%%%%%%%%%%%%%%%%%%%%%%%%%%%%%%%%%%%%%%%%%%%%%%%%%%%%%%%%%%%%%
\section{Rotating Probe: Radial Profile}
\label{sec_radial_profile}
In this section, the radial profile has been analysed for constantly rotating probe string whose one end is attached to the boundary and the other one is elongated up to the horizon of the spacetime. We consider the quark/antiquark corresponds to the attached end of the string rotating with constant angular frequency $\omega$ and radius $\mathcal{R}$ in two dimensional plane of the boundary gauge theory. The velocity and acceleration are given as $v= \mathcal{R} \omega$ and $a= \omega^2 \mathcal{R}$. Following the approach of \cite{Pokhrel_2025,Chakrabortty2016a}, we fix the gauge to describe the rotation of the string in the following way,
\begin{equation}
	X^\mu(\tau,\sigma) = (t=\tau, u = \sigma, x = \rho(\sigma)cos(\omega t + \theta(\sigma)), y = \rho(\sigma)sin(\omega t + \theta(\sigma)), z=0),
	\label{equation_energyloss_parameterization}
\end{equation}
where, $\theta(\sigma)$ and $\rho(\sigma)$ are the angular and radial profile of the rotating string with the boundary conditions,
\begin{equation}
	\theta(0) = 0\, {\rm and} \, \rho(0) = \mathcal{R}.
\end{equation}
Since, our interest is to study the rotating string's radial profile, here we find the Lagrangian equation of motion with respect to the variable $\rho$ for the rotating probe string in the AdS GB background with cloud of string and it is given as,
\begin{align}
	& 2(\Pi_\theta^2 - \omega^2 f^2 \rho^4) - \left[2 f h \rho^3 (h -\omega^2 \rho^2) f' - \rho \{\Pi_\theta^2 - f^2 ( 2 h \rho^2 - \omega^2 \rho^4)\}h'\right]\rho'\nonumber \\ & + 2h (\Pi_\theta^2 - \omega^2 f^2 \rho^4 ) \rho^{'2} - \rho^3 \left[ 2 f h^2 (h- \omega^2 \rho^2) f' - (\Pi_\theta^2 \omega^2 - f^2 h^2 )h'\right]\rho^{'3} \nonumber \\ & + 2 \rho(h-\omega^2 \rho^2) (\Pi_\theta^2 - f^2 h \rho^2) \rho^{''} = 0.
	\label{equation_energyloss_final_EOM_for_rho}
\end{align}
In the equation of motion, we remove the $\theta'$ dependency by using the equation of conjugate momentum $\Pi_\theta$ corresponding to the $\theta$ variable. In order to get the information about the radial profile, we solve the above equation numerically, since the analytical method to get the exact solution is very difficult. We use the boundary condition $h(u_c) -\rho(u_c)^2\omega^2 = 0$ and specify $\rho$ and $\rho'$ at critical point $u_c$ for certain fixed values of $\Pi_\theta, \omega, \alpha$ and temperature. To find the value of $\rho'$ at $u_c$, we expand about $u_c$ and retain terms linear in $\rho'$. 

The radial profile is obtained against the radial coordinate $u$ for different set of parameters as provided in the following figures (\ref{fig_radial_combined},\ref{fig_radial_3D}). From the plots of the radial profile, following things are observed. For each radial profile, there is a unique limit where $u\rightarrow 0$ fixes the radius of the rotating quark in the boundary and there is a unique intersection point $u_c$ between the black dotted profile (string moving with local speed of light) and the rotating radial profile $\rho(u)$. With the increase in the value of the conserved momenta, the radius $\rho(u_c)$ enhances but the corresponding value of $u_c$ decreases. Further, for angular frequency $\omega<1$, the rotating string radius is almost constant for fixed momentum. Whereas, a bending effect on the radial profile is observed for $\omega>1$, resulting in larger radius towards the larger value of $u$ compared to the radius at the boundary. In addition, for some fixed value of the temperature, angular frequency $\omega$ and conserved momenta $\Pi$, the radius of rotation and $u_c$ reduce due to the increase of string density $a$. However, the changes on radius of rotation and $u_c$ are almost negligible due to the change in GB coefficient which differs from \cite{Atashi_2020} for the lower value of angular frequency $\omega$. This difference may arises due to the effect of back reaction.
\begin{figure}[!h]
	\centering
	\subfigure[]{\includegraphics[width=0.4\linewidth]{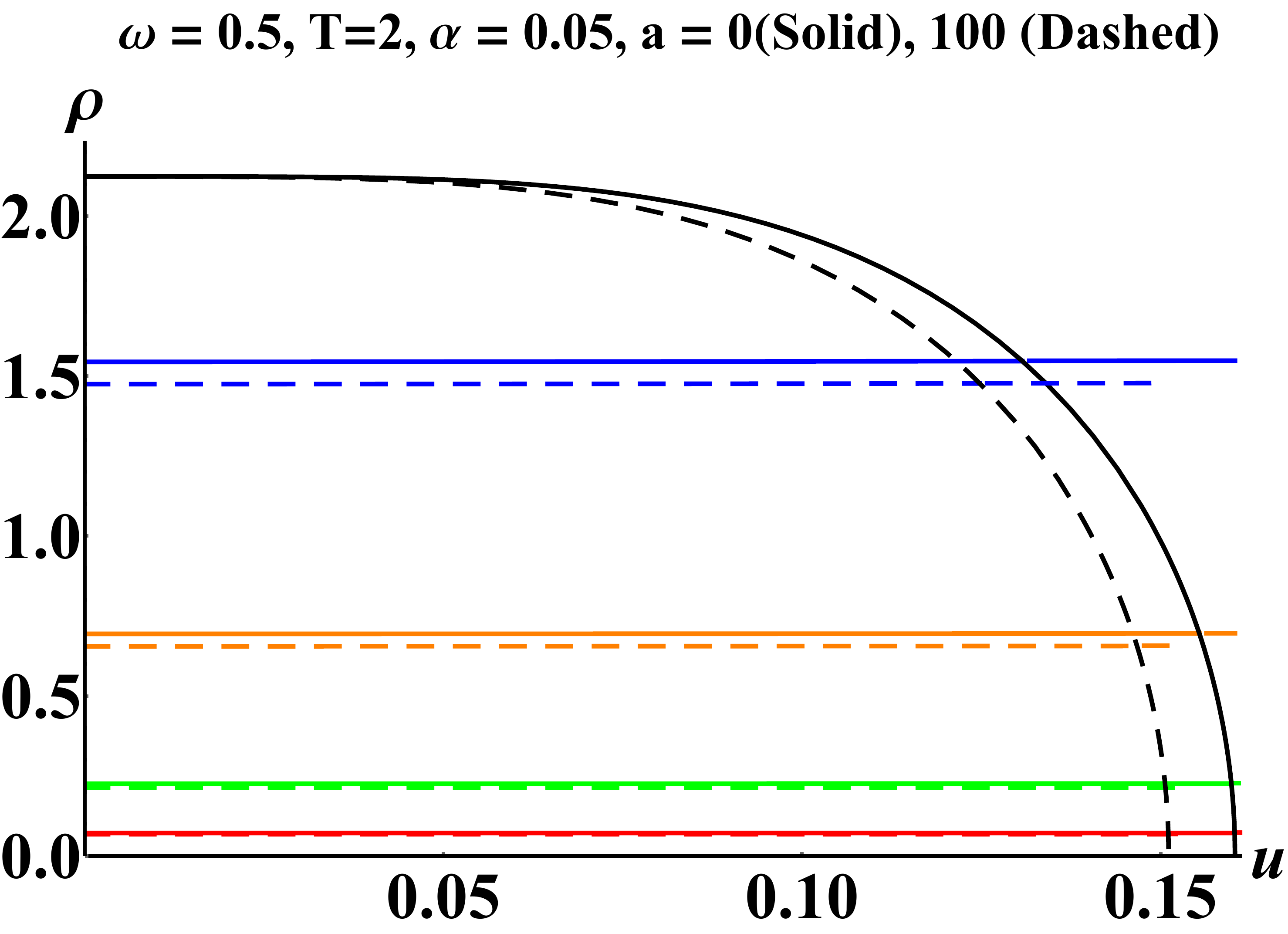}}
	\hspace{01cm}\subfigure[]{\includegraphics[width=0.4\linewidth]{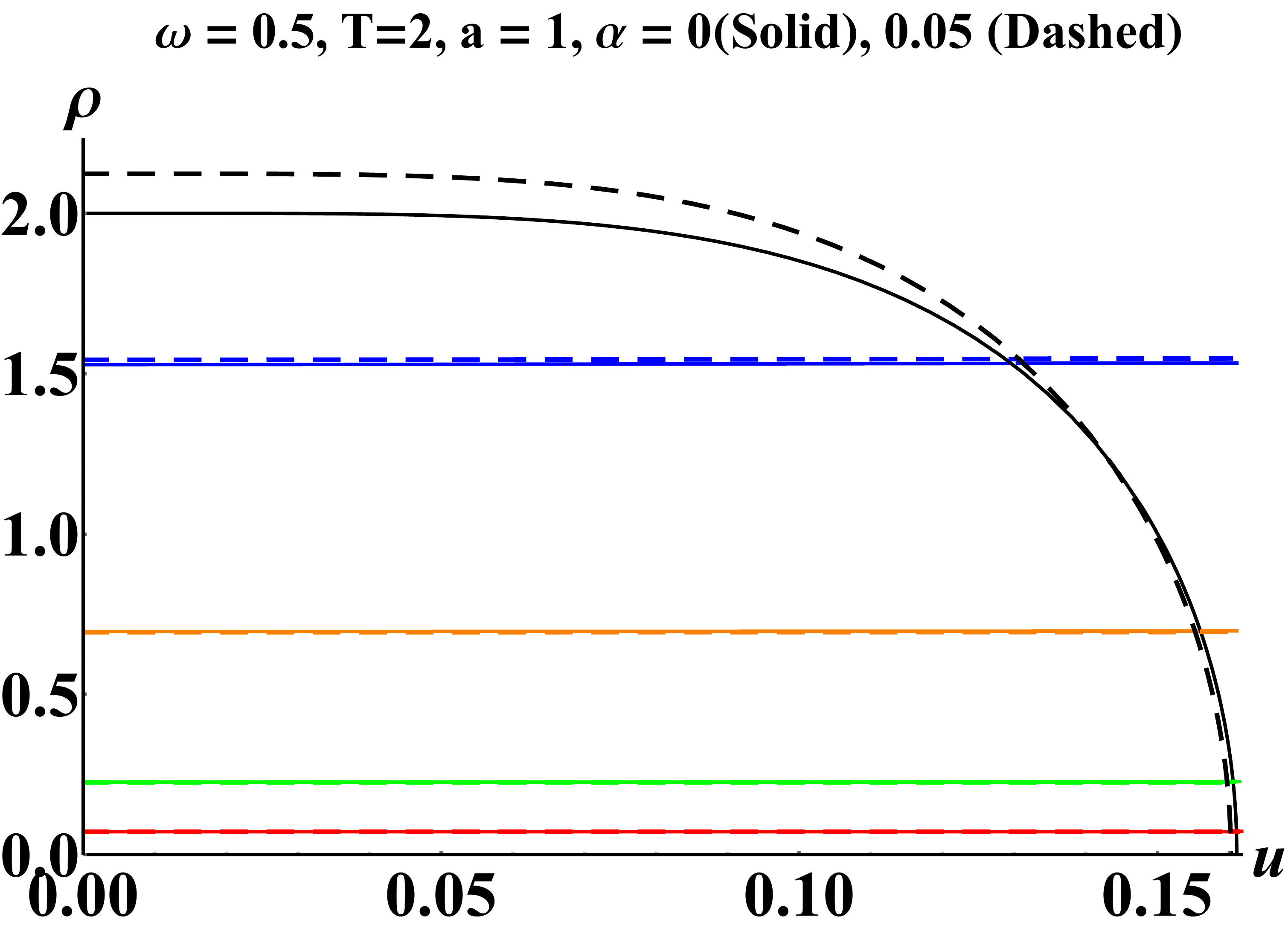}}
	\subfigure[]{\includegraphics[width=0.4\linewidth]{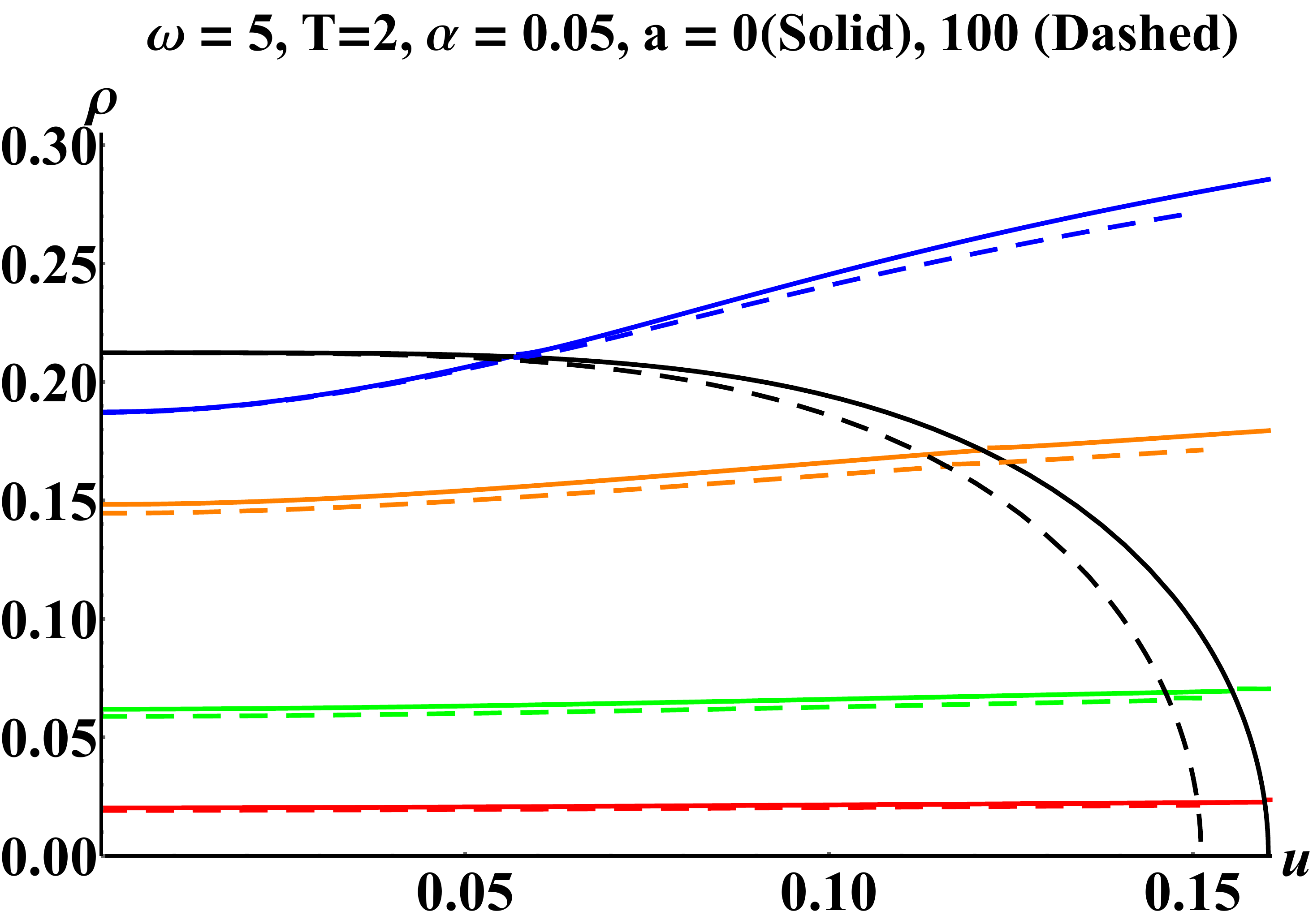}}
   \hspace{01cm} \subfigure[]{\includegraphics[width=0.4\linewidth]{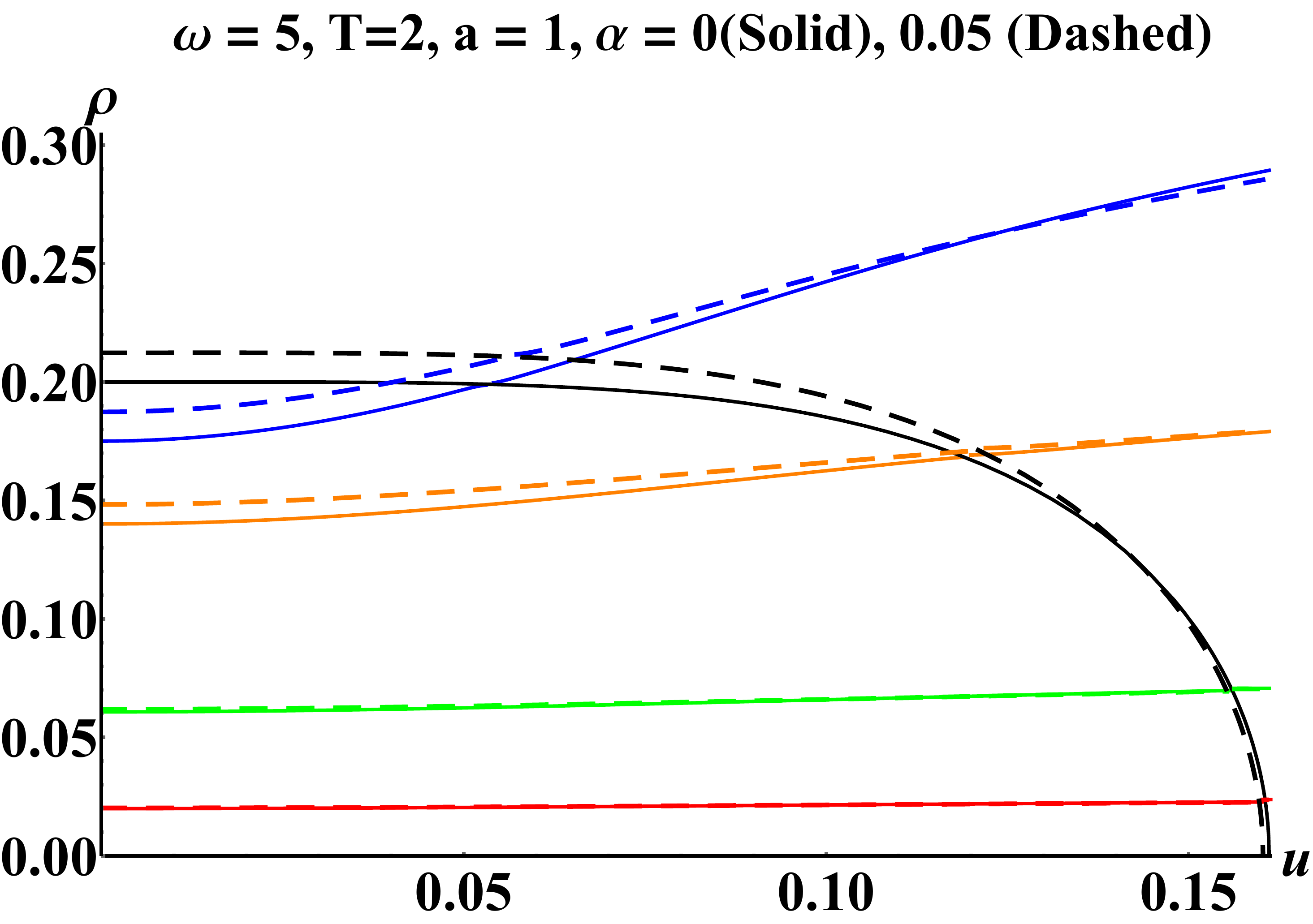}}
	\caption{$\rho$ vs $u$ at $T = 1$, $\omega = 0.5$ (upper row), $\omega=5$ (lower row). Plot (a) $\alpha=0.005$, $a=0$ (solid) and $a=100$ (dashed). Plot (b) $a=0.1$, $\alpha=0$ (solid) and $\alpha=0.01$ (dashed). Plot (c) $\alpha=0.005$, $a=0$ (solid) and $a=100$ (dashed). Plot (d) $a=0.1$, $\alpha=0$ (solid) and $\alpha=0.01$ (dashed). Different lines corresponds to $\Pi_\theta = .1$, $1$, $10$ and $70$ (bottom to top) respectively.}
	\label{fig_radial_combined}
\end{figure}
\begin{figure}[!h]
	\centering
	\subfigure[]{\includegraphics[width=0.31\linewidth]{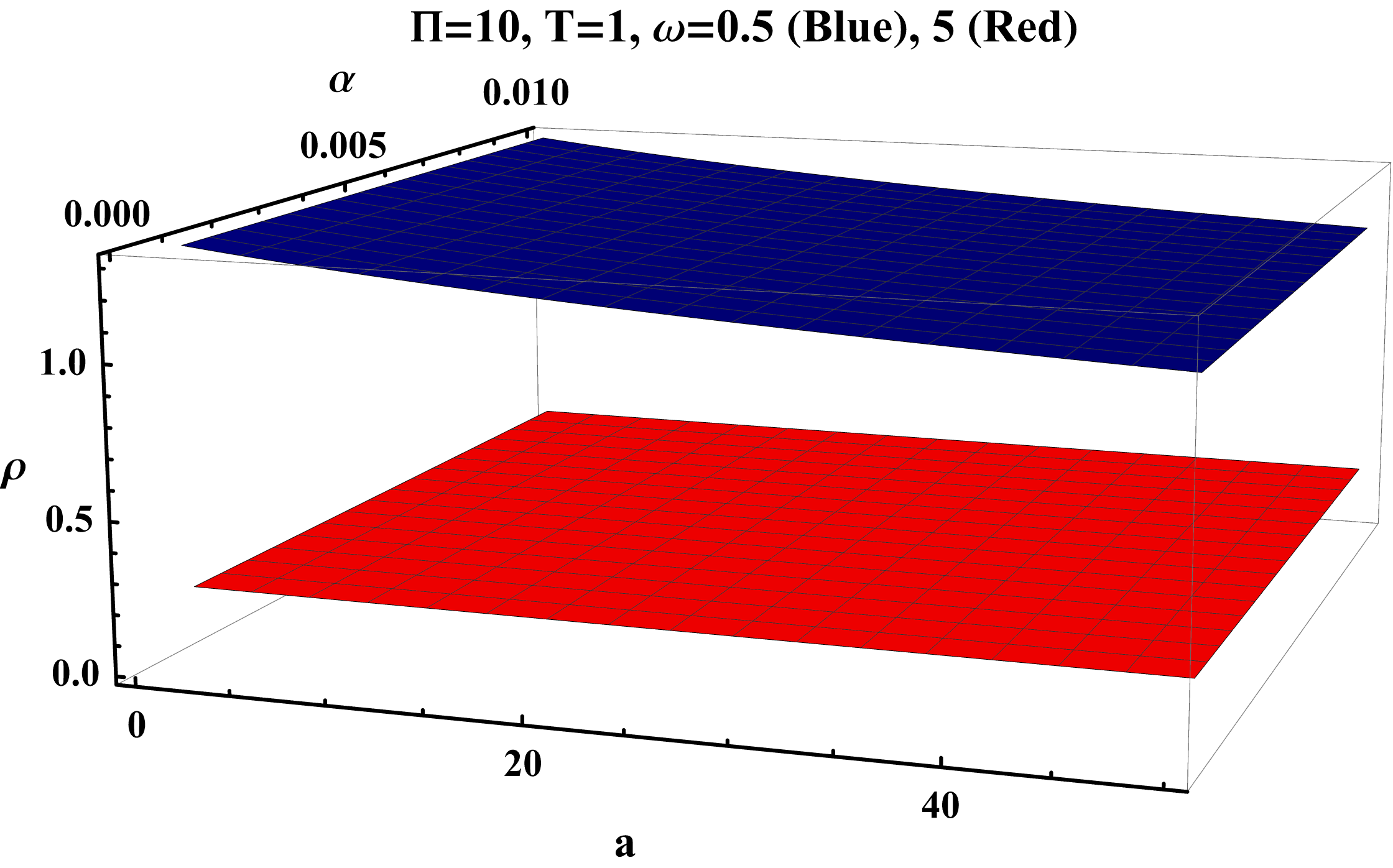}}
	\hspace{01cm}\subfigure[]{\includegraphics[width=0.31\linewidth]{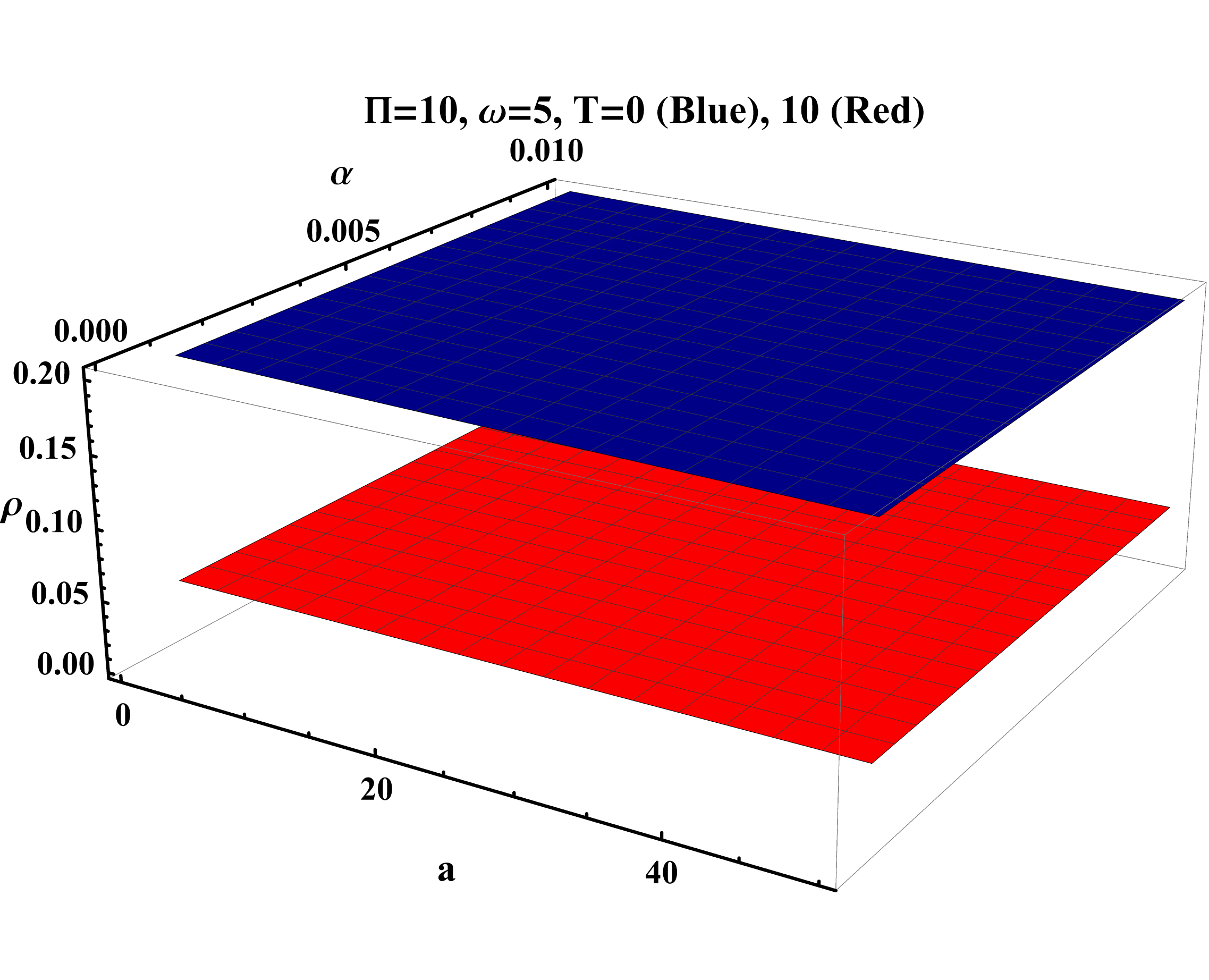}}
	\subfigure[]{\includegraphics[width=0.31\linewidth]{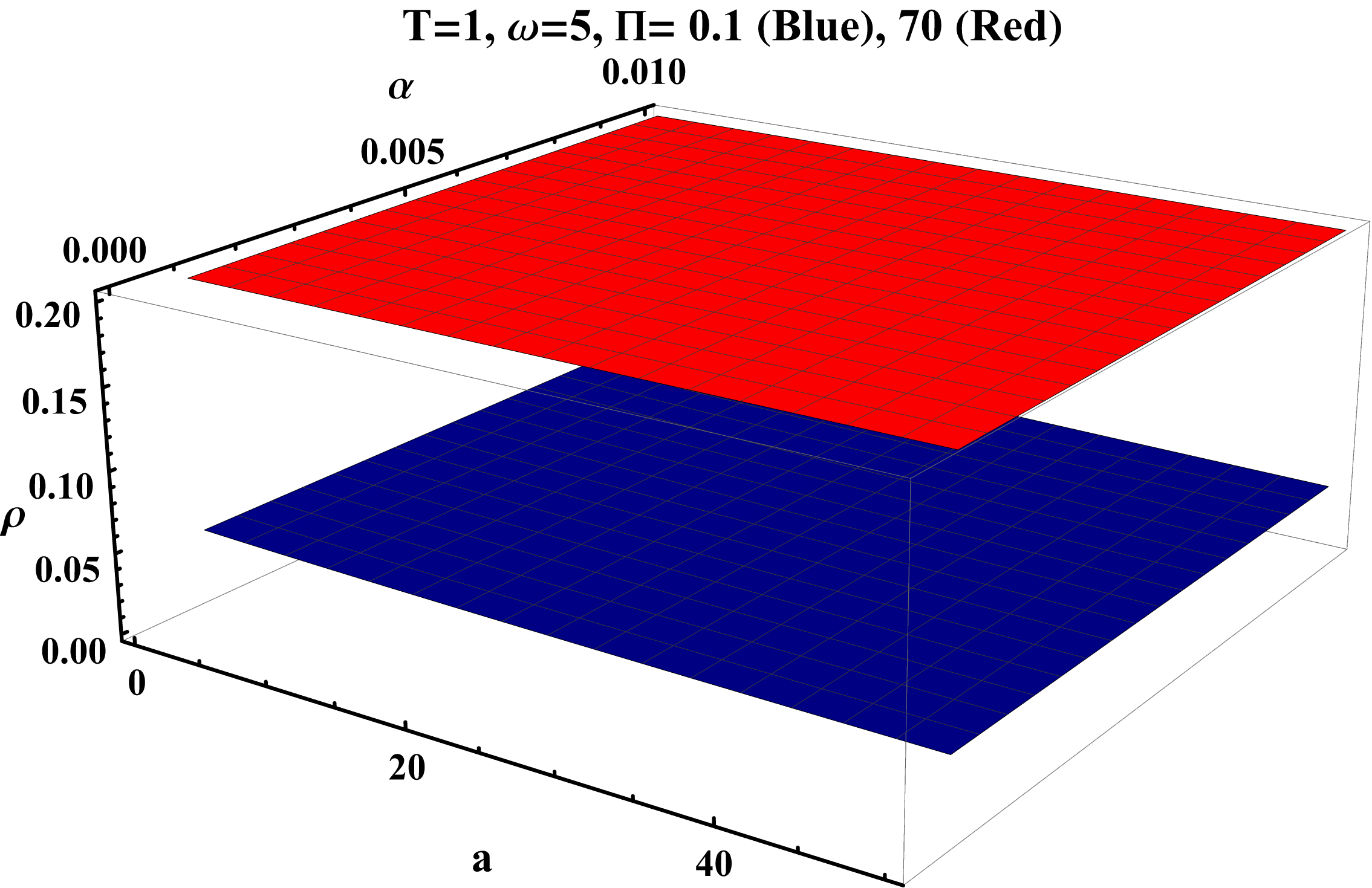}}
	\caption{Radial profile plot against string density and GB coefficient for different (a) $\omega$, (b) $T$ and (c) $\Pi$.}
	\label{fig_radial_3D}
\end{figure}

Now, with the knowledge on radial profile, we study the energy loss of the rotating probe quark in the next section.

%%%%%%%%%%%%%%%%%%%%%%%%%%%%%%%%%%%%%%%%%%%%%%%%%%%%%%%%%%%%%%%%%%%%%%%%%%%%%%%%%%%
\section{Rotating Probe: Energy Loss}\label{sec_energy_loss}
\begin{figure}[!h]
	\centering
	\subfigure[]{\includegraphics[width=0.32\linewidth]{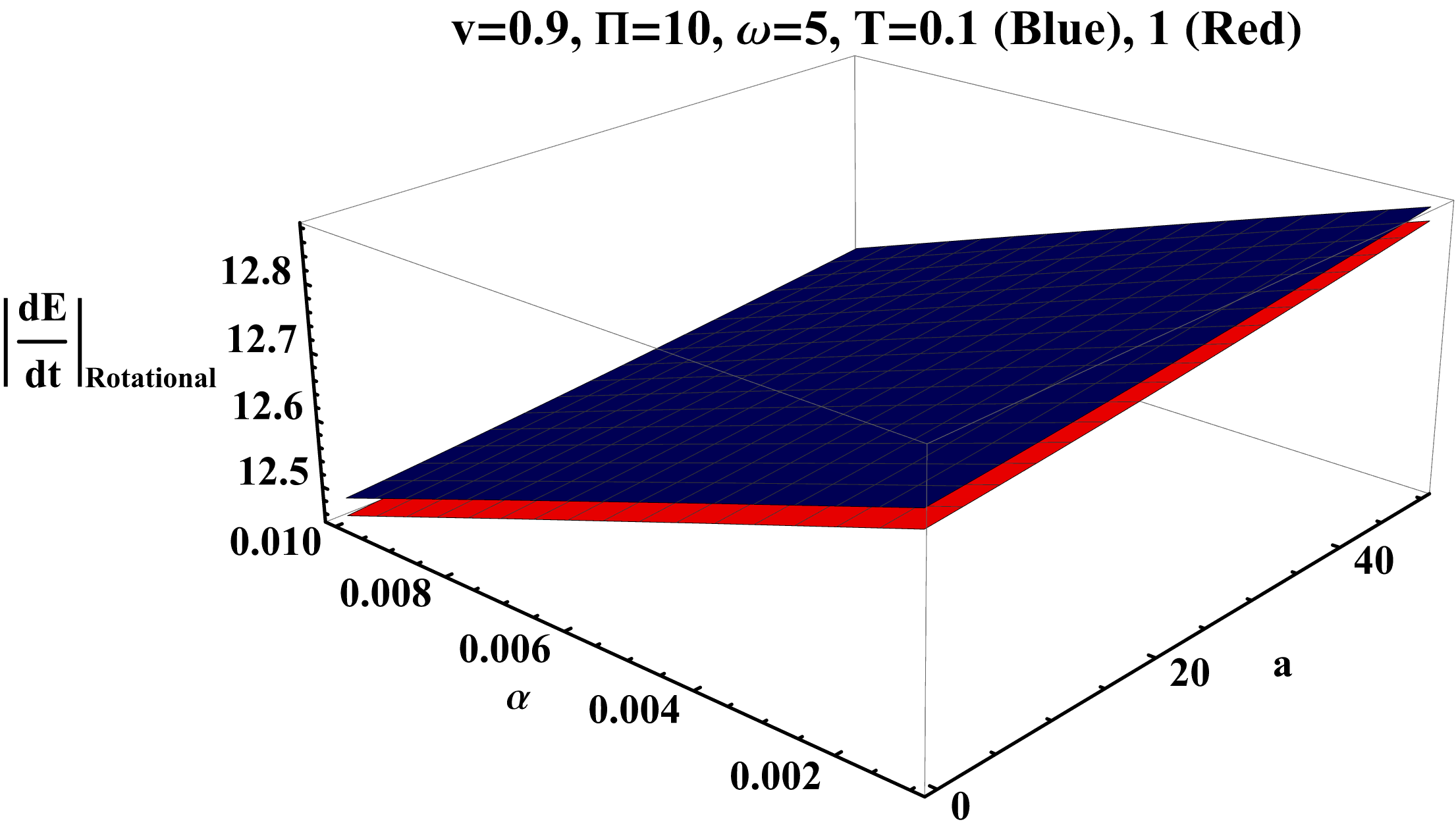}}
	\subfigure[]{\includegraphics[width=0.32\linewidth]{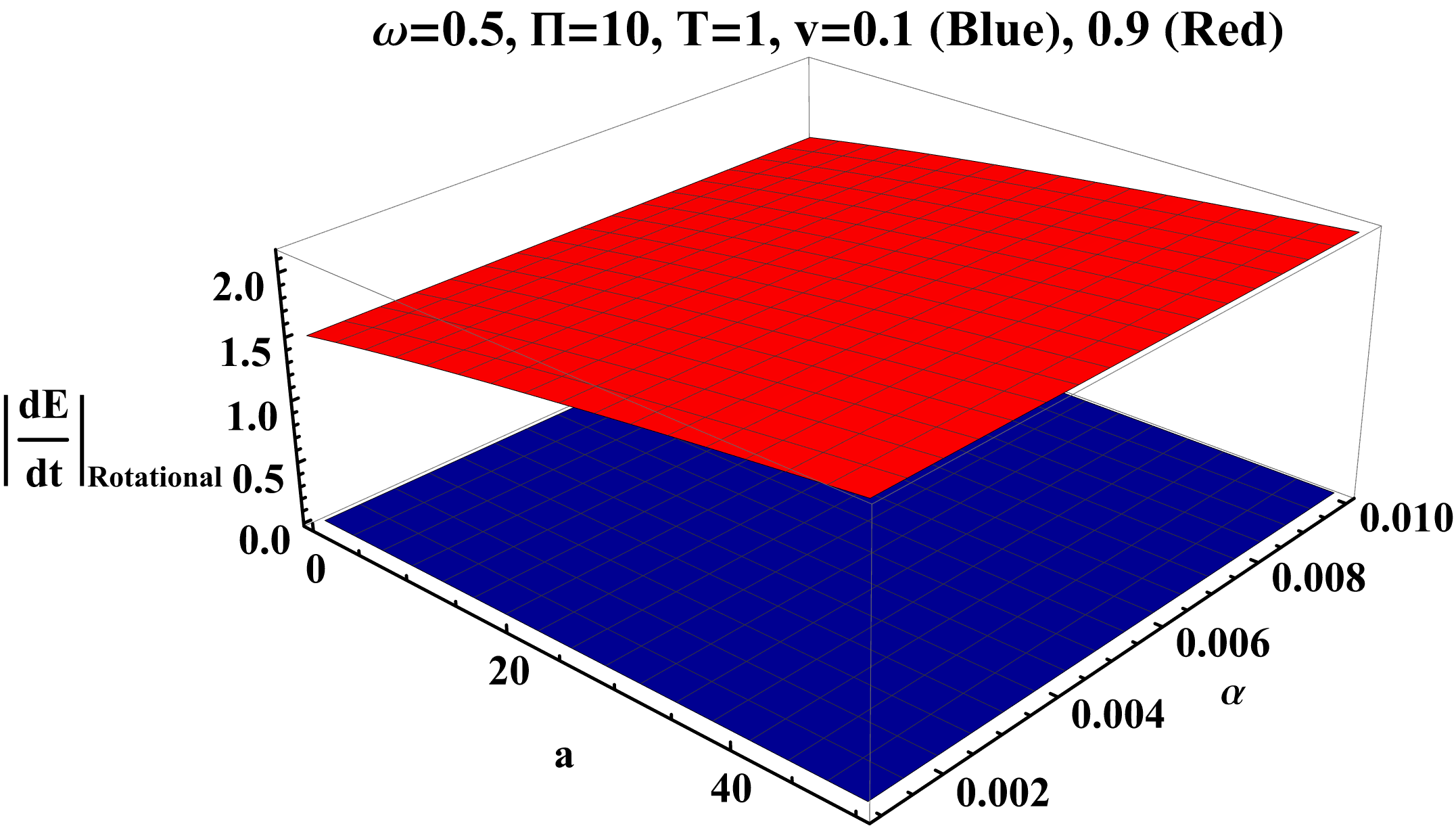}}
	\subfigure[]{\includegraphics[width=0.32\linewidth]{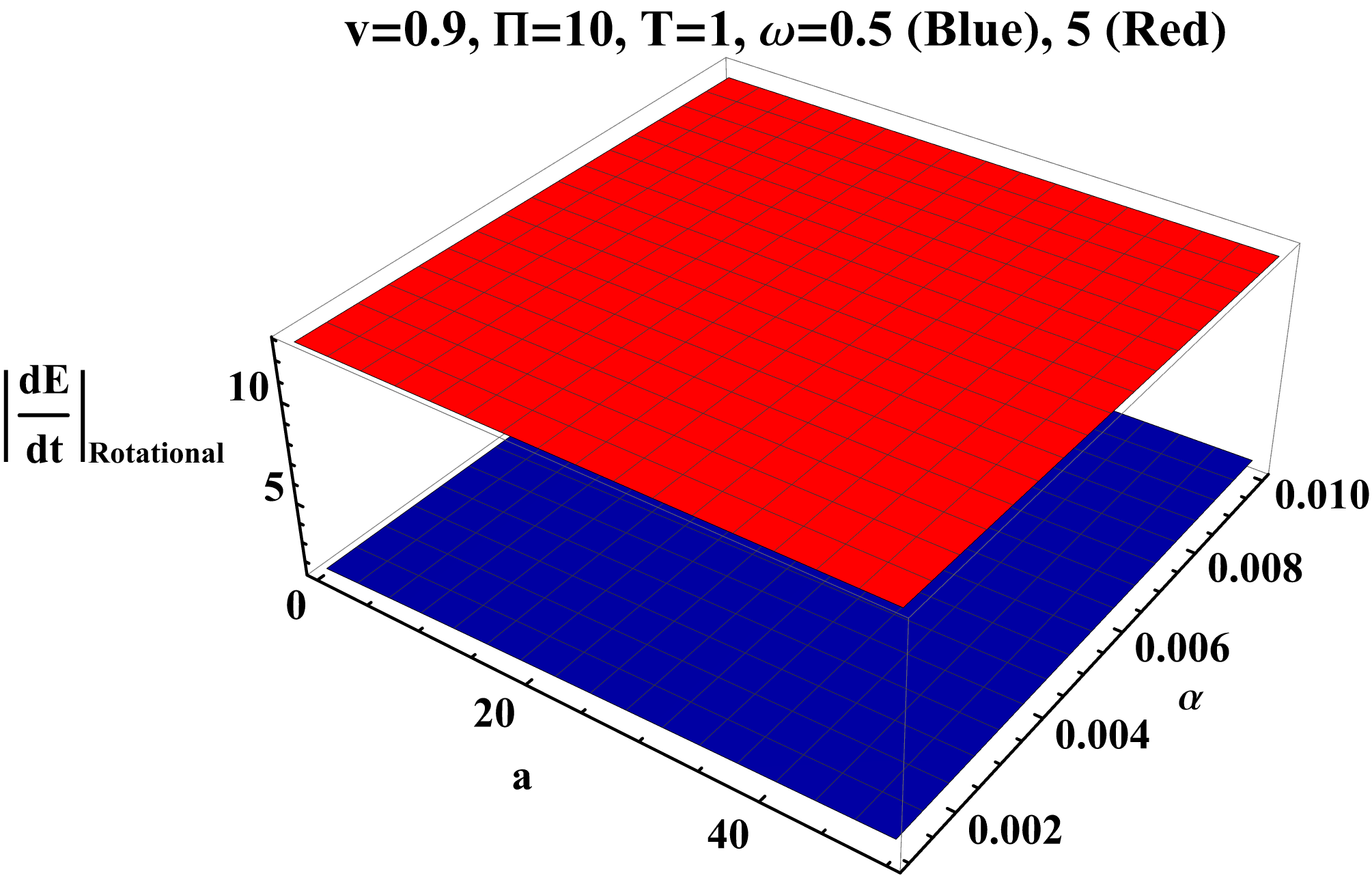}}
	\caption{Rotational energy loss 3D plot against $a$ and $\alpha$ for different (a) $T$, (b) $v$ and (c) $\omega$.}
	\label{fig_energyloss_rot3D}
\end{figure}

\begin{figure}[h]
	\centering
	\subfigure[]{\includegraphics[scale=0.105]{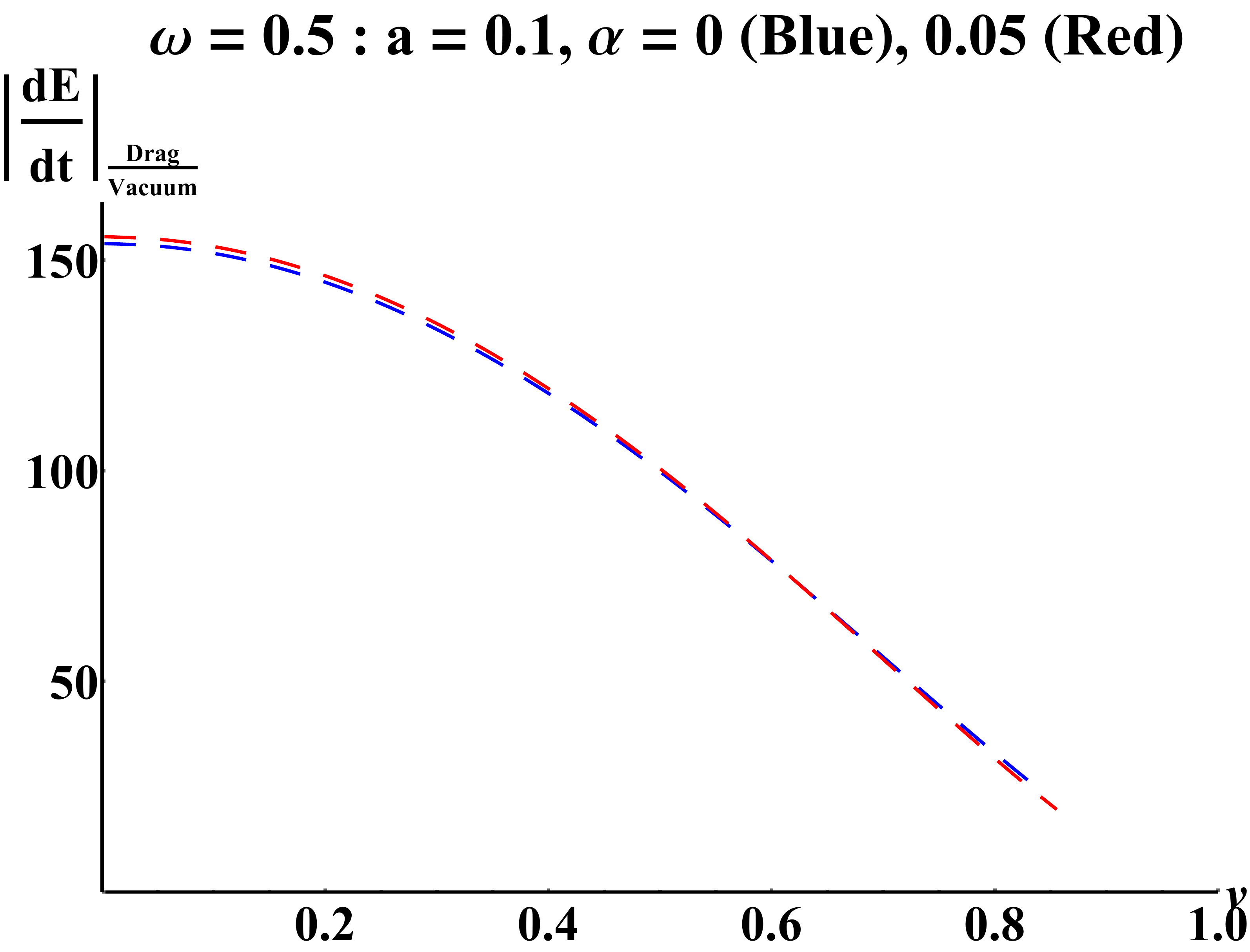}}
	\subfigure[]{\includegraphics[scale=0.105]{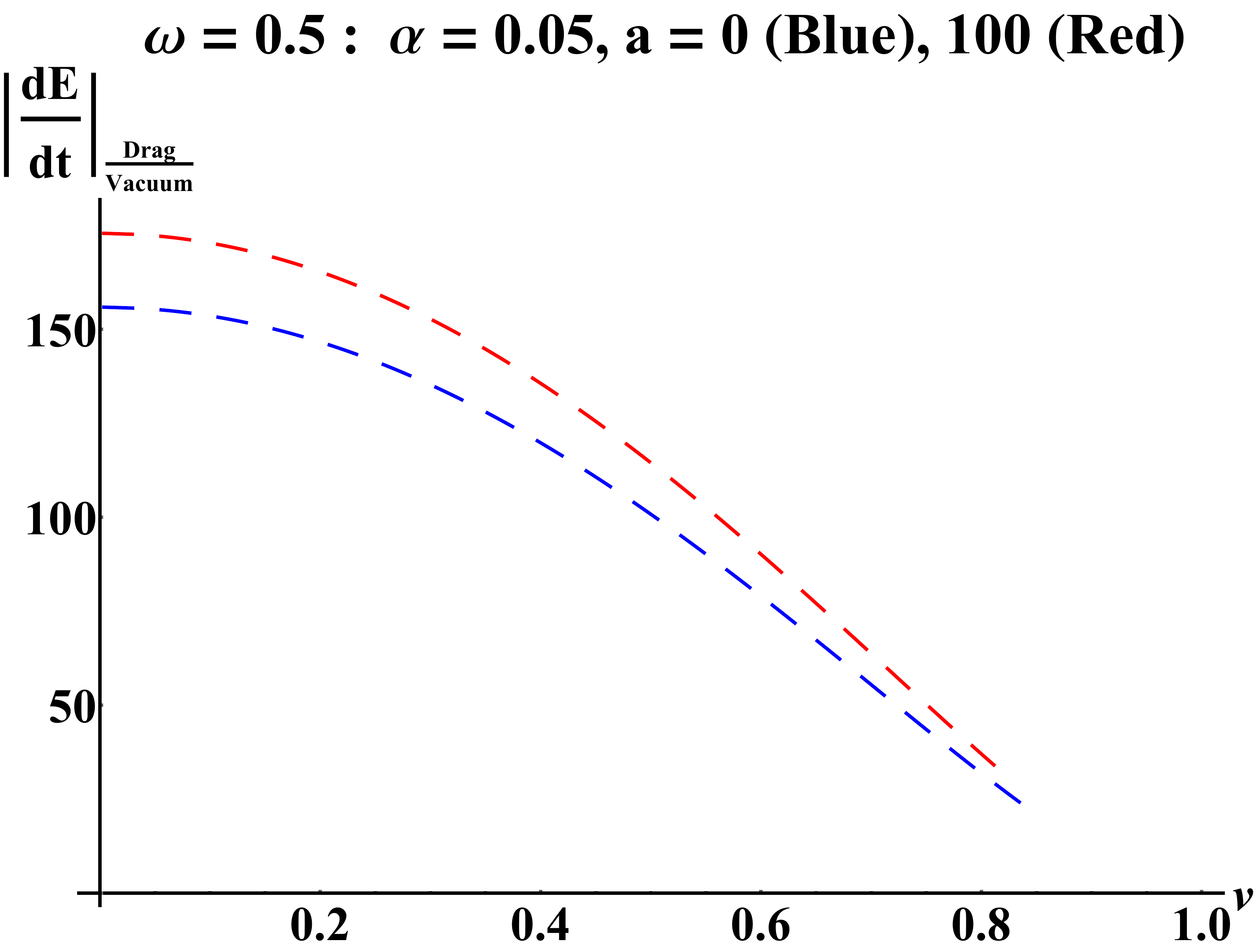}}
	\subfigure[]{\includegraphics[scale=0.095]{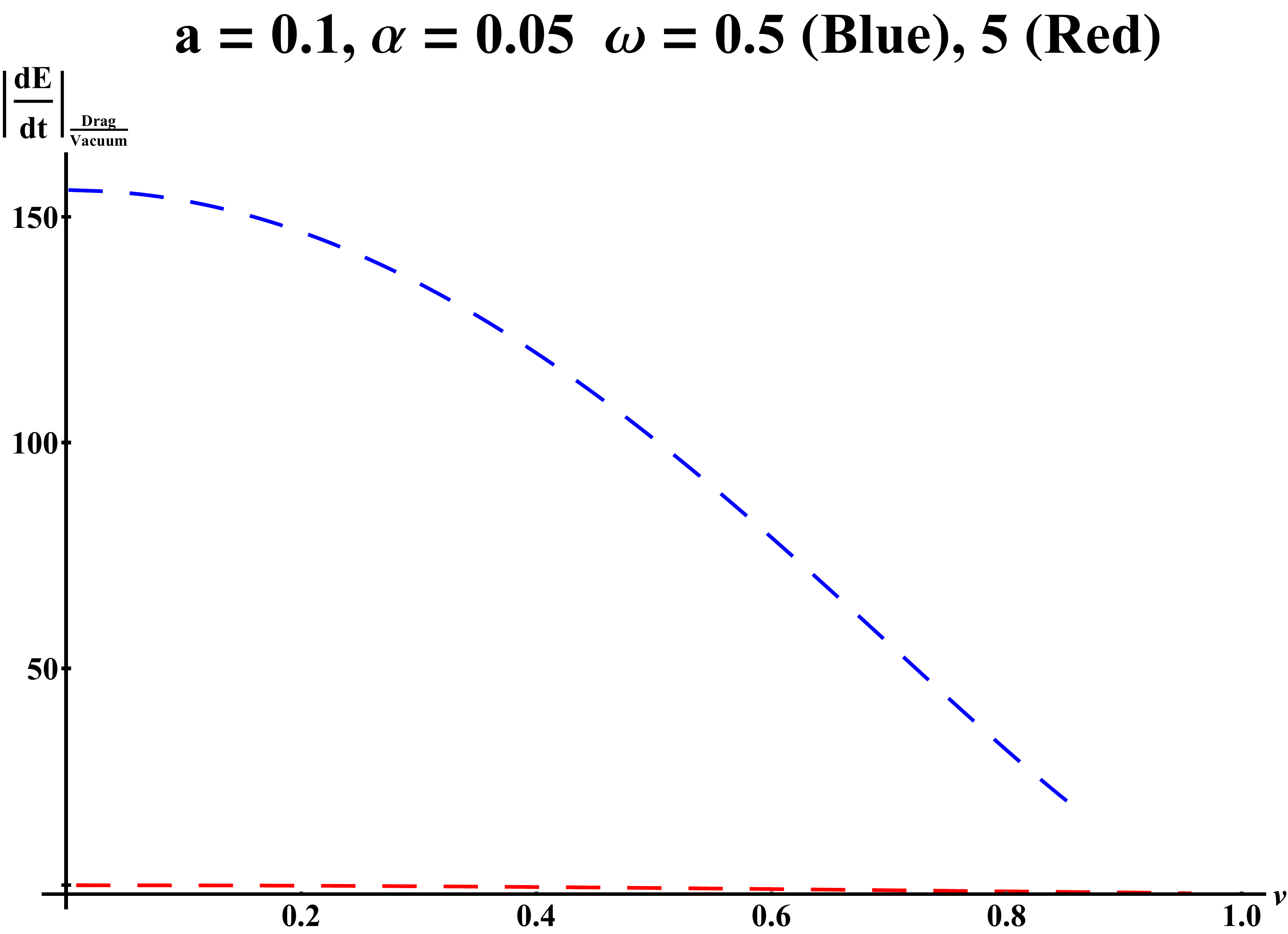}}
	\caption{ Ratio of drag to vacuum energy loss ($\frac{dE}{dt}$) against the velocity ($v$) for plot (a) different GB coefficient $\alpha$, plot (b) different string density $a$ and plot (c) different angular frequency $\omega$.}
	\label{fig_energyloss_DragByVac}
\end{figure}

\begin{figure}[h]
	\centering
	\subfigure[]{\includegraphics[scale=0.11]{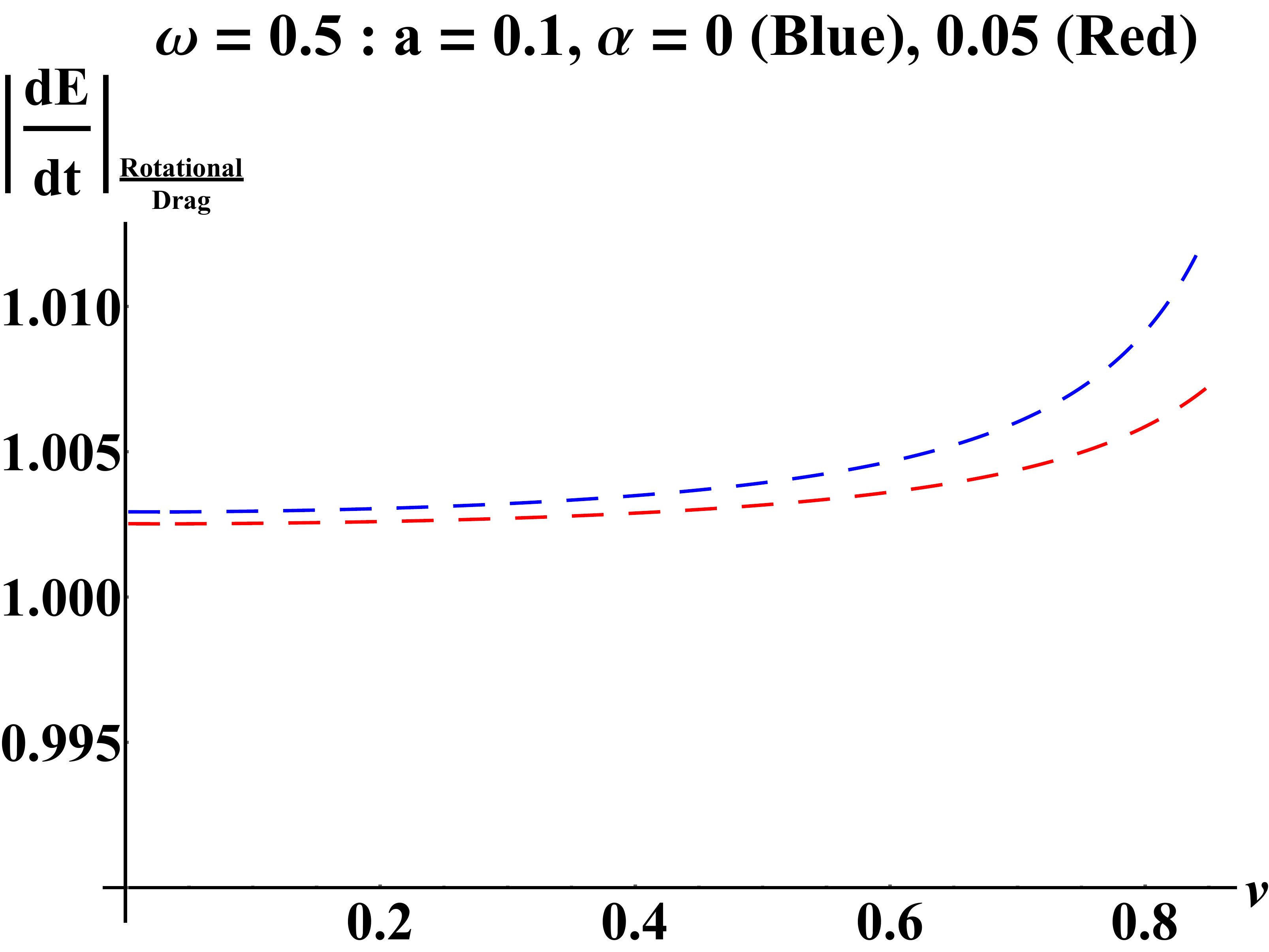}}
	\subfigure[]{\includegraphics[scale=0.11]{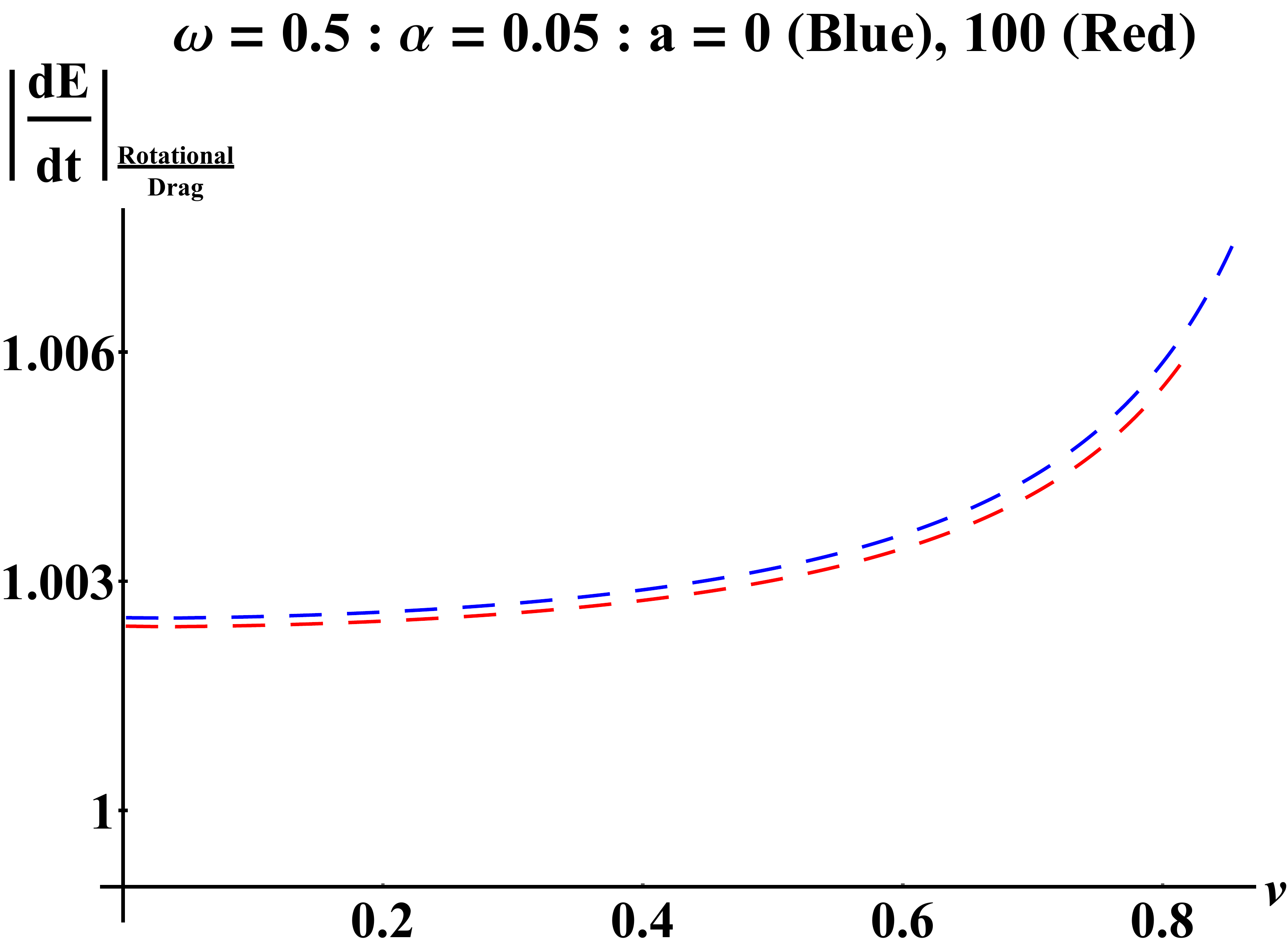}}
	\subfigure[]{\includegraphics[scale=0.11]{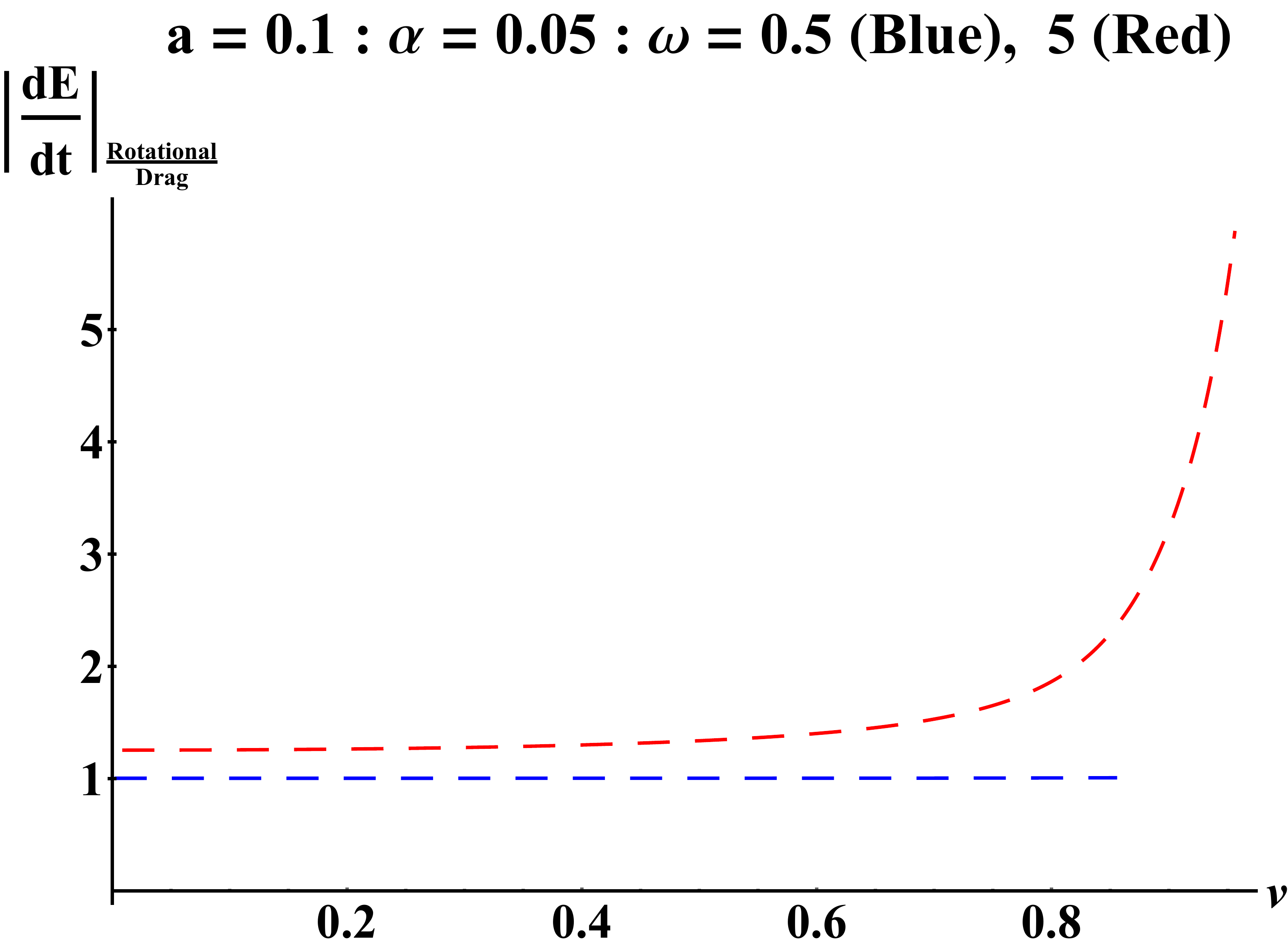}}
	\caption{ Plot of $\left.\frac{dE}{dt}\right|_{\frac{rotational}{drag}}$ against the velocity ($v$) for plot (a) different GB coefficient $\alpha$, plot (b) different string density $a$ and plot (c) different angular frequency $\omega$.}
	\label{fig_energyloss_RotByDrag}
\end{figure}

In this section, we study the energy loss of the rotating probe quark through the medium dissipation, radiation or vacuum energy loss. The interaction of the thermal medium with the probe quark in the presence of the higher derivative GB gravity and string cloud leads to the energy loss. Following \cite{Pokhrel_2025,Athanasiou_2010,Herzog_2007,Chakrabortty2016a}, holographically, the energy loss is studied by considering the rotating probe motion in the weakly coupled dual gravity background. Using the AdS/CFT duality, the rotational energy loss is expressed as,
\begin{equation}
	\left.\frac{dE}{dt}\right|_{rotational} = -\frac{\delta S}{\delta (\partial_\sigma X^0)} = \Pi_t^\sigma=\frac{h(u_c)}{2\pi \alpha' u_c^2},
\end{equation}
where $h(u_c)$ is the value of the metric function at the critical point $u_c$. From the above equation, the energy loss is studied relative to the probe string velocity $v$ for different set of parameters, such as the GB coefficient $\alpha$, string density $a$ and angular frequency $\omega$ keeping temperature $T$ fixed. The nature of the study is given in figure (\ref{fig_energyloss_rot3D}).
The energy loss enhances with increasing string density $a$ and angular frequency $\omega$, whereas the loss reduces slightly with the GB coefficient $\alpha$. Further, the energy loss increases with the increase of the velocity and for smaller $v$ the energy loss is negligible due to rotation.  
\begin{figure}[!h]
	\centering
	\subfigure[]{\includegraphics[width=0.32\linewidth]{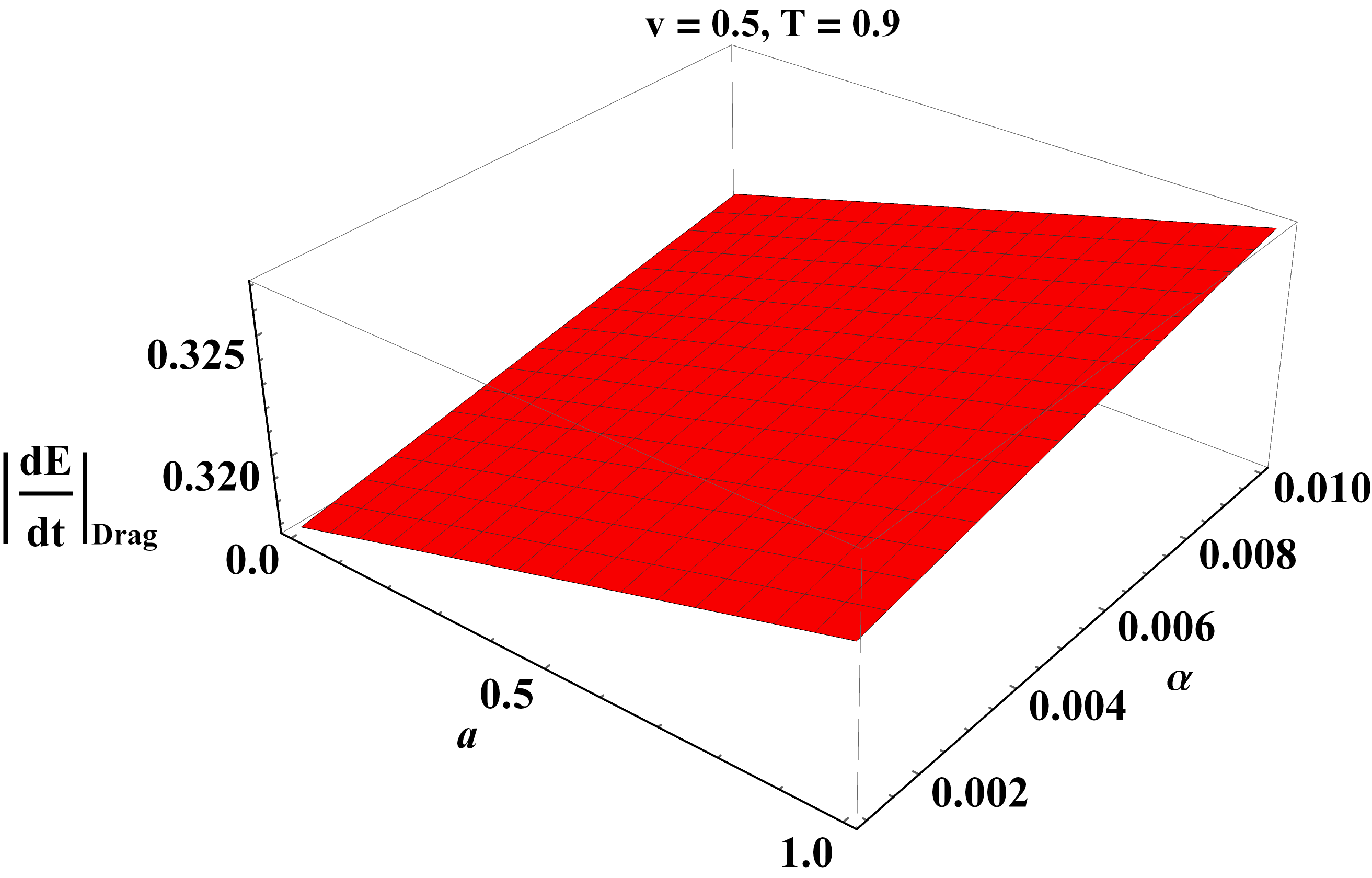}}
	\subfigure[]{\includegraphics[width=0.32\linewidth]{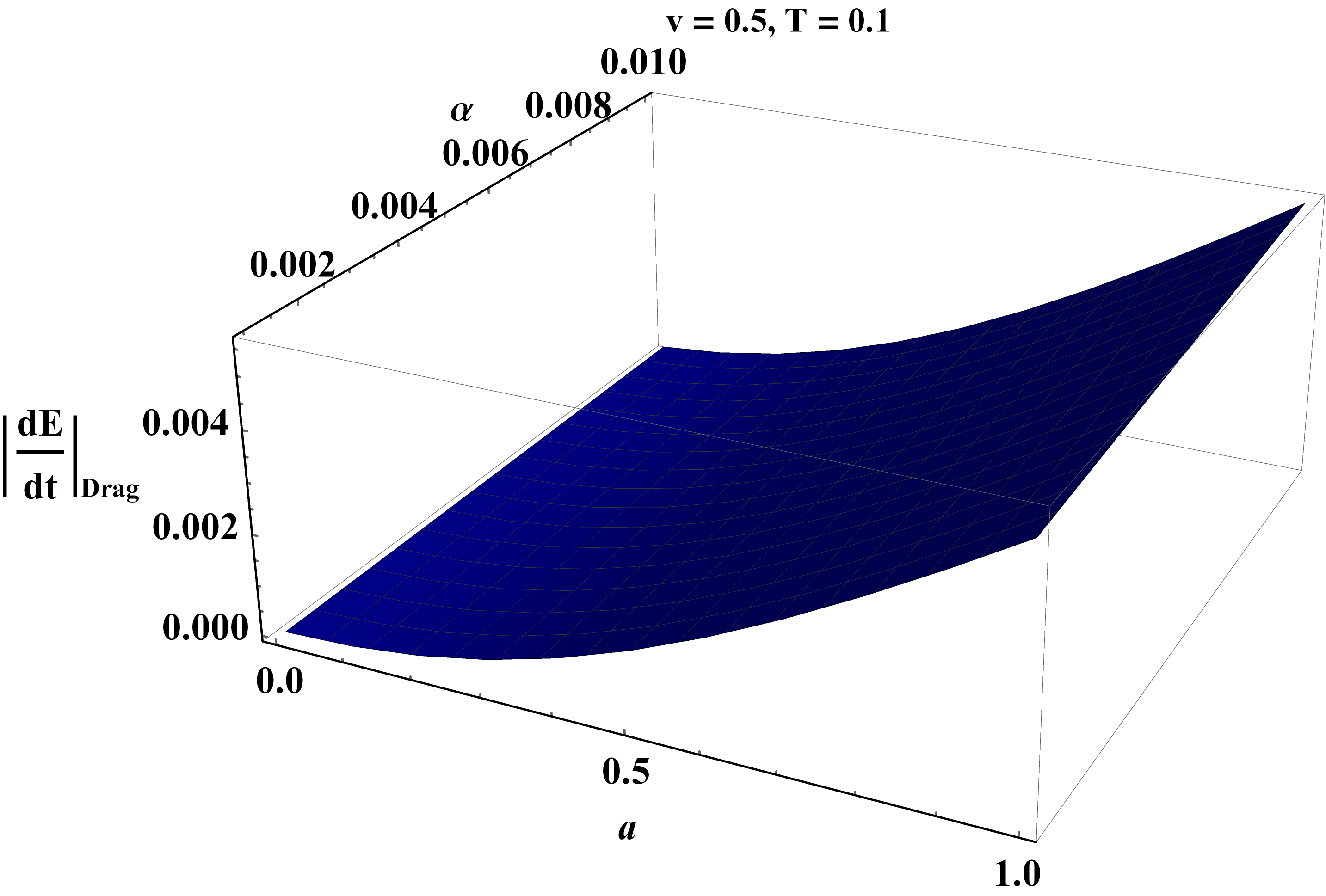}}
	\subfigure[]{\includegraphics[width=0.32\linewidth]{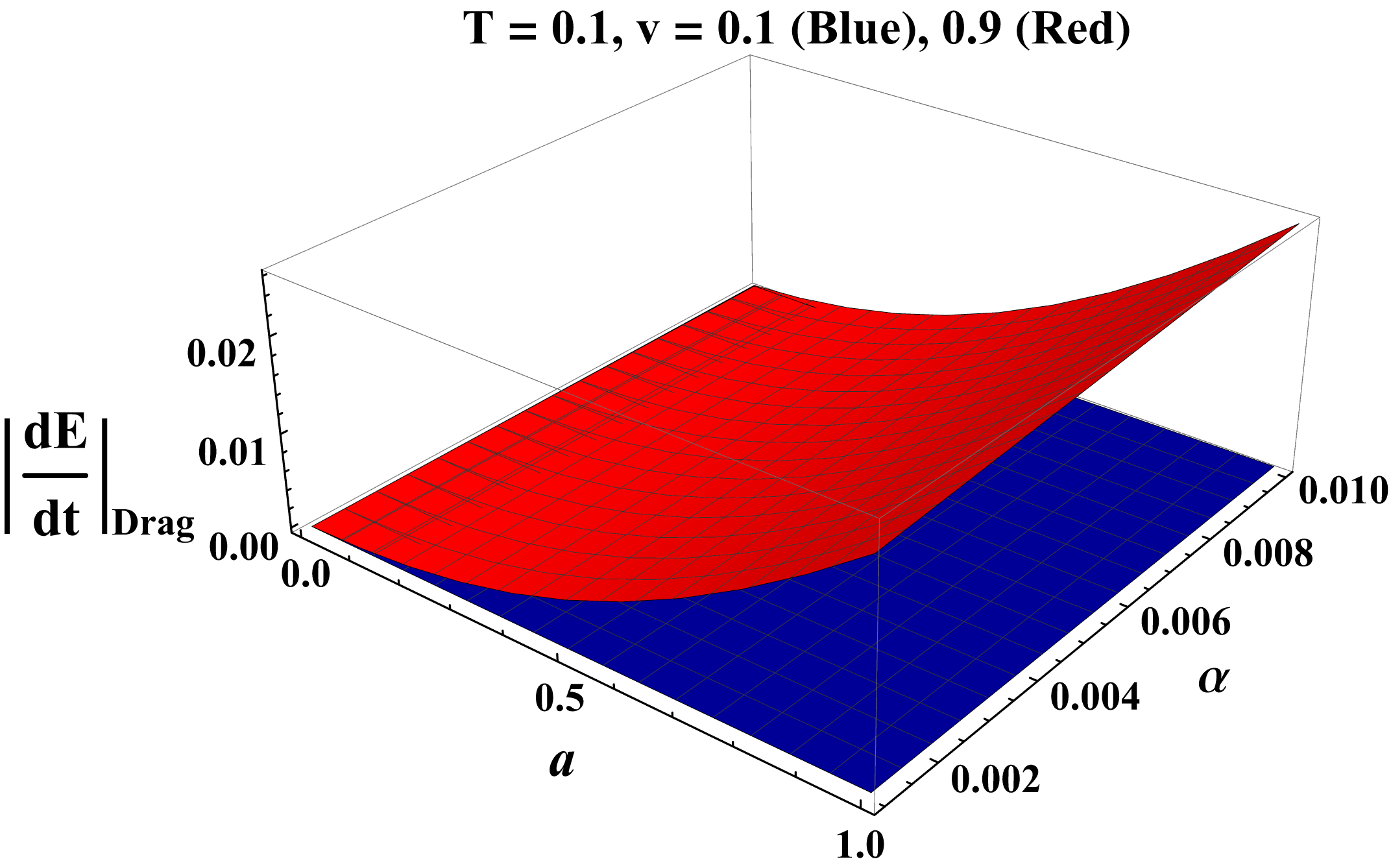}}
	\caption{Drag energy loss 3D plot against $a$ and $\alpha$ for (a) $T=0.9,\, v=0.5$, (b) $T=0.1,\,v=0.5$ and (c) $T=0.1,\,v=0.1,\, 0.9$.}
	\label{fig_energyloss_drag3D}
\end{figure}
Further, pure drag energy loss has been studied which is given by,
\begin{equation}
	\left.\frac{dE}{dt}\right|_{drag} = -\frac{\delta S}{\delta (\partial_\sigma X^1)} = \left.\frac{h(u_v)}{2\pi\alpha' u_v^2}\right|_{drag}.
\end{equation} 
In figure (\ref{fig_energyloss_drag3D}), the pure drag energy loss is plotted against the string density and GB coefficient. The nature of drag energy loss is similar to that of the rotational energy loss for various values of the GB coefficient and string density, except for high temperature and velocity region the drag energy loss slightly decreases with increasing GB coefficient. Additionally, the drag energy loss is plotted for different values of temperature and velocity and it is observed that with increase in temperature and velocity the drag energy loss increases.
\begin{figure}[!h]
	\centering
	\subfigure[]{\includegraphics[width=0.45\linewidth]{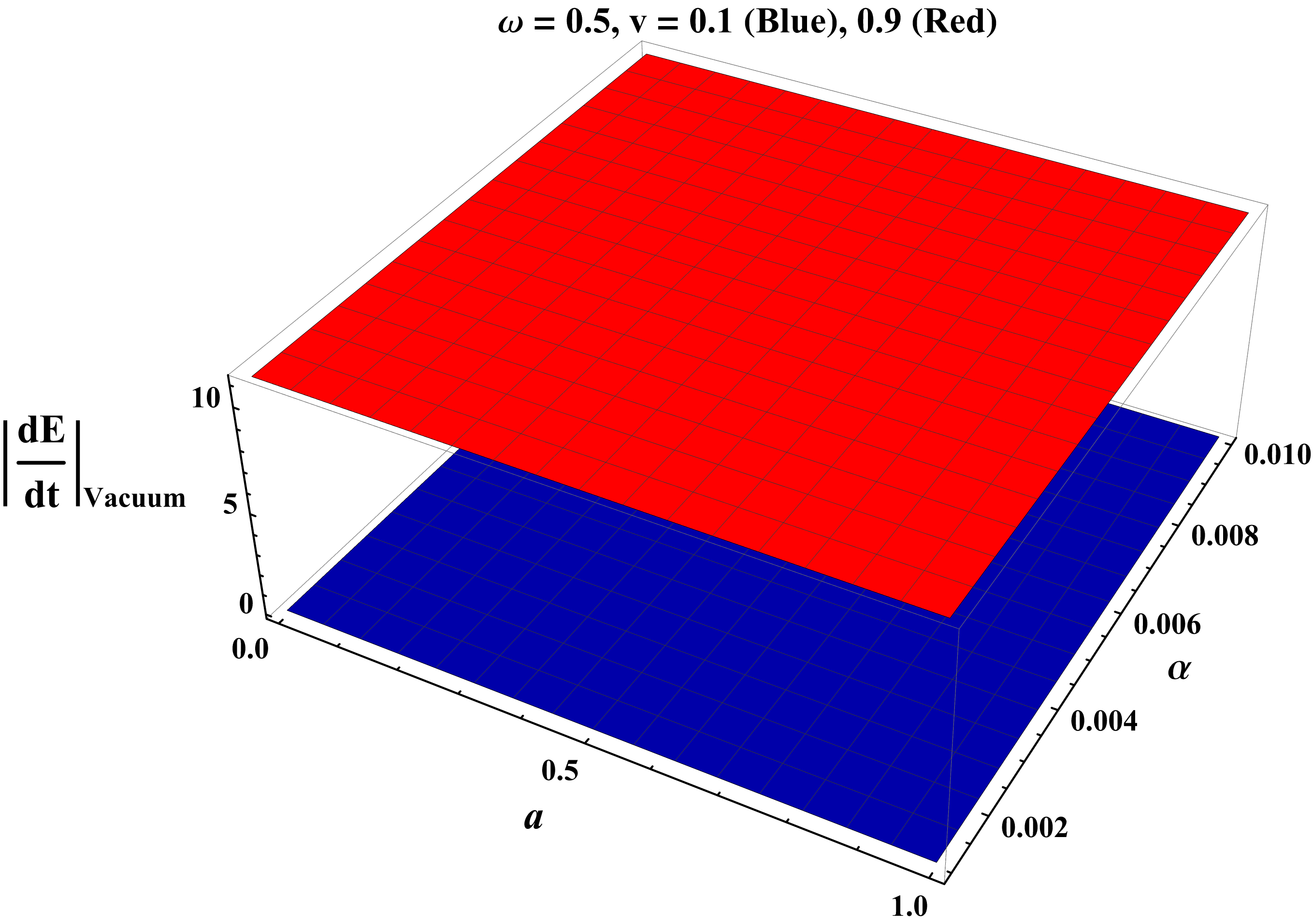}}
	\subfigure[]{\includegraphics[width=0.45\linewidth]{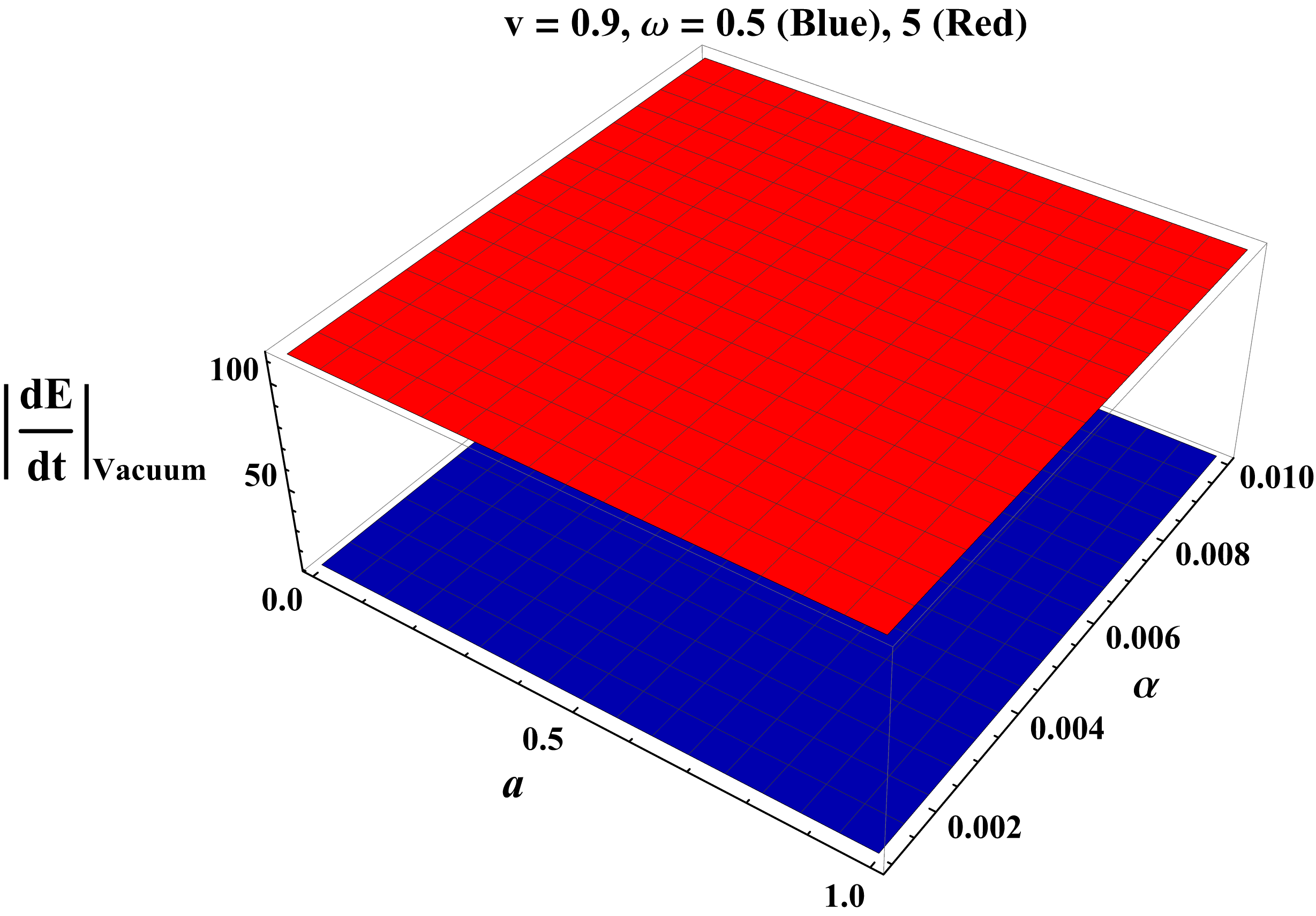}}
	\caption{Vacuum energy loss 3D plot against $a$ and $\alpha$ for (a) fixed $\omega$ and different $v$ and (b) fixed $v$ and different $\omega$.}
	\label{fig_energyloss_vac3D}
\end{figure}
The pure vacuum energy loss has been studied in \cite{Mikhailov2003}, which is given as,
\begin{equation}
	\left.\frac{dE}{dt}\right|_{vacuum} \approx \frac{v^2 \omega^2}{(1-v^2)^2}.
\end{equation}
It is found from figure (\ref{fig_energyloss_vac3D}), that the vacuum energy loss is independent of GB coefficient and flavour density, but it depends on the angular frequency $\omega$ and the velocity $v$. The vacuum energy loss enhances with angular frequency $\omega$ and velocity $v$.
\begin{figure}[h]
	\centering
	\subfigure[]{\includegraphics[scale=0.11]{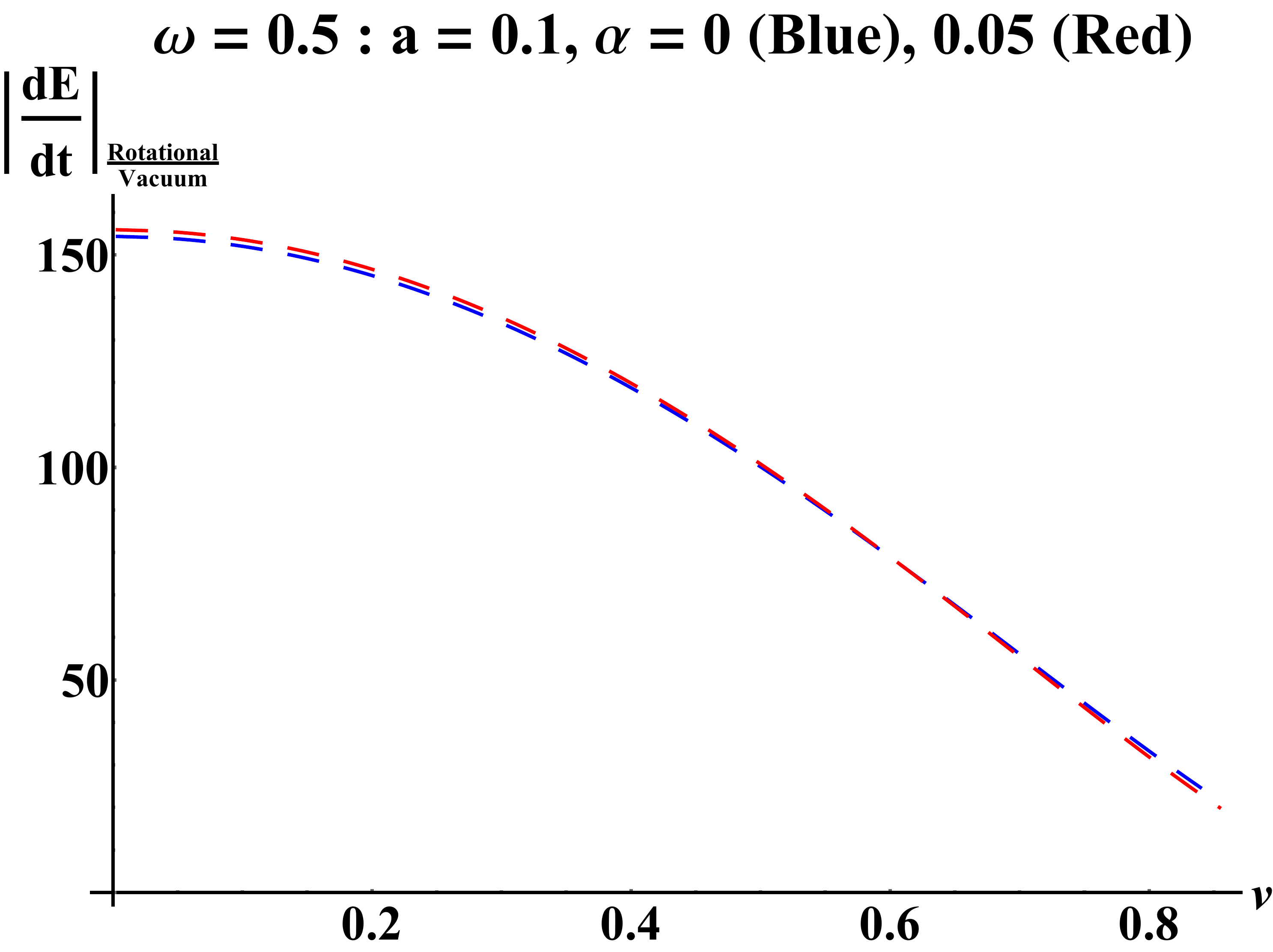}}
	\subfigure[]{\includegraphics[scale=0.11]{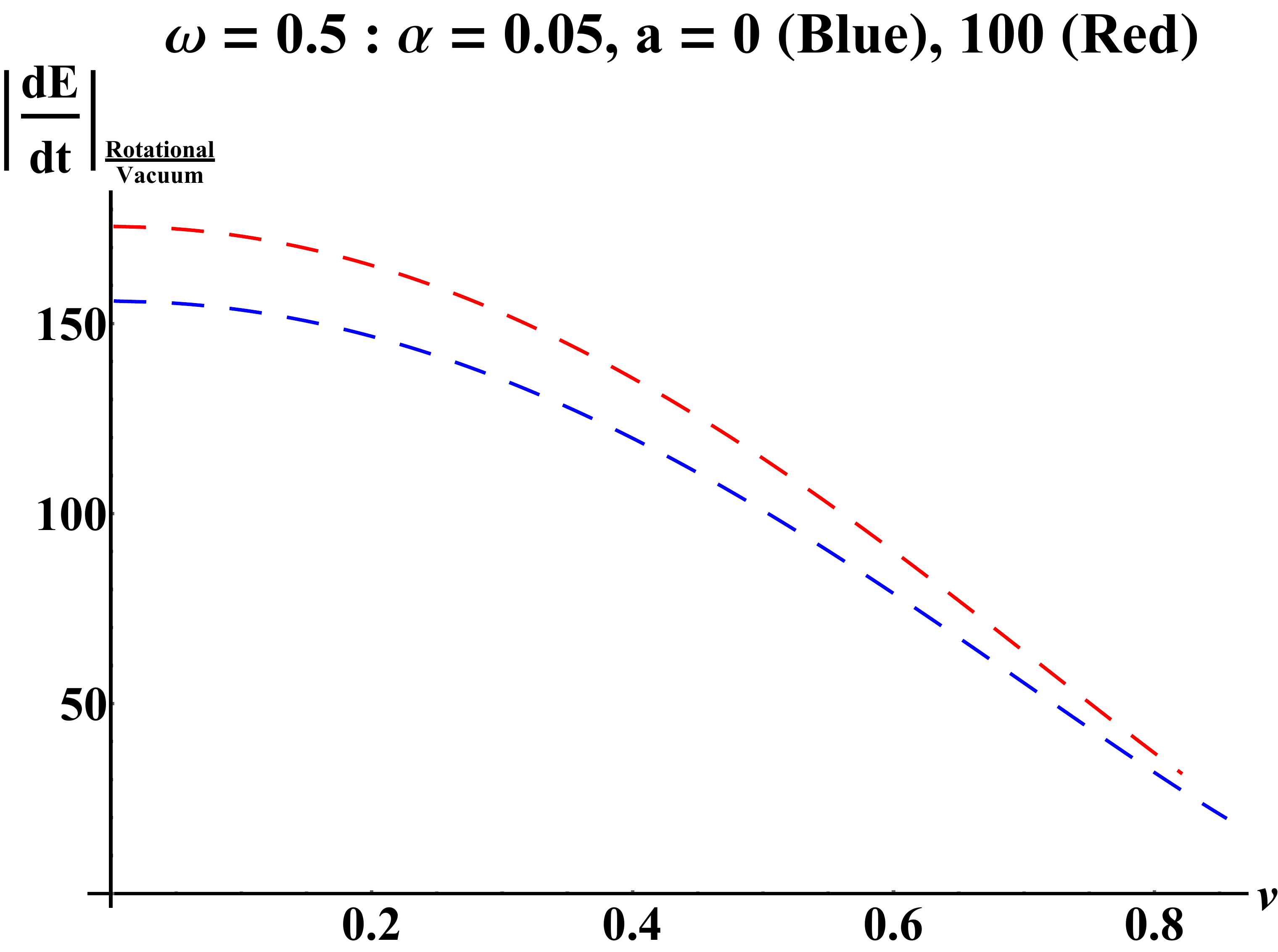}}
	\subfigure[]{\includegraphics[scale=0.11]{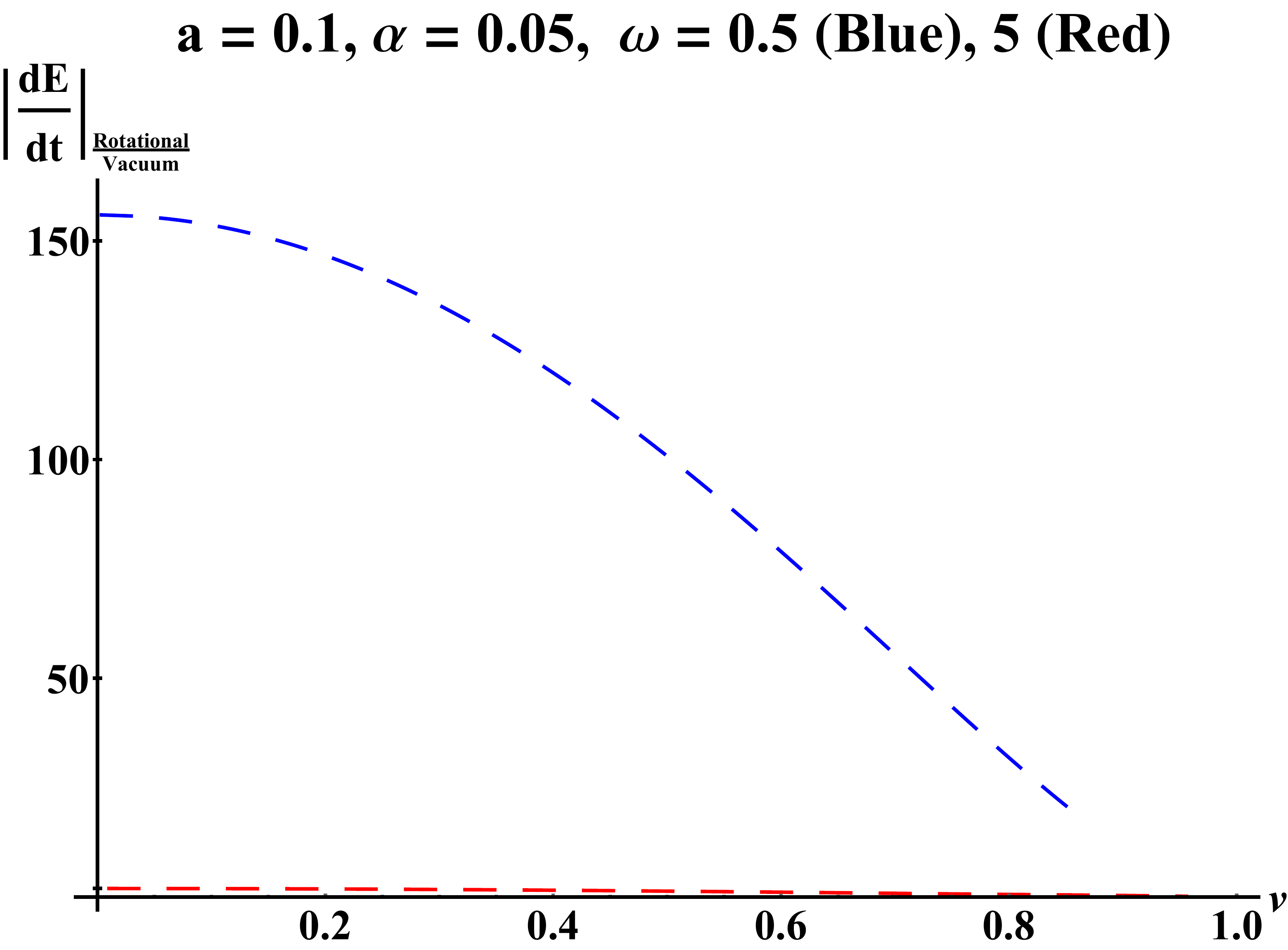}}
	\caption{ Plot of $\left.\frac{dE}{dt}\right|_{\frac{rotational}{vacuum}}$ against the velocity ($v$) for plot (a) different GB coefficient $\alpha$, plot (b) different string density $a$ and plot (c) different angular frequency $\omega$.}
	\label{fig_energyloss_RotByVac}
\end{figure}
In figure (\ref{fig_energyloss_RotByVac}), $\left.\frac{dE}{dt}\right|_{\frac{rotational}{vacuum}}$ is plotted against $v$. For the velocity of the probe $v$ less than the velocity of light,  the vacuum loss is dominated by the rotational loss. The ratio of dominance is large for larger value of string density, whereas it is small for larger value of $\omega$ and has negligible dependence on GB coefficient. When the probe velocity approaches the speed of light, both rotational and vacuum equally radiates but with opposite phase, like in the case of destructive interference. Furthermore, in figure (\ref{fig_energyloss_DragByVac}), the ratio of the drag to the vacuum loss is plotted, from which it is observed that the overall nature of the drag to vacuum energy loss is similar to that of the rotational to vacuum energy loss. It can be concluded from figure (\ref{fig_energyloss_RotByVac}) and (\ref{fig_energyloss_DragByVac}) that the rotational and drag radiation are in same phase like in the case of constructive interference. To reconfirm this, the ratio of rotational to drag loss is plotted in figure (\ref{fig_energyloss_RotByDrag}) and it is observed that, for small value of the probe velocity, the ratio approaches unity, which signifies that both the rotational and drag loss are equal.  Whereas for higher value of the probe velocity, specifically near to the speed of light, the ratio becomes greater than unity, which signifies that they are in same phase as if they are constructively interfering. 
Further, for higher value of the GB coefficient and flavour density, the rotational dominance is reduced, as the velocity approaches the speed of light.

%%%%%%%%%%%%%%%%%%%%%%%%%%%%%%%%%%%%%%%%%%%%%%%%%%%%%%%%%%%%%%%%%%%%%%%%%%%%%%%%%%%%%%%
\section{Conclusion and Discussion}\label{sec_conclusion}
In this work, employing the ideas of AdS/CFT duality, we studied the hydrodynamical properties of four dimensional $\mathcal{N}=4$ SYM thermal plasma with heavy flavour quarks and finite coupling correction dual to a five dimensional AdS Gauss-Bonnet gravity with string cloud background. The end points of the string cloud are attached to the boundary which corresponds to the heavy flavour quarks in the dual gauge theory and the GB correction term is said to correspond to the finite correction in 't Hooft coupling. Due to the presence of heavy flavour quark and finite correction, the strongly coupled QGP may be considered as perfect fluid as viscosity to entropy density ratio becomes less than the KSS bound and greater than zero if the GB coupling,  string density and temperature maintain a constraint relation which approaches to the heavy ion collision. With the consideration of an external probe quark traversing the thermal plasma, we have computed the drag force exerted on the probe quark and the corresponding energy loss. Considering, an external quark-antiquark pair as a probe, the screening length and jet quenching parameters has been studied. The external probe quark is represented in gravity dual by an external string, having one end attached to the boundary and whose body is hanging towards the horizon. The meson probe is realised as an external string with its end points attached to the boundary and the string's body hanging towards the horizon with a turning point in between.

The traversing probe experiences a drag force which has been studied and is found that the drag force increases with increasing value of velocity, temperature, string cloud density and GB coefficient. However, in high temperature and velocity region the drag experienced slightly decreases with increasing GB coefficient.

We have analysed the separation distance between $q\bar{q}$ pair, in both the perpendicular and parallel orientation of their axes with respect to its direction of motion. The separation increases monotonically, reaches a maximum (screening length $L_S$ of the pair) and then falls off relative to the constant of motion for both the configurations.  It is observed that in both the orientations, $L_S$ decreases and binding energy increases with increasing value of the GB coefficient, rapidity, string density and temperature, signifying rapid transition from bound to unbound state. Additionally, for the same set of parameters, $L_S$ in the parallel orientation is consistently greater than the perpendicular orientation suggesting a more stable quark–antiquark configuration when the alignment is parallel to the motion.

The jet quenching has been studied against parameters such as the coefficient of GB correction, temperature and string cloud density. The quenching parameter describes the suppression of the high transverse momentum in the thermal medium. We observed that the jet quenching increases with increasing values of the parameters suggesting enhancement in high transverse momentum suppression.

With the consideration of constantly rotating heavy probe quark in the boundary gauge theory, the radial profile of the probe string in the gravity with its corresponding energy loss is analysed. The radial profile remains constant for the value of angular frequency less than unity, whereas for angular frequency greater than unity, the radial profile starts bending and the radius becomes larger towards the horizon than in the boundary. For fixed value of frequency and temperature, the probe radius decreases with increasing value of the string cloud density, whereas the probe radius is almost same for the increasing value of the GB coefficient. The radial profile is closely associated to the energy loss. We have computed the rotational, drag and vacuum energy losses and further, we have also analysed their ratios with each other. For fixed value of temperature, the rotational energy loss increases at high velocity with increasing string density and angular frequency, whereas it decreases slightly with increasing GB coefficient. In the limit where velocity tends to zero, the energy loss becomes zero. Similarly, qualitative nature of drag energy loss is studied as that of the rotational energy loss. With the increase of temperature, the drag energy loss enhances. The pure vacuum energy loss depends only on the angular frequency and velocity $v$ but not on the values of string density and GB coefficient. The pure vacuum energy radiation is equivalent to the Lienard's electromagnetic radiation of the accelerating probe quark. The ratio of these energy losses are then studied. First of all, in the low velocity regime, the vacuum energy loss is dominated by the rotational energy loss. However in the high velocity regime, both radiate equally and the ratio tends toward unity, signifying opposite phase of these two radiation undergoing a destructive interference. Further, the rotational dominance is lower for larger value of angular frequency and greater for larger values of GB coefficient and string density. The ratio of the drag to the vacuum loss, suggests similar conclusion as that of the rotational to vacuum energy loss. Finally, we studied the ratio of the rotational to drag energy loss and found that the ratio is unity for smaller range of the velocity and angular frequency. As the velocity approaches the speed of light, the ratio increases and becomes more than unity, suggesting a constructive interference between rotational and drag energy radiations. The drag loss is dominated by the rotational loss for higher value of the angular frequency.

Overall, we have observed that drag force enhances with temperature, density of quark, the probe quark's velocity and finite correction in 't Hooft coupling whereas screening length reduces. The jet quenching parameter also increases with finite correction, density of quark and temperature.  When a probe quark rotates with a certain radius in the boundary, the corresponding string's radial profile does not depend on the finite correction, though it decreases with increasing string density in the bulk spacetime. The drag and rotational energy losses enhance with string density but reduces for finite correction when the velocity of the probe tends to speed of light. Finally, the vacuum energy loss does not depend on string density and finite correction.
%%%%%%%%%%%%%%%%%%%%%%%%%%%%%%%%%%%%%%%%%%%%%%%%%%%%%%%%%%%%%%%%%%%%%%%%%%%%%%%%%%%%%%%%%%%%
\section{Acknowledgment}
RP and IKPC would like to thank the T.M.A Pai research grant provided by the Sikkim Manipal Institute of Technology, Sikkim Manipal University (SMU). KPS would like to thank the Ministry of Tribal affairs, Government of India for providing the National Fellowship for Schedule Tribes(NFST). RP would also like to thank B. Sharma for providing valuable feedback on the manuscript.

\printbibliography

@article{Sadeghi:2022kgi,
		author = "Sadeghi, Mehdi",
		title = "{The effect of three matters on KSS bound}",
		eprint = "2203.16849",
		archivePrefix = "arXiv",
		primaryClass = "hep-th",
		doi = "10.1142/S021773232350181X",
		journal = "Mod. Phys. Lett. A",
		volume = "39",
		number = "01",
		pages = "2350181",
		year = "2024"
	}

@article{Son:2007vk,
    author = "Son, Dam T. and Starinets, Andrei O.",
    title = "{Viscosity, Black Holes, and Quantum Field Theory}",
    eprint = "0704.0240",
    archivePrefix = "arXiv",
    primaryClass = "hep-th",
    reportNumber = "INT-PUB-07-02",
    doi = "10.1146/annurev.nucl.57.090506.123120",
    journal = "Ann. Rev. Nucl. Part. Sci.",
    volume = "57",
    pages = "95--118",
    year = "2007"
}

@article{Kovtun:2004de,
    author = "Kovtun, P. and Son, Dan T. and Starinets, Andrei O.",
    title = "{Viscosity in strongly interacting quantum field theories from black hole physics}",
    eprint = "hep-th/0405231",
    archivePrefix = "arXiv",
    reportNumber = "INT-PUB-04-09, UW-PT-04-04",
    doi = "10.1103/PhysRevLett.94.111601",
    journal = "Phys. Rev. Lett.",
    volume = "94",
    pages = "111601",
    year = "2005"
}

@article{Romatschke:2007mq,
    author = "Romatschke, Paul and Romatschke, Ulrike",
    title = "{Viscosity Information from Relativistic Nuclear Collisions: How Perfect is the Fluid Observed at RHIC?}",
    eprint = "0706.1522",
    archivePrefix = "arXiv",
    primaryClass = "nucl-th",
    reportNumber = "INT-PUB-07-14",
    doi = "10.1103/PhysRevLett.99.172301",
    journal = "Phys. Rev. Lett.",
    volume = "99",
    pages = "172301",
    year = "2007"
}

@article{Song:2007fn,
    author = "Song, Huichao and Heinz, Ulrich W.",
    title = "{Suppression of elliptic flow in a minimally viscous quark-gluon plasma}",
    eprint = "0709.0742",
    archivePrefix = "arXiv",
    primaryClass = "nucl-th",
    reportNumber = "CERN-PH-TH-2007-154",
    doi = "10.1016/j.physletb.2007.11.019",
    journal = "Phys. Lett. B",
    volume = "658",
    pages = "279--283",
    year = "2008"
}

@article{PHENIX:2006iih,
    author = "Adare, A. and others",
    collaboration = "PHENIX",
    title = "{Energy Loss and Flow of Heavy Quarks in Au+Au Collisions at s(NN)**(1/2) = 200-GeV}",
    eprint = "nucl-ex/0611018",
    archivePrefix = "arXiv",
    doi = "10.1103/PhysRevLett.98.172301",
    journal = "Phys. Rev. Lett.",
    volume = "98",
    pages = "172301",
    year = "2007"
}

@article{Romatschke:2007eb,
    author = "Romatschke, Paul",
    editor = "Kunihiro, Teiji and Fukushima, Kenji and Hirano, Tetsufumi and Iida, Kei and Kitazawa, Masakiyo and Tachibana, Motoi and Iida, Hideaki and Takahashi, Toru T.",
    title = "{Fluid turbulence and eddy viscosity in relativistic heavy-ion collisions}",
    eprint = "0710.0016",
    archivePrefix = "arXiv",
    primaryClass = "nucl-th",
    reportNumber = "INT-PUB-07-26",
    doi = "10.1143/PTPS.174.137",
    journal = "Prog. Theor. Phys. Suppl.",
    volume = "174",
    pages = "137--144",
    year = "2008"
}

@article{Dusling:2007gi,
    author = "Dusling, K. and Teaney, D.",
    title = "{Simulating elliptic flow with viscous hydrodynamics}",
    eprint = "0710.5932",
    archivePrefix = "arXiv",
    primaryClass = "nucl-th",
    doi = "10.1103/PhysRevC.77.034905",
    journal = "Phys. Rev. C",
    volume = "77",
    pages = "034905",
    year = "2008"
}

@article{Pokhrel_2025,
    author = "Pokhrel, Rishi and Dey, Tanay K.",
    title = "{Hydrodynamical properties of baryon rich thermal plasma with flavour quarks}",
    eprint = "2410.14384",
    archivePrefix = "arXiv",
    primaryClass = "hep-th",
    doi = "10.1140/epjc/s10052-025-14039-7",
    journal = "Eur. Phys. J. C",
    volume = "85",
    number = "3",
    pages = "349",
    year = "2025"
}

@article{Fadafan_2008,
   title={R2curvature-squared corrections on drag force},
   volume={2008},
   ISSN={1029-8479},
   url={http://dx.doi.org/10.1088/1126-6708/2008/12/051},
   DOI={10.1088/1126-6708/2008/12/051},
   number={12},
   journal={Journal of High Energy Physics},
   publisher={Springer Science and Business Media LLC},
   author={Fadafan, Kazem Bitaghsir},
   year={2008},
   month=dec, pages={051–051} }

@article{Atashi_2020,
   title={Spiraling string in Gauss–Bonnet geometry},
   volume={800},
   ISSN={0370-2693},
   url={http://dx.doi.org/10.1016/j.physletb.2019.135090},
   DOI={10.1016/j.physletb.2019.135090},
   journal={Physics Letters B},
   publisher={Elsevier BV},
   author={Atashi, Mahdi and Bitaghsir Fadafan, Kazem},
   year={2020},
   month=jan, pages={135090} }

@article{Dey:2020yzl,
    author = "Dey, Tanay K. and Mukhopadhyay, Subir",
    title = "{AdS black holes with higher derivative corrections in presence of string cloud}",
    eprint = "2005.12054",
    archivePrefix = "arXiv",
    primaryClass = "hep-th",
    doi = "10.1140/epjc/s10052-020-08565-9",
    journal = "Eur. Phys. J. C",
    volume = "80",
    number = "11",
    pages = "1012",
    year = "2020"
}

@article{Hou_2021,
    author = "Hou, Defu and Atashi, Mahdi and Bitaghsir Fadafan, Kazem and Zhang, Zi-qiang",
    title = "{Holographic energy loss of a rotating heavy quark at finite chemical potential}",
    doi = "10.1016/j.physletb.2021.136279",
    journal = "Phys. Lett. B",
    volume = "817",
    pages = "136279",
    year = "2021"
}

@article{Chen_2024,
    author = "Chen, Bing and Chen, Xun and Li, Xiaohua and Zhu, Zhou-Run and Zhou, Kai",
    title = "{Exploring Transport Properties of Quark-Gluon Plasma with a Machine-Learning assisted Holographic Approach}",
    eprint = "2404.18217",
    archivePrefix = "arXiv",
    primaryClass = "hep-ph",
    month = "4",
    year = "2024"
}

@article{Bigazzi_2011,
   title={D3-D7 quark-gluon plasmas at finite baryon density},
   volume={2011},
   ISSN={1029-8479},
   url={http://dx.doi.org/10.1007/JHEP04(2011)060},
   DOI={10.1007/jhep04(2011)060},
   number={4},
   journal={Journal of High Energy Physics},
   publisher={Springer Science and Business Media LLC},
   author={Bigazzi, Francesco and Cotrone, Aldo L. and Mas, Javier and Mayerson, Daniel and Tarrío, Javier},
   year={2011},
   month=apr }

@article{Bigazzi_2009,
   title={D3-D7 quark-gluon plasmas},
   volume={2009},
   ISSN={1029-8479},
   url={http://dx.doi.org/10.1088/1126-6708/2009/11/117},
   DOI={10.1088/1126-6708/2009/11/117},
   number={11},
   journal={Journal of High Energy Physics},
   publisher={Springer Science and Business Media LLC},
   author={Bigazzi, Francesco and Cotrone, Aldo L and Mas, Javier and Paredes, Angel and Ramallo, Alfonso V and Tarrío, Javier},
   year={2009},
   month=nov, pages={117–117} }

@article{Cotrone_2007,
   title={Notes on a SQCD-like plasma dual and holographic renormalization},
   volume={2007},
   ISSN={1029-8479},
   url={http://dx.doi.org/10.1088/1126-6708/2007/11/034},
   DOI={10.1088/1126-6708/2007/11/034},
   number={11},
   journal={Journal of High Energy Physics},
   publisher={Springer Science and Business Media LLC},
   author={Cotrone, A.L and Pons, J.M and Talavera, P},
   year={2007},
   month=nov, pages={034–034} }

@article{Mikhailov2003,
    author = "Mikhailov, Andrei",
    title = "{Nonlinear waves in AdS / CFT correspondence}",
    eprint = "hep-th/0305196",
    archivePrefix = "arXiv",
    reportNumber = "NSF-KITP-03-38",
    month = "5",
    year = "2003"
}

@article{Caceres2006a,
archivePrefix = {arXiv},
arxivId = {hep-th/0606134},
author = {C{\'{a}}ceres, Elena and G{\"{u}}ijosa, Alberto},
doi = {10.1088/1126-6708/2006/12/068},
eprint = {0606134},
issn = {10298479},
journal = {J. High Energy Phys.},
number = {12},
primaryClass = {hep-th},
title = {{On drag forces and jet quenching in strongly-coupled plasmas}},
volume = {2006},
year = {2006}
}

@article{Chakrabortty2011a,
archivePrefix = {arXiv},
arxivId = {1108.0165},
author = {Chakrabortty, Shankhadeep},
doi = {10.1016/j.physletb.2011.09.112},
eprint = {1108.0165},
issn = {03702693},
journal = {Phys. Lett. Sect. B Nucl. Elem. Part. High-Energy Phys.},
number = {3},
pages = {244--250},
title = {{Dissipative force on an external quark in heavy quark cloud}},
volume = {705},
year = {2011}
}

@article{Maldacena1998,
archivePrefix = {arXiv},
arxivId = {hep-th/9803002},
author = {Maldacena, Juan},
doi = {10.1103/PhysRevLett.80.4859},
eprint = {9803002},
issn = {10797114},
journal = {Phys. Rev. Lett.},
number = {22},
pages = {4859--4862},
primaryClass = {hep-th},
title = {{Wilson Loops in Large N Field Theories}},
volume = {80},
year = {1998}
}

@article{Witten1998,
archivePrefix = {arXiv},
arxivId = {hep-th/9802150},
author = {Witten, Edward},
doi = {10.4310/atmp.1998.v2.n2.a2},
eprint = {9802150},
issn = {10950753},
journal = {Adv. Theor. Math. Phys.},
number = {2},
pages = {253--290},
primaryClass = {hep-th},
title = {{Anti de sitter space and holography}},
volume = {2},
year = {1998}
}

@article{Caceres2006,
archivePrefix = {arXiv},
arxivId = {hep-th/0605235},
author = {C{\'{a}}ceres, Elena and G{\"{u}}ijosa, Alberto},
doi = {10.1088/1126-6708/2006/11/077},
eprint = {0605235},
issn = {10298479},
journal = {J. High Energy Phys.},
number = {11},
primaryClass = {hep-th},
title = {{Drag force in a charged $ N = 4$ SYM plasma}},
volume = {2006},
year = {2006}
}

@article{Chakrabortty2016a,
archivePrefix = {arXiv},
arxivId = {1602.04761},
author = {Chakrabortty, Shankhadeep and Dey, Tanay K.},
doi = {10.1007/JHEP05(2016)094},
eprint = {1602.04761},
issn = {10298479},
journal = {J. High Energy Phys.},
number = {5},
title = {{Back reaction effects on the dynamics of heavy probes in heavy quark cloud}},
volume = {2016},
year = {2016}
}

@article{Maldacena1999,
archivePrefix = {arXiv},
arxivId = {hep-th/9711200},
author = {Maldacena, Juan},
doi = {10.1023/A:1026654312961},
eprint = {9711200},
issn = {00207748},
journal = {Int. J. Theor. Phys.},
number = {4},
pages = {1113--1133},
primaryClass = {hep-th},
title = {{The large-N limit of superconformal field theories and supergravity}},
volume = {38},
year = {1999}
}

@article{Chernicoff2008,
archivePrefix = {arXiv},
arxivId = {0803.3070},
author = {Chernicoff, Mariano and G{\"{u}}ijosa, Alberto},
doi = {10.1088/1126-6708/2008/06/005},
eprint = {0803.3070},
issn = {11266708},
journal = {J. High Energy Phys.},
number = {6},
pages = {1--66},
title = {{Acceleration, energy loss and screening in strongly-coupled gauge theories}},
volume = {2008},
year = {2008}
}

@article{Gubser_2006,
   title={Drag force in AdS/CFT},
   volume={74},
   ISSN={1550-2368},
   url={http://dx.doi.org/10.1103/PhysRevD.74.126005},
   DOI={10.1103/physrevd.74.126005},
   number={12},
   journal={Physical Review D},
   publisher={American Physical Society (APS)},
   author={Gubser, Steven S.},
   year={2006},
   month=dec }

@article{Herzog_2006,
	doi = {10.1088/1126-6708/2006/07/013},
  
	url = {https://doi.org/10.1088%2F1126-6708%2F2006%2F07%2F013},
  
	year = 2006,
	month = {jul},
  
	publisher = {Springer Science and Business Media {LLC}
},
  
	volume = {2006},
  
	number = {07},
  
	pages = {013--013},
  
	author = {Christopher P Herzog and Andreas Karch and Pavel Kovtun and Can Kozcaz and Laurence G Yaffe},
  
	title = {Energy loss of a heavy quark moving through $N = 4$ supersymmetric Yang-Mills plasma},
  
	journal = {Journal of High Energy Physics}
}

@article{Pokhrel2023,
    author = "Pokhrel, Rishi and Dey, Tanay K.",
    title = "{Charged AdS black holes in presence of string cloud and Cardy-Verlinde formula}",
    eprint = "2303.02702",
    archivePrefix = "arXiv",
    primaryClass = "hep-th",
    doi = "10.1016/j.nuclphysb.2024.116508",
    journal = "Nucl. Phys. B",
    volume = "1001",
    pages = "116508",
    year = "2024"
}

@article{Liu_2007a,
   title={Anti–de Sitter/Conformal-Field-Theory Calculation of Screening in a Hot Wind},
   volume={98},
   ISSN={1079-7114},
   url={http://dx.doi.org/10.1103/PhysRevLett.98.182301},
   DOI={10.1103/physrevlett.98.182301},
   number={18},
   journal={Physical Review Letters},
   publisher={American Physical Society (APS)},
   author={Liu, Hong and Rajagopal, Krishna and Wiedemann, Urs Achim},
   year={2007},
   month=may }

@article{Liu:2006ug,
    author = "Liu, Hong and Rajagopal, Krishna and Wiedemann, Urs Achim",
    title = "{Calculating the jet quenching parameter from AdS/CFT}",
    eprint = "hep-ph/0605178",
    archivePrefix = "arXiv",
    reportNumber = "MIT-CTP-3739, RBRC-601",
    doi = "10.1103/PhysRevLett.97.182301",
    journal = "Phys. Rev. Lett.",
    volume = "97",
    pages = "182301",
    year = "2006"
}

@article{Zajc_2008,
   title={The Fluid Nature of Quark-Gluon Plasma},
   volume={805},
   ISSN={0375-9474},
   url={http://dx.doi.org/10.1016/j.nuclphysa.2008.02.285},
   DOI={10.1016/j.nuclphysa.2008.02.285},
   number={1–4},
   journal={Nuclear Physics A},
   publisher={Elsevier BV},
   author={Zajc, W.A.},
   year={2008},
   month=jun, pages={283c–294c} }

@misc{muller2007,
      title={From Quark-Gluon Plasma to the Perfect Liquid}, 
      author={Berndt Müller},
      year={2007},
      eprint={0710.3366},
      archivePrefix={arXiv},
      primaryClass={nucl-th},
      url={https://arxiv.org/abs/0710.3366}, 
}

@article{Shuryak_2007,
   title={Toward the Theory of Strongly Coupled Quark-Gluon Plasma},
   volume={168},
   ISSN={0375-9687},
   url={http://dx.doi.org/10.1143/PTPS.168.320},
   DOI={10.1143/ptps.168.320},
   journal={Progress of Theoretical Physics Supplement},
   publisher={Oxford University Press (OUP)},
   author={Shuryak, Edward},
   year={2007},
   pages={320–329} }

@article{d_Enterria_2007,
   title={Quark–gluon matter},
   volume={34},
   ISSN={1361-6471},
   url={http://dx.doi.org/10.1088/0954-3899/34/7/S04},
   DOI={10.1088/0954-3899/34/7/s04},
   number={7},
   journal={Journal of Physics G: Nuclear and Particle Physics},
   publisher={IOP Publishing},
   author={d’Enterria, David},
   year={2007},
   month=may, pages={S53–S81} }

@misc{salgado2006,
      title={Heavy ions theory review}, 
      author={Carlos A. Salgado},
      year={2006},
      eprint={hep-ph/0609172},
      archivePrefix={arXiv},
      primaryClass={hep-ph},
      url={https://arxiv.org/abs/hep-ph/0609172}, 
}

@inproceedings{Shuryak_2007_report,
   title={STRONGLY COUPLED QUARK-GLUON PLASMA: THE STATUS REPORT},
   url={http://dx.doi.org/10.1142/9789812708267_0001},
   DOI={10.1142/9789812708267_0001},
   booktitle={Continuous Advances in QCD 2006},
   publisher={WORLD SCIENTIFIC},
   author={SHURYAK, E. V.},
   year={2007},
   month=mar }

@article{Tannenbaum_2006,
   title={Recent results in relativistic heavy ion collisions: from ‘a new state of matter’ to ‘the perfect fluid’},
   volume={69},
   ISSN={1361-6633},
   url={http://dx.doi.org/10.1088/0034-4885/69/7/R01},
   DOI={10.1088/0034-4885/69/7/r01},
   number={7},
   journal={Reports on Progress in Physics},
   publisher={IOP Publishing},
   author={Tannenbaum, M J},
   year={2006},
   month=jun, pages={2005–2059} }

@article{Muller_2006,
   title={Results from the Relativistic Heavy Ion Collider},
   volume={56},
   ISSN={1545-4134},
   url={http://dx.doi.org/10.1146/annurev.nucl.56.080805.140556},
   DOI={10.1146/annurev.nucl.56.080805.140556},
   number={1},
   journal={Annual Review of Nuclear and Particle Science},
   publisher={Annual Reviews},
   author={Müller, Berndt and Nagle, James L.},
   year={2006},
   month=nov, pages={93–135} }

@article{Gyulassy_2005,
   title={New forms of QCD matter discovered at RHIC},
   volume={750},
   ISSN={0375-9474},
   url={http://dx.doi.org/10.1016/j.nuclphysa.2004.10.034},
   DOI={10.1016/j.nuclphysa.2004.10.034},
   number={1},
   journal={Nuclear Physics A},
   publisher={Elsevier BV},
   author={Gyulassy, Miklos and McLerran, Larry},
   year={2005},
   month=mar, pages={30–63} }

@article{Adcox_2005,
   title={Formation of dense partonic matter in relativistic nucleus–nucleus collisions at RHIC: Experimental evaluation by the PHENIX Collaboration},
   volume={757},
   ISSN={0375-9474},
   url={http://dx.doi.org/10.1016/j.nuclphysa.2005.03.086},
   DOI={10.1016/j.nuclphysa.2005.03.086},
   number={1–2},
   journal={Nuclear Physics A},
   publisher={Elsevier BV},
   author={Adcox, K. and Adler, S.S. and Afanasiev, S. and Aidala, C. and Ajitanand, N.N. and Akiba, Y. and Al-Jamel, A. and Alexander, J. and Amirikas, R. and Aoki, K. and Aphecetche, L. and Arai, Y. and Armendariz, R. and Aronson, S.H. and Averbeck, R. and Awes, T.C. and Azmoun, R. and Babintsev, V. and Baldisseri, A. and Barish, K.N. and Barnes, P.D. and Barrette, J. and Bassalleck, B. and Bathe, S. and Batsouli, S. and Baublis, V. and Bauer, F. and Bazilevsky, A. and Belikov, S. and Bellaiche, F.G. and Belyaev, S.T. and Bennett, M.J. and Berdnikov, Y. and Bhagavatula, S. and Bjorndal, M.T. and Boissevain, J.G. and Borel, H. and Borenstein, S. and Botelho, S. and Brooks, M.L. and Brown, D.S. and Bruner, N. and Bucher, D. and Buesching, H. and Bumazhnov, V. and Bunce, G. and Burward-Hoy, J.M. and Butsyk, S. and Camard, X. and Carey, T.A. and Chai, J.-S. and Chand, P. and Chang, J. and Chang, W.C. and Chavez, L.L. and Chernichenko, S. and Chi, C.Y. and Chiba, J. and Chiu, M. and Choi, I.J. and Choi, J. and Choudhury, R.K. and Christ, T. and Chujo, T. and Chung, M.S. and Chung, P. and Cianciolo, V. and Cleven, C.R. and Cobigo, Y. and Cole, B.A. and Comets, M.P. and Constantin, P. and Csanád, M. and Csörgő, T. and Cussonneau, J.P. and d’Enterria, D. and Dahms, T. and Das, K. and David, G. and Deák, F. and Delagrange, H. and Denisov, A. and Deshpande, A. and Desmond, E.J. and Devismes, A. and Dietzsch, O. and Dinesh, B.V. and Drachenberg, J.L. and Drapier, O. and Drees, A. and Dubey, A.K. and du Rietz, R. and Durum, A. and Dutta, D. and Dzhordzhadze, V. and Ebisu, K. and Efremenko, Y.V. and Egdemir, J. and El Chenawi, K. and Enokizono, A. and En’yo, H. and Espagnon, B. and Esumi, S. and Ewell, L. and Ferdousi, T. and Fields, D.E. and Finck, C. and Fleuret, F. and Fokin, S.L. and Forestier, B. and Fox, B.D. and Fraenkel, Z. and Frantz, J.E. and Franz, A. and Frawley, A.D. and Fukao, Y. and Fung, S.-Y. and Gadrat, S. and Garpman, S. and Gastineau, F. and Germain, M. and Ghosh, T.K. and Glenn, A. and Godoi, A.L. and Gogiberidze, G. and Gonin, M. and Gosset, J. and Goto, Y. and Granier de Cassagnac, R. and Grau, N. and Greene, S.V. and Grosse Perdekamp, M. and Gunji, T. and Gupta, S.K. and Guryn, W. and Gustafsson, H.-Å. and Hachiya, T. and Hadjhenni, A. and Haggerty, J.S. and Hagiwara, M.N. and Hamagaki, H. and Hansen, A.G. and Hara, H. and Harada, H. and Hartouni, E.P. and Haruna, K. and Harvey, M. and Haslum, E. and Hasuko, K. and Hayano, R. and Hayashi, N. and He, X. and Heffner, M. and Hemmick, T.K. and Heuser, J.M. and Hibino, M. and Hidas, P. and Hiejima, H. and Hill, J.C. and Ho, D.S. and Hobbs, R. and Holmes, M. and Holzmann, W. and Homma, K. and Hong, B. and Hoover, A. and Horaguchi, T. and Hur, H.M. and Ichihara, T. and Ikonnikov, V.V. and Imai, K. and Inaba, M. and Inuzuka, M. and Ippolitov, M.S. and Isenhower, D. and Isenhower, L. and Ishihara, M. and Isobe, T. and Issah, M. and Isupov, A. and Jacak, B.V. and Jang, W.Y. and Jeong, Y. and Jia, J. and Jin, J. and Jinnouchi, O. and Johnson, B.M. and Johnson, S.C. and Joo, K.S. and Jouan, D. and Kajihara, F. and Kametani, S. and Kamihara, N. and Kaneta, M. and Kang, J.H. and Kann, M. and Kapoor, S.S. and Katou, K. and Kawabata, T. and Kawagishi, T. and Kazantsev, A.V. and Kelly, S. and Khachaturov, B. and Khanzadeev, A. and Kikuchi, J. and Kim, D.H. and Kim, D.J. and Kim, D.W. and Kim, E. and Kim, G.-B. and Kim, H.J. and Kim, S.Y. and Kim, Y.-S. and Kim, Y.G. and Kinney, E. and Kinnison, W.W. and Kiss, A. and Kistenev, E. and Kiyomichi, A. and Kiyoyama, K. and Klein-Boesing, C. and Klinksiek, S. and Kobayashi, H. and Kochenda, L. and Kochetkov, V. and Koehler, D. and Kohama, T. and Kohara, R. and Komkov, B. and Konno, M. and Kopytine, M. and Kotchetkov, D. and Kozlov, A. and Kroon, P.J. and Kuberg, C.H. and Kunde, G.J. and Kurihara, N. and Kurita, K. and Kuroki, Y. and Kweon, M.J. and Kwon, Y. and Kyle, G.S. and Lacey, R. and Ladygin, V. and Lajoie, J.G. and Lauret, J. and Le Bornec, Y. and Lebedev, A. and Leckey, S. and Lee, D.M. and Lee, M.K. and Lee, S. and Leitch, M.J. and Leite, M.A.L. and Li, X.H. and Li, Z. and Lim, D.J. and Lim, H. and Litvinenko, A. and Liu, M.X. and Liu, X. and Liu, Y. and Liu, Z. and Maguire, C.F. and Mahon, J. and Makdisi, Y.I. and Malakhov, A. and Malik, M.D. and Manko, V.I. and Mao, Y. and Mark, S.K. and Markacs, S. and Martinez, G. and Marx, M.D. and Masaike, A. and Masui, H. and Matathias, F. and Matsumoto, T. and McCain, M.C. and McGaughey, P.L. and Melnikov, E. and Merschmeyer, M. and Messer, F. and Messer, M. and Miake, Y. and Milan, J. and Miller, T.E. and Milov, A. and Mioduszewski, S. and Mischke, R.E. and Mishra, G.C. and Mitchell, J.T. and Mohanty, A.K. and Morrison, D.P. and Moss, J.M. and Moukhanova, T.V. and Mühlbacher, F. and Mukhopadhyay, D. and Muniruzzaman, M. and Murata, J. and Nagamiya, S. and Nagasaka, Y. and Nagata, Y. and Nagle, J.L. and Naglis, M. and Nakada, Y. and Nakamura, T. and Nandi, B.K. and Nara, M. and Newby, J. and Nguyen, M. and Nikkinen, L. and Nilsson, P. and Nishimura, S. and Norman, B. and Nyanin, A.S. and Nystrand, J. and O’Brien, E. and Ogilvie, C.A. and Ohnishi, H. and Ojha, I.D. and Okada, H. and Okada, K. and Omiwade, O.O. and Ono, M. and Onuchin, V. and Oskarsson, A. and Österman, L. and Otterlund, I. and Oyama, K. and Ozawa, K. and Paffrath, L. and Pal, D. and Palounek, A.P.T. and Pantuev, V. and Papavassiliou, V. and Park, J. and Park, W.J. and Parmar, A. and Pate, S.F. and Pei, H. and Peitzmann, T. and Penev, V. and Peng, J.-C. and Pereira, H. and Peresedov, V. and Peressounko, D.Yu. and Petridis, A.N. and Pierson, A. and Pinkenburg, C. and Pisani, R.P. and Pitukhin, P. and Plasil, F. and Pollack, M. and Pope, K. and Purschke, M.L. and Purwar, A.K. and Qu, H. and Qualls, J.M. and Rak, J. and Ravinovich, I. and Read, K.F. and Reuter, M. and Reygers, K. and Riabov, V. and Riabov, Y. and Roche, G. and Romana, A. and Rosati, M. and Rose, A.A. and Rosendahl, S.S.E. and Rosnet, P. and Rukoyatkin, P. and Rykov, V.L. and Ryu, S.S. and Sadler, M.E. and Sahlmueller, B. and Saito, N. and Sakaguchi, A. and Sakaguchi, T. and Sakai, M. and Sakai, S. and Sako, H. and Sakuma, T. and Samsonov, V. and Sanfratello, L. and Sangster, T.C. and Santo, R. and Sato, H.D. and Sato, S. and Sawada, S. and Schlei, B.R. and Schutz, Y. and Semenov, V. and Seto, R. and Sharma, D. and Shaw, M.R. and Shea, T.K. and Shein, I. and Shibata, T.-A. and Shigaki, K. and Shiina, T. and Shimomura, M. and Shin, Y.H. and Shohjoh, T. and Shoji, K. and Sibiriak, I.G. and Sickles, A. and Silva, C.L. and Silvermyr, D. and Sim, K.S. and Simon-Gillo, J. and Singh, C.P. and Singh, V. and Sivertz, M. and Skutnik, S. and Smith, W.C. and Soldatov, A. and Soltz, R.A. and Sondheim, W.E. and Sorensen, S.P. and Sourikova, I.V. and Staley, F. and Stankus, P.W. and Starinsky, N. and Steinberg, P. and Stenlund, E. and Stepanov, M. and Ster, A. and Stoll, S.P. and Sugioka, M. and Sugitate, T. and Suire, C. and Sullivan, J.P. and Sumi, Y. and Sun, Z. and Suzuki, M. and Sziklai, J. and Tabaru, T. and Takagi, S. and Takagui, E.M. and Taketani, A. and Tamai, M. and Tanaka, K.H. and Tanaka, Y. and Tanida, K. and Taniguchi, E. and Tannenbaum, M.J. and Taranenko, A. and Tarján, P. and Tepe, J.D. and Thomas, J. and Thomas, J.H. and Thomas, T.L. and Tian, W. and Togawa, M. and Tojo, J. and Torii, H. and Towell, R.S. and Tram, V.-N. and Tserruya, I. and Tsuchimoto, Y. and Tsuruoka, H. and Tsvetkov, A.A. and Tuli, S.K. and Tydesjö, H. and Tyurin, N. and Uam, T.J. and Ushiroda, T. and Valle, H. and van Hecke, H.W. and Velissaris, C. and Velkovska, J. and Velkovsky, M. and Vertesi, R. and Veszprémi, V. and Villatte, L. and Vinogradov, A.A. and Volkov, M.A. and Vorobyov, A. and Vznuzdaev, E. and Wagner, M. and Wang, H. and Wang, X.R. and Watanabe, Y. and Wessels, J. and White, S.N. and Willis, N. and Winter, D. and Witzig, C. and Wohn, F.K. and Woody, C.L. and Wysocki, M. and Xie, W. and Yagi, K. and Yang, Y. and Yanovich, A. and Yokkaichi, S. and Young, G.R. and Younus, I. and Yushmanov, I.E. and Zajc, W.A. and Zaudkte, O. and Zhang, C. and Zhang, Z. and Zhou, S. and Zhou, S.J. and Zimányi, J. and Zolin, L. and Zong, X.},
   year={2005},
   month=aug, pages={184–283} }

@article{Arsene_2005,
   title={Quark–gluon plasma and color glass condensate at RHIC? The perspective from the BRAHMS experiment},
   volume={757},
   ISSN={0375-9474},
   url={http://dx.doi.org/10.1016/j.nuclphysa.2005.02.130},
   DOI={10.1016/j.nuclphysa.2005.02.130},
   number={1–2},
   journal={Nuclear Physics A},
   publisher={Elsevier BV},
   author={Arsene, I. and Bearden, I.G. and Beavis, D. and Besliu, C. and Budick, B. and Bøggild, H. and Chasman, C. and Christensen, C.H. and Christiansen, P. and Cibor, J. and Debbe, R. and Enger, E. and Gaardhøje, J.J. and Germinario, M. and Hansen, O. and Holm, A. and Holme, A.K. and Hagel, K. and Ito, H. and Jakobsen, E. and Jipa, A. and Jundt, F. and Jørdre, J.I. and Jørgensen, C.E. and Karabowicz, R. and Kim, E.J. and Kozik, T. and Larsen, T.M. and Lee, J.H. and Lee, Y.K. and Lindahl, S. and Løvhøiden, G. and Majka, Z. and Makeev, A. and Mikelsen, M. and Murray, M.J. and Natowitz, J. and Neumann, B. and Nielsen, B.S. and Ouerdane, D. and Płaneta, R. and Rami, F. and Ristea, C. and Ristea, O. and Röhrich, D. and Samset, B.H. and Sandberg, D. and Sanders, S.J. and Scheetz, R.A. and Staszel, P. and Tveter, T.S. and Videbæk, F. and Wada, R. and Yin, Z. and Zgura, I.S.},
   year={2005},
   month=aug, pages={1–27} }

@article{Back_2005,
   title={The PHOBOS perspective on discoveries at RHIC},
   volume={757},
   ISSN={0375-9474},
   url={http://dx.doi.org/10.1016/j.nuclphysa.2005.03.084},
   DOI={10.1016/j.nuclphysa.2005.03.084},
   number={1–2},
   journal={Nuclear Physics A},
   publisher={Elsevier BV},
   author={Back, B.B. and Baker, M.D. and Ballintijn, M. and Barton, D.S. and Becker, B. and Betts, R.R. and Bickley, A.A. and Bindel, R. and Budzanowski, A. and Busza, W. and Carroll, A. and Chai, Z. and Decowski, M.P. and García, E. and Gburek, T. and George, N.K. and Gulbrandsen, K. and Gushue, S. and Halliwell, C. and Hamblen, J. and Harrington, A.S. and Hauer, M. and Heintzelman, G.A. and Henderson, C. and Hofman, D.J. and Hollis, R.S. and Hołyński, R. and Holzman, B. and Iordanova, A. and Johnson, E. and Kane, J.L. and Katzy, J. and Khan, N. and Kucewicz, W. and Kulinich, P. and Kuo, C.M. and Lee, J.W. and Lin, W.T. and Manly, S. and McLeod, D. and Mignerey, A.C. and Nouicer, R. and Olszewski, A. and Pak, R. and Park, I.C. and Pernegger, H. and Reed, C. and Remsberg, L.P. and Reuter, M. and Roland, C. and Roland, G. and Rosenberg, L. and Sagerer, J. and Sarin, P. and Sawicki, P. and Seals, H. and Sedykh, I. and Skulski, W. and Smith, C.E. and Stankiewicz, M.A. and Steinberg, P. and Stephans, G.S.F. and Sukhanov, A. and Tang, J.-L. and Tonjes, M.B. and Trzupek, A. and Vale, C.M. and van Nieuwenhuizen, G.J. and Vaurynovich, S.S. and Verdier, R. and Veres, G.I. and Wenger, E. and Wolfs, F.L.H. and Wosiek, B. and Woźniak, K. and Wuosmaa, A.H. and Wysłouch, B. and Zhang, J.},
   year={2005},
   month=aug, pages={28–101} }

@article{Adams_2005,
   title={Experimental and theoretical challenges in the search for the quark–gluon plasma: The STAR Collaboration’s critical assessment of the evidence from RHIC collisions},
   volume={757},
   ISSN={0375-9474},
   url={http://dx.doi.org/10.1016/j.nuclphysa.2005.03.085},
   DOI={10.1016/j.nuclphysa.2005.03.085},
   number={1–2},
   journal={Nuclear Physics A},
   publisher={Elsevier BV},
   author={Adams, J. and Aggarwal, M.M. and Ahammed, Z. and Amonett, J. and Anderson, B.D. and Arkhipkin, D. and Averichev, G.S. and Badyal, S.K. and Bai, Y. and Balewski, J. and Barannikova, O. and Barnby, L.S. and Baudot, J. and Bekele, S. and Belaga, V.V. and Bellingeri-Laurikainen, A. and Bellwied, R. and Berger, J. and Bezverkhny, B.I. and Bharadwaj, S. and Bhasin, A. and Bhati, A.K. and Bhatia, V.S. and Bichsel, H. and Bielcik, J. and Bielcikova, J. and Billmeier, A. and Bland, L.C. and Blyth, C.O. and Bonner, B.E. and Botje, M. and Boucham, A. and Bouchet, J. and Brandin, A.V. and Bravar, A. and Bystersky, M. and Cadman, R.V. and Cai, X.Z. and Caines, H. and Calderón de la Barca Sánchez, M. and Castillo, J. and Catu, O. and Cebra, D. and Chajecki, Z. and Chaloupka, P. and Chattopadhyay, S. and Chen, H.F. and Chen, Y. and Cheng, J. and Cherney, M. and Chikanian, A. and Christie, W. and Coffin, J.P. and Cormier, T.M. and Cramer, J.G. and Crawford, H.J. and Das, D. and Das, S. and de Moura, M.M. and Dedovich, T.G. and Derevschikov, A.A. and Didenko, L. and Dietel, T. and Dogra, S.M. and Dong, W.J. and Dong, X. and Draper, J.E. and Du, F. and Dubey, A.K. and Dunin, V.B. and Dunlop, J.C. and Dutta Mazumdar, M.R. and Eckardt, V. and Edwards, W.R. and Efimov, L.G. and Emelianov, V. and Engelage, J. and Eppley, G. and Erazmus, B. and Estienne, M. and Fachini, P. and Faivre, J. and Fatemi, R. and Fedorisin, J. and Filimonov, K. and Filip, P. and Finch, E. and Fine, V. and Fisyak, Y. and Fu, J. and Gagliardi, C.A. and Gaillard, L. and Gans, J. and Ganti, M.S. and Geurts, F. and Ghazikhanian, V. and Ghosh, P. and Gonzalez, J.E. and Gos, H. and Grachov, O. and Grebenyuk, O. and Grosnick, D. and Guertin, S.M. and Guo, Y. and Gupta, A. and Gutierrez, T.D. and Hallman, T.J. and Hamed, A. and Hardtke, D. and Harris, J.W. and Heinz, M. and Henry, T.W. and Hepplemann, S. and Hippolyte, B. and Hirsch, A. and Hjort, E. and Hoffmann, G.W. and Huang, H.Z. and Huang, S.L. and Hughes, E.W. and Humanic, T.J. and Igo, G. and Ishihara, A. and Jacobs, P. and Jacobs, W.W. and Jedynak, M. and Jiang, H. and Jones, P.G. and Judd, E.G. and Kabana, S. and Kang, K. and Kaplan, M. and Keane, D. and Kechechyan, A. and Khodyrev, V.Yu. and Kiryluk, J. and Kisiel, A. and Kislov, E.M. and Klay, J. and Klein, S.R. and Koetke, D.D. and Kollegger, T. and Kopytine, M. and Kotchenda, L. and Kramer, M. and Kravtsov, P. and Kravtsov, V.I. and Krueger, K. and Kuhn, C. and Kulikov, A.I. and Kumar, A. and Kutuev, R.Kh. and Kuznetsov, A.A. and Lamont, M.A.C. and Landgraf, J.M. and Lange, S. and Laue, F. and Lauret, J. and Lebedev, A. and Lednicky, R. and Lehocka, S. and LeVine, M.J. and Li, C. and Li, Q. and Li, Y. and Lin, G. and Lindenbaum, S.J. and Lisa, M.A. and Liu, F. and Liu, H. and Liu, L. and Liu, Q.J. and Liu, Z. and Ljubicic, T. and Llope, W.J. and Long, H. and Longacre, R.S. and Lopez-Noriega, M. and Love, W.A. and Lu, Y. and Ludlam, T. and Lynn, D. and Ma, G.L. and Ma, J.G. and Ma, Y.G. and Magestro, D. and Mahajan, S. and Mahapatra, D.P. and Majka, R. and Mangotra, L.K. and Manweiler, R. and Margetis, S. and Markert, C. and Martin, L. and Marx, J.N. and Matis, H.S. and Matulenko, Yu.A. and McClain, C.J. and McShane, T.S. and Meissner, F. and Melnick, Yu. and Meschanin, A. and Miller, M.L. and Minaev, N.G. and Mironov, C. and Mischke, A. and Mishra, D.K. and Mitchell, J. and Mohanty, B. and Molnar, L. and Moore, C.F. and Morozov, D.A. and Munhoz, M.G. and Nandi, B.K. and Nayak, S.K. and Nayak, T.K. and Nelson, J.M. and Netrakanti, P.K. and Nikitin, V.A. and Nogach, L.V. and Nurushev, S.B. and Odyniec, G. and Ogawa, A. and Okorokov, V. and Oldenburg, M. and Olson, D. and Pal, S.K. and Panebratsev, Y. and Panitkin, S.Y. and Pavlinov, A.I. and Pawlak, T. and Peitzmann, T. and Perevoztchikov, V. and Perkins, C. and Peryt, W. and Petrov, V.A. and Phatak, S.C. and Picha, R. and Planinic, M. and Pluta, J. and Porile, N. and Porter, J. and Poskanzer, A.M. and Potekhin, M. and Potrebenikova, E. and Potukuchi, B.V.K.S. and Prindle, D. and Pruneau, C. and Putschke, J. and Rakness, G. and Raniwala, R. and Raniwala, S. and Ravel, O. and Ray, R.L. and Razin, S.V. and Reichhold, D. and Reid, J.G. and Reinnarth, J. and Renault, G. and Retiere, F. and Ridiger, A. and Ritter, H.G. and Roberts, J.B. and Rogachevskiy, O.V. and Romero, J.L. and Rose, A. and Roy, C. and Ruan, L. and Russcher, M. and Sahoo, R. and Sakrejda, I. and Salur, S. and Sandweiss, J. and Sarsour, M. and Savin, I. and Sazhin, P.S. and Schambach, J. and Scharenberg, R.P. and Schmitz, N. and Seger, J. and Seyboth, P. and Shahaliev, E. and Shao, M. and Shao, W. and Sharma, M. and Shen, W.Q. and Shestermanov, K.E. and Shimanskiy, S.S. and Sichtermann, E. and Simon, F. and Singaraju, R.N. and Smirnov, N. and Snellings, R. and Sood, G. and Sorensen, P. and Sowinski, J. and Speltz, J. and Spinka, H.M. and Srivastava, B. and Stadnik, A. and Stanislaus, T.D.S. and Stock, R. and Stolpovsky, A. and Strikhanov, M. and Stringfellow, B. and Suaide, A.A.P. and Sugarbaker, E. and Suire, C. and Sumbera, M. and Surrow, B. and Swanger, M. and Symons, T.J.M. and Szanto de Toledo, A. and Tai, A. and Takahashi, J. and Tang, A.H. and Tarnowsky, T. and Thein, D. and Thomas, J.H. and Timoshenko, S. and Tokarev, M. and Trentalange, S. and Tribble, R.E. and Tsai, O.D. and Ulery, J. and Ullrich, T. and Underwood, D.G. and Van Buren, G. and van Leeuwen, M. and Vander Molen, A.M. and Varma, R. and Vasilevski, I.M. and Vasiliev, A.N. and Vernet, R. and Vigdor, S.E. and Viyogi, Y.P. and Vokal, S. and Voloshin, S.A. and Waggoner, W.T. and Wang, F. and Wang, G. and Wang, G. and Wang, X.L. and Wang, Y. and Wang, Y. and Wang, Z.M. and Ward, H. and Watson, J.W. and Webb, J.C. and Westfall, G.D. and Wetzler, A. and Whitten, C. and Wieman, H. and Wissink, S.W. and Witt, R. and Wood, J. and Wu, J. and Xu, N. and Xu, Z. and Xu, Z.Z. and Yamamoto, E. and Yepes, P. and Yurevich, V.I. and Zborovsky, I. and Zhang, H. and Zhang, W.M. and Zhang, Y. and Zhang, Z.P. and Zoulkarneev, R. and Zoulkarneeva, Y. and Zubarev, A.N.},
   year={2005},
   month=aug, pages={102–183} }

@article{Baier_1997,
   title={Radiative energy loss of high energy quarks and gluons in a finite-volume quark-gluon plasma},
   volume={483},
   ISSN={0550-3213},
   url={http://dx.doi.org/10.1016/S0550-3213(96)00553-6},
   DOI={10.1016/s0550-3213(96)00553-6},
   number={1–2},
   journal={Nuclear Physics B},
   publisher={Elsevier BV},
   author={Baier, R. and Dokshitzer, Yu.L. and Mueller, A.H. and Peigné, S. and Schiff, D.},
   year={1997},
   month=jan, pages={291–320} }

@article{Eskola_2005,
   title={The fragility of high- hadron spectra as a hard probe},
   volume={747},
   ISSN={0375-9474},
   url={http://dx.doi.org/10.1016/j.nuclphysa.2004.09.070},
   DOI={10.1016/j.nuclphysa.2004.09.070},
   number={2–4},
   journal={Nuclear Physics A},
   publisher={Elsevier BV},
   author={Eskola, K.J. and Honkanen, H. and Salgado, C.A. and Wiedemann, U.A.},
   year={2005},
   month=jan, pages={511–529} }

@article{Casalderrey_Solana_2006,
   title={Heavy quark diffusion in strongly coupled $\mathcal{N}=4$ Yang-Mills theory},
   volume={74},
   ISSN={1550-2368},
   url={http://dx.doi.org/10.1103/PhysRevD.74.085012},
   DOI={10.1103/physrevd.74.085012},
   number={8},
   journal={Physical Review D},
   publisher={American Physical Society (APS)},
   author={Casalderrey-Solana, Jorge and Teaney, Derek},
   year={2006},
   month=oct }

@article{Matsuo_2006,
   title={Drag force in SYM plasma with B field from AdS/CFT},
   volume={2006},
   ISSN={1029-8479},
   url={http://dx.doi.org/10.1088/1126-6708/2006/10/055},
   DOI={10.1088/1126-6708/2006/10/055},
   number={10},
   journal={Journal of High Energy Physics},
   publisher={Springer Science and Business Media LLC},
   author={Matsuo, Toshihiro and Tomino, Dan and Wen, Wen-Yu},
   year={2006},
   month=oct, pages={055–055} }

@article{Talavera_2007,
   title={Drag force in a string model dual to large-NQCD},
   volume={2007},
   ISSN={1029-8479},
   url={http://dx.doi.org/10.1088/1126-6708/2007/01/086},
   DOI={10.1088/1126-6708/2007/01/086},
   number={01},
   journal={Journal of High Energy Physics},
   publisher={Springer Science and Business Media LLC},
   author={Talavera, Pere},
   year={2007},
   month=jan, pages={086–086} }

@article{Nakano_2007,
   title={Drag force, jet quenching, and an AdS/QCD model},
   volume={75},
   ISSN={1550-2368},
   url={http://dx.doi.org/10.1103/PhysRevD.75.085016},
   DOI={10.1103/physrevd.75.085016},
   number={8},
   journal={Physical Review D},
   publisher={American Physical Society (APS)},
   author={Nakano, Eiji and Teraguchi, Shunsuke and Wen, Wen-Yu},
   year={2007},
   month=apr }

@article{Bertoldi_2007,
   title={Holography and unquenched quark-gluon plasmas},
   volume={76},
   ISSN={1550-2368},
   url={http://dx.doi.org/10.1103/PhysRevD.76.065007},
   DOI={10.1103/physrevd.76.065007},
   number={6},
   journal={Physical Review D},
   publisher={American Physical Society (APS)},
   author={Bertoldi, G. and Bigazzi, F. and Cotrone, A. L. and Edelstein, J. D.},
   year={2007},
   month=sep }

@article{Chernicoff_2006,
   title={The energy of a moving quark-antiquark pair in an $ N = 4$ SYM plasma},
   volume={2006},
   ISSN={1029-8479},
   url={http://dx.doi.org/10.1088/1126-6708/2006/09/068},
   DOI={10.1088/1126-6708/2006/09/068},
   number={09},
   journal={Journal of High Energy Physics},
   publisher={Springer Science and Business Media LLC},
   author={Chernicoff, Mariano and García, J. Antonio and Güijosa, Alberto},
   year={2006},
   month=sep, pages={068–068} }

@article{Zhu_2024,
    author = "Zhu, Xiangrong and Wu, Ping-ping and Zhang, Zi-qiang",
    title = "{Screening length in a soft wall AdS/QCD model}",
    doi = "10.1140/epja/s10050-024-01249-y",
    journal = "Eur. Phys. J. A",
    volume = "60",
    number = "2",
    pages = "35",
    year = "2024"
}

@article{Fadafan_2009,
	title={Stirring strongly coupled plasma},
	volume={61},
	ISSN={1434-6052},
	url={http://dx.doi.org/10.1140/epjc/s10052-009-0885-6},
	DOI={10.1140/epjc/s10052-009-0885-6},
	number={4},
	journal={The European Physical Journal C},
	publisher={Springer Science and Business Media LLC},
	author={Fadafan, Kazem Bitaghsir and Liu, Hong and Rajagopal, Krishna and Wiedemann, Urs Achim},
	year={2009},
	month=feb, pages={553–567} }

@article{Athanasiou_2010,
   title={Synchrotron radiation in strongly coupled conformal field theories},
   volume={81},
   ISSN={1550-2368},
   url={http://dx.doi.org/10.1103/PhysRevD.81.126001},
   DOI={10.1103/physrevd.81.126001},
   number={12},
   journal={Physical Review D},
   publisher={American Physical Society (APS)},
   author={Athanasiou, Christiana and Chesler, Paul M. and Liu, Hong and Nickel, Dominik and Rajagopal, Krishna},
   year={2010},
   month=jun }

@article{Herzog_2007,
   title={Spinning dragging strings},
   volume={2007},
   ISSN={1029-8479},
   url={http://dx.doi.org/10.1088/1126-6708/2007/10/087},
   DOI={10.1088/1126-6708/2007/10/087},
   number={10},
   journal={Journal of High Energy Physics},
   publisher={Springer Science and Business Media LLC},
   author={Herzog, Christopher P and Vuorinen, Aleksi},
   year={2007},
   month=oct, pages={087–087} }

@article{DeWolfe:2010he,
    author = "DeWolfe, Oliver and Gubser, Steven S. and Rosen, Christopher",
    title = "{A holographic critical point}",
    eprint = "1012.1864",
    archivePrefix = "arXiv",
    primaryClass = "hep-th",
    reportNumber = "COLO-HEP-559, PUPT-2360",
    doi = "10.1103/PhysRevD.83.086005",
    journal = "Phys. Rev. D",
    volume = "83",
    pages = "086005",
    year = "2011"
}
\end{document}